%% file: geo_spin_ecc_plunge_WF.tex
\documentclass[aps, prd, reprint, notitlepage, a4paper, floats, amsmath, amssymb, amsfonts,superscriptaddress,showpacs, showkeys,nofootinbib,longbibliography]{revtex4-2}

\pdfoutput=1

\usepackage{graphicx}
\usepackage[usenames,dvipsnames]{xcolor}
\usepackage[normalem]{ulem}

\definecolor{dodgerblue}{HTML}{1E90FF}
\definecolor{viennared}{HTML}{DA0A14}
\usepackage[
	bookmarksnumbered, bookmarksopen, bookmarksopenlevel=2,
	breaklinks=true,
	colorlinks=true, filecolor=viennared, citecolor=dodgerblue,
	linkcolor=viennared, urlcolor=dodgerblue, menucolor=viennared,
]{hyperref}
\usepackage[english]{babel}
\allowdisplaybreaks[1]
\usepackage[utf8]{inputenc}
\usepackage{lmodern}
\usepackage{booktabs}
\usepackage{diagbox}
\usepackage{dcolumn}
	\newcolumntype{.}{D{.}{.}{13}}
	\newcolumntype{d}[1]{D{.}{.}{#1}}
\allowdisplaybreaks

\graphicspath{{fig/}}

\newcommand{\AEI}{\affiliation{Max Planck Institute for Gravitational Physics (Albert Einstein Institute), Am M\"uhlenberg 1, Potsdam 14476, Germany}}
\newcommand{\Maryland}{\affiliation{Department of Physics, University of Maryland, College Park, MD 20742, USA}}
\newcommand{\COG}{\affiliation{Center of Gravity, Niels Bohr Institute, Blegdamsvej 17, 2100 Copenhagen, Denmark}}
\newcommand{\URI}{\affiliation{Department of Physics and Center for Computational Research, University of Rhode Island, Kingston, RI 02881, USA}}  
\newcommand{\UMassDPhy}{\affiliation{Department of Physics and Center for Scientific Computing and Data Science Research, University of Massachusetts, Dartmouth, MA 02747, USA}}
\begin{document}

\title{
	Peaking into the abyss: Characterizing the merger of equatorial-eccentric-geodesic plunges in rotating black holes
}
%%%%%%%%%%%%%%%%%%%%%%%%%%%%%%%%%%%%%%%%%%%%%%%%%%%% Title page %%%%%%%%%%%%%%%%%%%%%%%%%%%%%%%%%%%%%%%%%%%%%%%%%%%%

\author{Guglielmo Faggioli}\email{guglielmo.faggioli@aei.mpg.de}
\AEI
\author{Maarten van de Meent}
\COG
\AEI
\author{Alessandra Buonanno}
\AEI
\Maryland
\author{Gaurav Khanna}
\URI
\UMassDPhy
%\CSCVRUMass

\date{\today}

%%%%%%%%%%%%%%%%%%%%%%%%%%%%%%%%%%%%%%%%%%%%%%%%%%% Abstract %%%%%%%%%%%%%%%%%%%%%%%%%%%%%%%%%%%%%%%%%%%%%%%%%%%%%%
\begin{abstract}
	We study the gravitational waveforms generated by critical, equatorial plunging geodesics of the Kerr metric that start from an unstable-circular-orbit, which describe the test-mass limit of spin-aligned eccentric black-hole mergers.
	The waveforms are generated employing a time-domain Teukolsky code. We span different values of the Kerr spin $-0.99 \le a \le 0.99 $ and of the critical eccentricity $e_c$, for bound ($0 \le e_c<1$) and unbound plunges ($e_c \ge 1$). 
	We find that, contrary to expectations, the waveform modes $h_{\ell m}$ do not always manifest a peak for high eccentricities or spins.
	In case of the dominant $h_{22}$ mode, we determine the precise region
	of the parameter space in which its peak exists.
	In this region, we provide a characterization of the merger quantities of the $h_{22}$ mode and of the higher-order modes, providing the merger structure of the equatorial eccentric plunges of the Kerr spacetime in the test-mass limit. 
\end{abstract}

\maketitle

%%%%%%%%%%%%%%%%%%%%%%%%%%%%%%%%%%%%%%%%%%%%%%%%%%%%%%%

\section{Introduction}

The first detection of gravitational waves (GWs)~\cite{LIGOScientific:2016aoc} made by the LIGO-Virgo collaboration~\cite{LIGOScientific:2018mvr, LIGOScientific:2019lzm, LIGOScientific:2020ibl, LIGOScientific:2021usb} in 2015 represented a breakthrough in GWs astronomy. Since this event the LIGO-Virgo-KAGRA (LVK) collaboration~\cite{KAGRA:2013rdx, LIGOScientific:2021djp} detected about a hundred signals from the coalescences of binary compact objects, like stellar-origin black holes (BHs) and neutron stars~\cite{KAGRA:2021vkt} and is expected to add more than $ \sim 200$ signals by the end of the ongoing fourth observing run (O4).

Upcoming observational runs~\cite{KAGRA:2013rdx} as well as the forthcoming next-generation experiments, such as the Einstein Telescope~\cite{Punturo_2010,Abac:2025saz}, Cosmic Explorer~\cite{Evans:2021gyd} and LISA~\cite{LISA,LISA:2024hlh} will be characterized by higher sensitivity and by a number of detections increased by a factor $\sim 10^3$ and will be able to probe binary populations in lower frequency bands and for
smaller mass ratios, which can exhibit larger eccentricities.
Eccentricity plays a crucial role in understanding BH binaries (BBHs) formation channels~\cite{Mandel:2009nx, Rodriguez:2018rmd, Fragione:2018vty, Zevin:2021rtf} such as dynamical interactions in dense stellar environments~\cite{Kozai:1962zz, LIDOV1962719} and dynamical capture mechanisms~\cite{Samsing:2013kua, Zevin:2018kzq, Zevin:2021rtf,PortegiesZwart:1999nm, Miller:2001ez}. Moreover, it has been shown that the effects of eccentricity in GWs cannot be ignored since even when the eccentricities are small (e.g., $e \sim 0.1$ at 20 Hz~\cite{Divyajyoti:2023rht}) systematic biases arise in the extraction of source properties in inference studies~\cite{Gupte:2024jfe, Divyajyoti:2023rht, Favata:2013rwa, Ramos-Buades:2019uvh, Cho:2022cdy, Guo:2022ehk, GilChoi:2022waq, Das:2024zib} and tests of General Relativity~\cite{Saini:2022igm, Saini:2023rto, Narayan:2023vhm, Gupta:2024gun, Shaikh:2024wyn, Bhat:2022amc, Bhat:2024hyb}. These facts motivate the development of fast and accurate waveform models that incorporate eccentricity as an inference parameter.

A common approach in GW modeling is to employ results from numerical-relativity (NR) simulations to construct models by interpolating NR data to either fully describe the waveform~\cite{Blackman:2015pia,Varma:2018mmi,Varma:2019csw,Yoo:2023spi} or to model specific portions of it, in particular the merger and ringdown parts, across the parameter space. This procedure allows to incorporate the complex dynamics of binary systems, particularly in regions where analytical methods are less effective. Other approaches blend NR and BH perturbation theory data~\cite{Rifat:2019ltp, Islam:2022laz, Rink:2024swg} to develop models that can handle a wide range of mass-ratios.

While the quasi-circular (QC) inspiral-merger-ringdown morphology of the waveforms has been extensively studied and modeled in the last decades, a strong effort is currently on going to characterize and model the features of the gravitational waveforms emitted by eccentric BBHs. In particular, an extension from QC to eccentric merger and post-merger parts of the waveforms are of interest for models that combine analytical and numerical methods to model the entire GW signals. Among them the models based on the effective-one-body (EOB) formalism~\cite{Buonanno:1998gg, Buonanno:2000ef} that stand out for their ability to combine analytical approximation methods with NR results, achieving both high accuracy and computational efficiency for the QC systems~\cite{Damour:2008gu, Pan:2010hz, Pan:2013rra, Taracchini:2013rva, Bohe:2016gbl, Nagar:2018zoe, Cotesta:2018fcv, Babak:2016tgq, Ossokine:2020kjp, Nagar:2018gnk, Nagar:2020pcj, Riemenschneider:2021ppj, Pompili:2023tna}, and recently for eccentric systems~\cite{Khalil:2021txt, Ramos-Buades:2021adz, Gamboa:2024hli, Gamboa:2024imd, Bini:2012ji, Chiaramello:2020ehz, Nagar:2021gss, Albanesi:2021rby, Placidi:2021rkh, Nagar:2021xnh, Albanesi:2022ywx, Albanesi:2022xge, Nagar:2022fep, Albanesi:2023bgi, Placidi:2023ofj, Nagar:2024dzj, Nagar:2024oyk, Hinderer:2017jcs, Cao:2017ndf, Liu:2019jpg, Liu:2021pkr, Liu:2023dgl}. These models currently incorporate eccentricity only in the inspiral part and employ QC merger-ringdown models.

In this family of models, specifically for the QC case, the ringdown portion of the waveform is modelled by fitting an ansatz to the NR ringdown and then by performing a hierarchical fit of the resulting coefficients as functions of the system's intrinsic parameters, like the symmetric mass ratio and the spins of the BBH. To accurately model the final part of the waveform, the ringdown must be properly connected to the inspiral segment, which is computed using resummed post-Newtonian (PN) expressions of the waveform modes~\cite{Damour:2007xr}. This attachment procedure is performed by applying continuity conditions around the merger part of the waveform in order to ensure continuity between the inspiral waveform modes and the ringdown model fitted from NR data. 

For spinning BBHs with small mass ratios, the merger properties can be studied using BH perturbation theory (BHPT) results. This is done by considering waveforms generated by solving the Teukolsky equation~\cite{Teukolsky:1973ha} for a test mass (TM) orbiting a Kerr BH. Results from the TM limit have already been applied to inform and improve the accuracy of waveform models on the entire mass ratio regimes, especially for the family of EOB models~\cite{Nagar:2006xv, Damour:2007xr, Damour:2008gu, Barausse:2011kb, Taracchini:2013wfa, Taracchini:2014zpa, Albanesi:2021rby, Albanesi:2022ywx, Albertini:2022rfe, Albertini:2022dmc, vandeMeent:2023ols, Albertini:2023aol, Albanesi:2023bgi, Albertini:2024rrs, Faggioli:2024ugn, Albanesi:2024fts, Leather:2025nhu}. 
%\comm{AB: I would erase the rest of this pararaph because we are not very specific on what those papers did with respect to the others. It seems to me a repetition.}
Among these works, Refs.~\cite{Barausse:2011kb, Taracchini:2013wfa, Taracchini:2014zpa, Albanesi:2021rby, Albanesi:2022ywx, Albanesi:2023bgi} shown the importance of modeling the merger and post-merger stages of the waveforms in the TM limit. In particular, in the QC case, Refs.~\cite{Barausse:2011kb, Taracchini:2014zpa} studied the merger and post-merger properties of waveforms generated with equatorial QC trajectories of a small mass orbiting a Kerr BH. These studies enriched the merger-ringdown parts of EOB models in the small mass ratio regimes.

To further improve eccentric waveform models, recent works started extending the characterization and modeling of the merger-ringdown features of eccentric systems both in the comparable and small mass-ratio regimes. Among these studies we mention Refs.~\cite{Carullo:2023kvj, Carullo:2024smg}, which characterized the merger properties of eccentric waveforms and the ringdown amplitudes in the comparable mass case using eccentric NR waveforms of the RIT catalog~\cite{Healy:2022wdn} and of the SXS collaboration~\cite{Chu:2009md, Lovelace:2010ne, Lovelace:2011nu, Buchman:2012dw, Hemberger:2013hsa, Scheel:2014ina, Blackman:2015pia, Lovelace:2014twa, Mroue:2013xna, Kumar:2015tha, Chu:2015kft, Boyle:2019kee}, together with numerical results from BHPT~\cite{Albanesi:2023bgi}. More recently, Ref.~\cite{Nee:2025zdy} investigated the impact of the relativistic anomaly on the merger features for the comparable-mass regime.
Effort has also been pursued in the small mass-ratio regime. In particular, Ref.~\cite{Albanesi:2023bgi} developed a non-spinning EOB ringdown model in the TM limit that takes into account residual eccentricity at merger. This work uses EOB trajectories to source the Teukolsky equation and generate the waveforms. A more recent study~\cite{Becker:2024xdi} investigated the merger properties of the waveforms generating the trajectories with numerical fluxes computed through the frequency-domain Teukolsky code~\cite{Hughes:2021exa}. 

In these EOB models, the standard procedure has been to identify the merger with the peak of the gravitational dominant mode. More specifically, the time at which the inpiral-plunge waveform is attached to the merger-ringdown one is identified with the mode-peak's time, both in the comparable and small mass-ratio regimes~\cite{Cotesta:2018fcv, Pompili:2023tna, Riemenschneider:2021ppj} (see also Refs.~\cite{Baker:2008mj, Kelly:2011bp, Healy:2014eua, Healy:2017mvh, Healy:2016lce, Keitel:2016krm}). However, it has been shown that for QC small mass-ratio scenarios with high prograde spin of the central Kerr BH (e.g., $a \ge 0.95$), these peaks are not resolvable~\cite{Taracchini:2014zpa} or, in the near-extremal Kerr scenarios, they happen during the inspiral part of the waveform, hence many cycles before the plunge and merger part~\cite{Gralla:2016qfw}. These results show that the waveform modes' peak is not necessarily associated with the merger part of the GWs signal in all the parameter space.

Inspired by these works, in this article, we characterize the late-stage properties of the waveforms of eccentric align-spin orbits of Kerr by introducing a model-independent methodology. Since we are interested in the TM limit of the final segment of these waveforms, we focus on the signal produced during the plunge phase and, thus, study waveforms associated with plunges and mergers in the TM limit. As the mass ratio vanishes, plunges become increasingly geodesic~\cite{Buonanno:2000ef, Ori:2000zn}. Recent works extended the features of the inspiral to plunge transition for small mass-ratio eccentric systems~\cite{Becker:2024xdi, Lhost:2024jmw}. In particular, Ref.~\cite{Lhost:2024jmw} provided analytical expressions for the energy and the angular-momentum of a TM orbiting a Schwarzschild BH for the eccentric transition to plunge, when the system is near the last-stable-orbit (LSO). Interestingly, these expressions show that the plunges approach a specific class of geodesics that start from an unstable-circular-orbit (UCO) in the infinite past. Analytical, closed-form expressions for this class of geodesic plunges have been derived in Refs.~\cite{Mummery:2023hlo,Dyson:2023fws}.
In this work, we use these analytical expressions to compute a set of this class of geodesic eccentric plunges for different values of the radius of the UCO from which they start and for different spins of the central Kerr BH. With these trajectories, we source a time-domain (TD) Teukolsky code~\cite{Sundararajan:2007jg,Sundararajan:2008zm,Sundararajan:2010sr,Zenginoglu:2011zz,Field:2020rjr} and produce the gravitational waveforms emitted by the trajectories. This allows us to analyze and characterize the properties at the merger in the parameter space of the TM eccentric aligned Kerr waveforms.

This article is structured as follows. In Sec.~\ref{Sec.: methodology}, we describe the methodology of our work. In particular, in Sec.~\ref{Sec.:critical plunge geodesics}, we introduce and describe the class of critical plunge geodesics we consider for generating the waveforms. Section~\ref{Sec.: waveforms computation} describes how we generate the gravitational waveforms using the TD Teukolsky code. In Sec.~\ref{Subsec: Waveforms characterization}, we study the existence of the peak of the $(2,2)$ spherical harmonic mode of the strain $h$ and provide an overview of the characterization of the waveforms. Section~\ref{Sec: results} is dedicated to the main results of our work. In particular, in Sec.~\ref{Subsec: ell = 2 m = 2 mode characterization}, we provide a characterization of the quantities, commonly considered in GW models, evaluated at the peak of the amplitude of the $(2,2)$ mode. In Sec.~\ref{Subsec: higher order modes characterization}, we show similar results for the higher-order modes, focusing in particular on the $(3,3)$ mode, and in Sec.~\ref{Sec.: Bondi news and psi_4 peaks characterization} we comment on the phenomenology of the peak of the amplitude of other two quantities related to the strain, the Bondi news and the Weyl scalar. Finally, in Sec.~\ref{Sec: conclusions}, we discuss the main conclusions of our study. 
In Appendix~\ref{Sec.: appendix why critical plunges?} we further motivate restricting our attention to critical plunge geodesics 
while in Appendix~\ref{Sec.: convergence with mass ratio}, we show how the waveforms generated through these geodesics represent the TM limit of the generic eccentric plunge. Appendix~\ref{Sec.: Comparison of Delta t_22 with previous works} connects our work to the QC case studied in a previous study~\cite{Taracchini:2014zpa} 
while in Appendix~\ref{Sec.: circularization behaviour} we show how the critical plunge geodesics behave more circular as their critical eccentricity increases. 
In Appendix~\ref{Sec.: ell = m higher order modes similar behaviour}, we show the features at merger of the higher order modes, 
and in Appendix~\ref{Sec.: l=2 m=2 merger values} we list some of the values at merger of the quantities commonly considered in waveform modeling  for the $(2,2)$ mode.

\subsection*{Notation}
In this work we adopt geometric units $G = c = 1$ and consider a Kerr BH of mass $M$ with dimensionless spin $a = J/M^2$. The Kerr metric is expressed in Boyer-Lindquist coordinates $\{\tilde{t}, \tilde{r}, \theta, \varphi \}$ and we consider scaled dimensionless variables $t=\tilde{t}/M$ and $r=\tilde{r}/M$. We restrict our analysis to the equatorial plane, $\theta = \pi/2$. For the ease of notation we set the mass of the Kerr BH to $M=1$. 

\section{Methodology} \label{Sec.: methodology}
\begin{figure}[tp!]
  	\includegraphics[width=1.\linewidth]{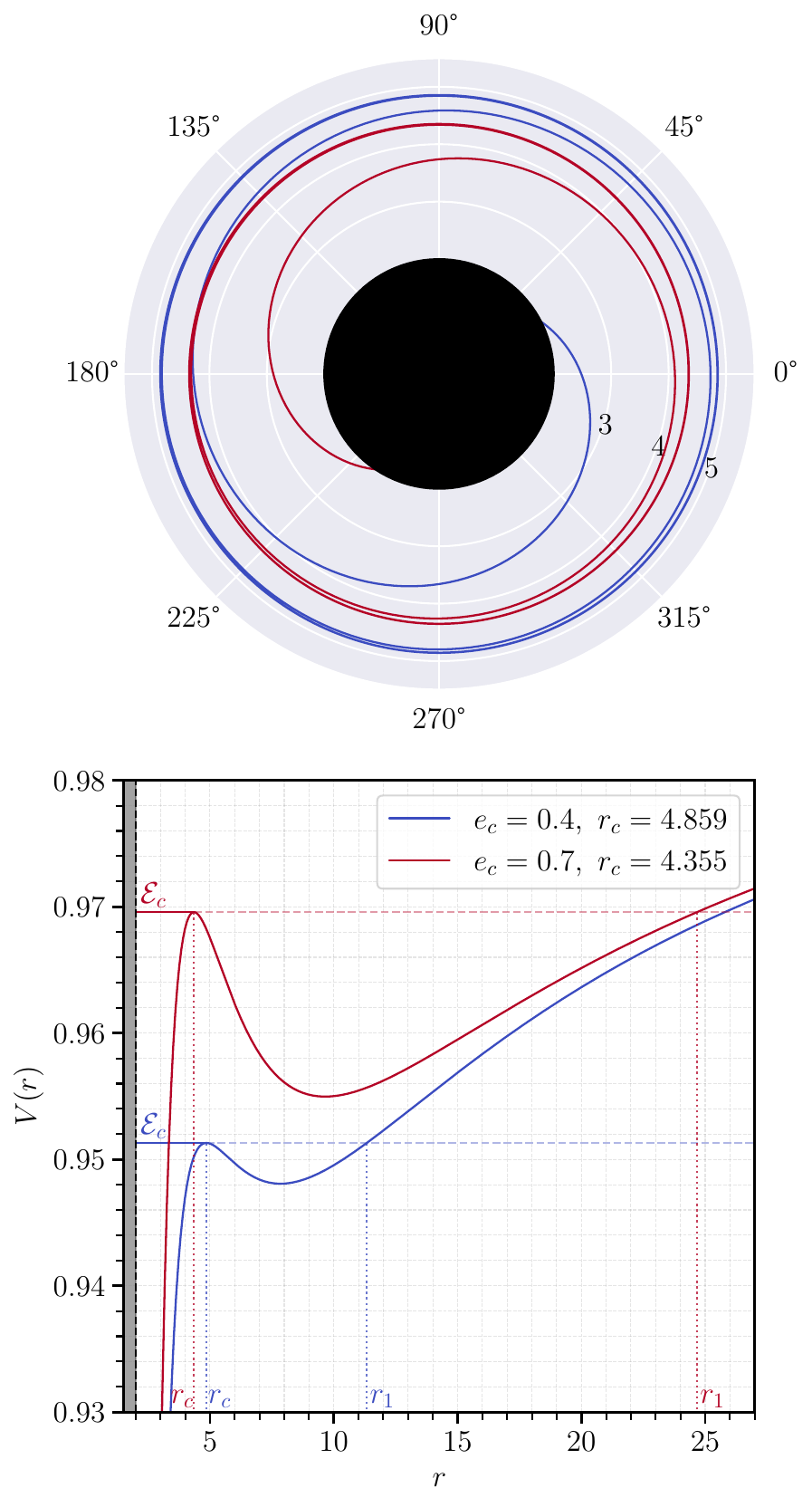}
  	\caption{In the upper panel we show two critical plunge geodesics for a Kerr BH with spin $a=0$. This class of geodesics is characterized by the value of the critical eccentricity $e_c$, as described in Eq.~\eqref{eq: e_LSO def}. In the figure we show two geodesics with $e_c=\{ 0.4, \ 0.7 \}$. The bottom panel shows the radial potentials $V(r)$ of the two geodesics and their energies (horizontal lines), which correspond to the energies $\mathcal{E}_c = \{ 0.95130, \ 0.96957 \}$ of their respective UCOs. We also mark the positions of the critical radii $r_c = \{4.859, \ 4.355 \}$ and of the outer turning point $r_1 = \{ 11.3, \ 24.7 \}$ with vertical dotted lines. The vertical gray area is the area inside the BH horizon.}
  	\label{fig:inst_plot_1}
\end{figure}
This section describes the methodology of our work. We first provide a review of the plunging geodesics that we consider to generate the waveforms and we describe how these waveforms are computed at future-null infinity by sourcing the time-domain Teukolsky code with the plunging geodesics. We also describe how we characterize the merger part of the different waveforms.
\subsection{Critical plunge geodesics} \label{Sec.:critical plunge geodesics}
In the equatorial plane, the Kerr metric~\cite{Kerr:1963ud} expressed in dimensionless Boyer-Lindquist coordinates takes the form
\begin{equation}
% & ds^2 = g_{\mu \nu} dx^{\mu}dx^{\nu} \ , 
    ds^2 = g_{tt} dt^2 + g_{rr} dr^2 + g_{\varphi \varphi} d\varphi^2 + g_{t \varphi} dt d\varphi \ , 
\end{equation}
with
	\begin{subequations}
    \begin{align}
    g_{tt} &= -1 + \frac{2}{r} \ , \\
    g_{rr} &= \frac{r^2}{\Delta} \ , \\
    g_{\varphi \varphi} &= r^2 + a^2 + \frac{2 a^2}{r} \ , \\
    g_{t \varphi} &= - \frac{4 a}{r} \ ,
    %g_{tt} &= -1 + \frac{2}{r} \ , \quad g_{rr} = \frac{r^2}{\Delta} \ , \\
    %g_{\varphi \varphi} &= r^2 + a^2 + \frac{2 a^2}{r} \ , \quad g_{t \varphi} = - \frac{4 a}{r} \ ,
    \end{align}
    \end{subequations}
and 
	\begin{equation}
    \Delta = r^2 - 2r + a^2 \ .
    \end{equation}
The evolution of equatorial geodesics in Kerr spacetime is commonly described by leveraging the existence of three constants of motion: the particle’s mass $\mu$, the orbital energy $\mathcal{E}$, and the orbital angular momentum $\mathcal{L}$. When considering these constants the equatorial geodesics in Kerr spacetime satisfy the following equations~\cite{Carter:1968rr}
\begin{subequations}\label{eq:geoeom}
\begin{align}
	\label{eq:radialeom}
		\left(\frac{dr}{d\lambda}\right)^2 & = (\mathcal{E}(r^2+a^2)-a\mathcal{L})^2-\Delta(r^2+(a\mathcal{E}-\mathcal{L})^2), \\
		% &= (1-\mathcal{E}^2)(r_1-r)(r_2-r)(r_3-r)r \\
		% &= R(r),\\
\label{eq:timeeom}
		\frac{dt}{d\lambda} & = \frac{(r^2+a^2)}{\Delta}(\mathcal{E}(r^2+a^2) a\mathcal{L})+a(\mathcal{L}-a\mathcal{E}),
\intertext{and}
\label{eq:azimuthaleom}
		\frac{d\varphi}{d\lambda} & = \frac{a}{\Delta}(\mathcal{E}(r^2+a^2)-a\mathcal{L})+\mathcal{L}-a\mathcal{E},
\end{align}
\end{subequations}
where $\lambda$ is the Mino time variable, which is related to proper time $\tau$ through
\begin{equation}
	\frac{d\tau}{d\lambda} = r^2.
\end{equation}
The right hand side of Eq.~\eqref{eq:radialeom} is a 4th order polynomial in the variable $r$, and can always be written in the form~\cite{Fujita:2009bp}
\begin{equation}
	\label{eq:radialeom with roots}
\left(\frac{dr}{d\lambda}\right)^2=(1-\mathcal{E}^2)(r_1-r)(r_2-r)(r_3-r)r = R(r),
\end{equation}
where $r_1$, $r_2$ and $r_3$ are roots of $R(r)$. 
In any situation where these roots of $R(r)$ are real and satisfy
\begin{equation}
	\frac{1}{r_1} < \frac{1}{r_2} < \frac{1}{r_3} < \frac{1}{r_{+}},
\end{equation}
where $r_{+} = 1+\sqrt{1-a^2}$ is the outer event horizon of the Kerr metric , then, there will be a geodesic solution satisfying
\begin{equation} 
	% \frac{1}{r_1} < \frac{1}{r} < \frac{1}{r_2},
	r_2 < r(\lambda) < r_1.
\end{equation}
In these situations, the roots $r_1$ and $r_2$ are respectively named \textit{apocenter} and \textit{pericenter}
and the geodesic solution can be described with the Keplerian parametrization in terms of the eccentricity $e$, the semilatus rectum $p$, and the relativistic anomaly $\xi$, \cite{Darwin:1959,Darwin:1961},  
\begin{align} \label{eq.: e, p, xi def}
	e &= \frac{r_1 - r_2}{r_1 + r_2}, \quad  
	p = \frac{2r_1 r_2}{r_1 + r_2}, \quad  
	\cos \xi = \frac{p - r}{e r}. 
\end{align}
Note, that when $r_1>0$ these geodesics describe bound orbits with $0\le e < 1$, and when $r_1<0$ they describe scattering orbits with $e>1$.

In this work we are interested in the study of the plunge-merger properties of GWs. As we will show in Sec.~\ref{Sec.: waveforms computation}, we generate these waveforms by sourcing and numerically solving the Teukolsky equation in the time-domain with plunge geodesics of the Kerr metric in the equatorial eccentric case.
While plunge geodesics (i.e., geodesics crossing the future (outer) event horizon at $r = r_{+}$) exist for any value of $\mathcal{E}$ and $\mathcal{L}$, in this work, we will be interested in a special class of orbits in which all four roots of $R(r)$ are real and satisfy
\begin{equation}\label{eq:LSO}
	\frac{1}{r_1} < \frac{1}{r_2} = \frac{1}{r_3} < \frac{1}{r_{+}}.
\end{equation}

When this is true, the radial equation allows three distinct type of solutions: (i) an UCO at $r=r_2=r_3=r_c$; (ii) so called ``homoclinic'' orbits \cite{Levin:2008yp,Perez-Giz:2008ajn} approaching $r_{c}=r_2=r_3$ from the outside in the infinite future and/or past; and (iii) critical plunge geodesics approaching $r_c=r_2=r_3$ from the inside in the infinite past and crossing the future outer event horizon at $r=r_{+}$.\footnote{Technically, there also exists the time reverse of a critical plunge: a geodesic crossing the past event horizon and approaching $r_2=r_3=r_c$ in the infinite future.} These solutions are of interest because they describe the TM limit of the transition from (bound) inspiral motion to plunging motion. Consequently, the locus of points in the configuration space of geodesics satisfying Eq.~\eqref{eq:LSO} is variably known as the separatrix (between bound and plunge orbits) or the LSO.
In this work we are particularly interested in the critical plunge geodesics since they describe the final part of the trajectory of an extreme mass-ratio inspiral~\cite{Buonanno:2000ef, Ori:2000zn, Lhost:2024jmw}. In Appendix~\ref{Sec.: appendix why critical plunges?} we phenomenologically illustrate this.

The repeated root $r=r_2=r_3$ in $R(r)$ guarantees that the equations for critical plunge geodesics can be solved explicitly in closed form in terms of elementary functions \cite{Mummery:2023hlo,Dyson:2023fws}.
The critical plunge geodesics in a Kerr spacetime with fixed spin $a$ form a one parameter family. A possible choice is to parametrize them with the radius $r_{c}$ of the associated UCO (henceforth critical radius), which can take any value between the light ring radius~\cite{Bardeen:1972fi}
\begin{align}
%r_{\rm LR} &= 2+\cos\alpha -\sqrt{3}\sin\alpha,\quad\text{with}\\
%\alpha &= \frac{2}{3}\arcsin a,
r_{\rm LR} &= 2+\cos\left( \frac{2}{3}\arcsin a \right) -\sqrt{3}\sin\left(\frac{2}{3}\arcsin a\right),
\end{align}
 and the radius of the inner most stable circular orbit (ISCO) \cite{Bardeen:1972fi},
\begin{subequations}
\begin{align}
%r_{\rm isco} &= 3+Z_2- a\sqrt{\frac{(3-Z_1)(3+Z_1+Z_2)}{a^2}} ,
%\quad\text{with}\\
%Z_1 &= 1+\sqrt[3]{1-a^2}\left(\sqrt[3]{1+a}+\sqrt[3]{1-a}\right),
%\quad\text{and}\\
%Z_2 &= \sqrt{3a^2+Z_1^2}.
r_{\rm isco} &= 3+Z_2- a\sqrt{\frac{(3-Z_1)(3+Z_1+2Z_2)}{a^2}} , \\
Z_1 &= 1+\sqrt[3]{1-a^2}\left(\sqrt[3]{1+a}+\sqrt[3]{1-a}\right),\\
Z_2 &= \sqrt{3a^2+Z_1^2}.
\end{align}
\end{subequations}
That is
\begin{equation}
r_{\rm LR} < r_c \le r_{\rm isco}.
\end{equation}
Keeping in mind that these plunge geodesics have the same constants of motion as the corresponding UCO~\cite{Bardeen:1972fi}, the energy and angular momentum of critical plunge geodesics can be expressed in terms of the parameter $r_c$,
\begin{subequations}
\begin{align}
\label{eq:energy critical}
	\mathcal{E}_c &= \frac{(r_c-2)\sqrt{r_c}+a}{\sqrt{(r_c-3)r_c^2+2a r_c^{3/2}}}, \\
\label{eq:angular momentum critical}
	\mathcal{L}_c &= \frac{r_c^2-2 a \sqrt{r_c}+a^2}{\sqrt{(r_c-3)r_c^2+2a r_c^{3/2}}}.
\end{align}
\end{subequations}
Instead of using $r_c$ to parametrize the critical plunge geodesics, it can be more physically insightful to parametrize them using the eccentricity of the corresponding homoclinic orbit outside the UCO~\cite{Levin:2008yp}. If we consider the definition of the eccentricity parameter $e$ in Eq.~\eqref{eq.: e, p, xi def} and we consider the critical limit in Eq.~\eqref{eq:LSO}, for which $r_2=r_3=r_c$, we can define the critical eccentricity
\begin{equation} \label{eq: e_LSO def}
% e_c = \frac{r_1-r_c}{r_1+r_c},
e_c = \frac{r_1-r_c}{r_1+r_c},
\end{equation}
which corresponds to the eccentricity of the homoclinic orbit outside of the UCO \footnote{In Ref.~\cite{Levin:2008yp} the quantity that we define as $e_c$ corresponds to $e^{\rm hc}$. 
In our work we prefer to use the nomenclature $e_c$ to connect with the fact that we consider the class of critical plunge geodesics rather than homoclinic orbits.}.
Using the expressions for $\mathcal{E}_c$ and $\mathcal{L}_c$ in Eqs.~\eqref{eq:energy critical}-~\eqref{eq:angular momentum critical}, we can solve for the roots of Eq.~\eqref{eq:radialeom} to find
\begin{equation}
r_1 = \frac{2r_c (\sqrt{r_c}-a)^2}{(r_c-4)r_c+ 4a\sqrt{r_c}-a^2},
\end{equation}
and get
\begin{equation} \label{Eq.: e_LSO vs rc_c}
%e_c = \frac{r_c(6-r_c)- 8a\sqrt{r_c}+3a^2}{r_c(r_c-2)+a^2}.
e_c = \frac{r_c(6-r_c)- 8a\sqrt{r_c}+3a^2}{r_c(r_c-2)+a^2}.
\end{equation}

For critical radii $r_c$ between the radius of the innermost bound circular orbit (IBCO) \cite{Bardeen:1972fi}
\begin{equation}
r_{\rm ibco} = 2 -a +2\sqrt{1-a},
\end{equation}
and $r_{\rm isco}$, the homoclinic orbit outside the UCO is a bound orbit and $e_c$ varies between 0 (at $r_{\rm isco}$) and 1 (at $r_{\rm ibco}$). For $r_{\rm LR}< r <r_{\rm ibco}$ it is a scattering orbit with $1<e_c<3$. In this work we also consider configurations for which $e_c>1$, as we will show in Sec.~\ref{Sec: results}. In the upper panel of Fig.~\ref{fig:inst_plot_1} we show two examples of critical plunge geodesics on the equatorial plane of a Kerr BH with $a=0$. The two geodesics have $e_c=0.4$ (blue curve) and $e_c=0.7$ (red curve). The bottom panel of Fig.~\ref{fig:inst_plot_1} shows the radial potentials $V(r)$ of the geodesics. The radial potential $V(r)$ is defined by inverting Eq.~\eqref{eq:radialeom} for the energy $\mathcal{E}$ at the turning points, i.e. when $dr/d\lambda = 0$. Hence it is $V(r)=\mathcal{E}(r, dr/d\lambda = 0, \mathcal{L}, a)$. In the bottom panel we also show the energies of the geodesics (horizontal lines), that are the same of the respective energies of the UCOs. From this panel it is possible to notice that the critical radius $r_c$ decreases as $e_c$ increases, as we will make explicit in Fig.~\ref{ecc_threshold_values_extraction_methods_compared} for a particular spin configuration.

To understand the behaviour of the critical plunge geodesics near the critical radius $r_{c}$ we can combine Eq.~\eqref{eq:radialeom} and Eq.~\eqref{eq:timeeom} with the expressions for $\mathcal{E}$ and $\mathcal{L}$ and expand around $r_{c}$ to find,

\begin{equation}
\left(\frac{d(r_{c}-r)}{d\lambda}\right)^2 = \Lambda_c^2 (r_{c}-r)^2 + \mathcal{O}(r_{c}-r)^3,
\end{equation}
with
\begin{align}
\Lambda_c &=  \frac{\sqrt{r_c(6-r_c)-4 a \sqrt{r_c}+3a^2}}{r_c(r_c^{3/2}+a)}.
\end{align}
For a plunging solution we should have $r < r_c$ and $d(r_{c}-r)/d\lambda >0$, imposing this we obtain the approximate solution
\begin{equation} \label{Eq.: asymptotically rc}
	r_{c}-r \propto \exp\left(\Lambda_c t\right),
\end{equation}
i.e., the solution asymptotically approaches $r_c$ as the coordinate time $t\to -\infty$.

For any geodesic the instantaneous angular velocity is given by
\begin{equation} \label{Eq.: orbitalfreq}
	\Omega = \frac{d\varphi}{dt} = \frac{b(r-2)+2a}{r^3+a^2(r +2) - 2 a b},
\end{equation}
where $b=\mathcal{L}/\mathcal{E}$ is the generalized impact parameter, which for critical plunge geodesics is given by
\begin{equation}
	b_{c} = \frac{r_{c}^2 -2a\sqrt{r_{c}} + a^2}{(r_{c}-2)\sqrt{r_{c}} +a }.
\end{equation} 
In the infinite past $\Omega$ asymptotes to the angular velocity of the UCO,
\begin{equation}
\Omega_c = \frac{1}{r_c^{3/2}+a}.
\end{equation}

The angular velocity $\Omega$ will take its maximal value at
\begin{align}
	r_{\Omega_{\rm peak}} &= \Bigl(1-\frac{a}{b_c}\Bigr)\Bigl(1+2\cos\psi \Bigr),\quad\text{with}\\
	\psi &= \frac{1}{3}\arccos\Bigl[1-\frac{
		a b_c^2}{2(b_c-a)}\Bigr].
\end{align}
One can show that this is always smaller than the light ring radius $r_{\rm LR}$ and therefore always smaller than $r_c$; critical plunge geodesics always achieve a maximal orbital frequency along the way down.

\subsection{Waveforms computation} \label{Sec.: waveforms computation}
The class of plunge geodesics we consider in this article can be interpreted as the plunges of a small mass $\mu$ into a Kerr BH of mass $M$ in the limit $\mu/M \rightarrow 0$. Hence, we can use BH perturbation theory as the framework to produce the gravitational waveforms sourced by critical plunge geodesics, numerically.
We compute the waveforms by solving the Teukolsky master equation~\cite{Teukolsky:1973ha}, which in Boyer-Lindquist coordinates reads
  \begin{equation} \label{Teukolsky equation}
    \begin{aligned}
      % & - \left[ \frac{(r^2 + a^2)^2}{\Delta} -a^2\sin{\theta}^2 \right]\partial_{tt}\Psi - \frac{4Mar}{\Delta} \partial_{t \varphi} \Psi \\
      % & - 2s \left[ r - \frac{M(r^2 - a^2)}{\Delta} + ia\cos{\theta} \right] \partial_{t} \Psi + \Delta^{-s} \partial_{r}(\Delta^{s+1} \partial_r \Psi) \\
      % & + \frac{1}{\sin{\theta}} \partial_{\theta} (\sin{\theta} \partial_{\theta} \Psi) + \left[ \frac{1}{\sin{\theta}^2} - \frac{a^2}{\Delta} \right]\partial_{\varphi \varphi} \Psi \\
      % & + 2s \left[ \frac{a(r - M)}{\Delta} + \frac{i\cos{\theta}}{\sin{\theta}^2} \right] \partial_{\varphi} \Psi - (s^2 \cot{\theta}^2 - s) \Psi \\
      % & = -4 \pi (r^2 + a^2 \cos{\theta}^2) T \ .
      & - \left[ \frac{(r^2 + a^2)^2}{\Delta} -a^2\sin{\theta}^2 \right]\partial_{tt}\Psi - \frac{4ar}{\Delta} \partial_{t \varphi} \Psi \\
      & - 2s \left[ r - \frac{(r^2 - a^2)}{\Delta} + ia\cos{\theta} \right] \partial_{t} \Psi + \Delta^{-s} \partial_{r}(\Delta^{s+1} \partial_r \Psi) \\
      & + \frac{1}{\sin{\theta}} \partial_{\theta} (\sin{\theta} \partial_{\theta} \Psi) + \left[ \frac{1}{\sin{\theta}^2} - \frac{a^2}{\Delta} \right]\partial_{\varphi \varphi} \Psi \\
      & + 2s \left[ \frac{a(r - 1)}{\Delta} + \frac{i\cos{\theta}}{\sin{\theta}^2} \right] \partial_{\varphi} \Psi - (s^2 \cot{\theta}^2 - s) \Psi \\
      & = -4 \pi (r^2 + a^2 \cos{\theta}^2) T \ .
    \end{aligned}
  \end{equation}
  This equation describes the evolution of scalar, vector, and tensor perturbations of a Kerr BH.
The parameter $s$ is the \textit{spin weight} of the field. In particular, when $s = \pm 2$ the equation describes radiative degrees of freedom, and for $s = -2$ it is $\Psi = (r - ia\cos{\theta})^4\psi_4$, where 
  $\psi_4$ is the Weyl curvature scalar that describes outgoing GWs.
  
  A system composed of a TM orbiting a Kerr BH is interpreted as a perturbed Kerr metric, and within this interpretation the source term $T$ on the right-hand side of Eq.~\eqref{Teukolsky equation}
  describes a TM moving in the Kerr spacetime. 
  The source term $T$ of a TM orbiting a Kerr BH is constructed from Dirac-delta functions of the variables $r$ and $\theta$, as well as first and second derivatives of the delta functions in these variables. 
  These terms are sourced at the location of the TM, hence the source $T$ depends on the trajectory that the TM follows in the Kerr spacetime. Full details on construction of the TD source term can be found in Refs.~\cite{Sundararajan:2008zm, Sundararajan:2010sr}.
  In this analysis, the trajectories used to source the term $T$ are the critical plunge geodesics introduced in Sec.~\ref{Sec.:critical plunge geodesics}.
  
To solve Eq.~\eqref{Teukolsky equation}, we make use of the finite difference TD code developed in Refs.~\cite{Sundararajan:2007jg, Sundararajan:2008zm,Sundararajan:2010sr,Zenginoglu:2011zz,Field:2020rjr}, which employs hyperboloidal time slicing to extract the GW modes at future-null infinity.  
  At future null infinity, the Weyl scalar $\psi_4$ and the waveform strain $h = h_{+} - i h_{\times}$ are related by the expression
  \begin{equation} \label{Eq.: psi_4 definition}
    \psi_4 = \frac{1}{2}\frac{\partial^2 h}{\partial t^2} = \frac{1}{2} \left( \frac{\partial^2 h_{+}}{\partial t^2} - i\frac{\partial^2 h_{\times}}{\partial t^2} \right) \ ,
  \end{equation}
  and following standard practices, the waveform strain is decomposed in $-2$-spin-weighted spherical harmonics
  \begin{equation} \label{Eq.: strain spherical harmonic decomposition}
    h = \sum_{\ell, m} {}_{-2}Y_{\ell m}(\theta, \varphi)\,h_{\ell m}  \ ,
  \end{equation} 
  where $h_{\ell m}$ are the (spherical harmonic) gravitational-waveform  modes.
More details on the numerical accuracy of the TD Teukolsky code can be found in Refs.~\cite{Sundararajan:2007jg, Sundararajan:2008zm, Sundararajan:2010sr,Zenginoglu:2011zz, Barausse:2011kb, Field:2020rjr}.

\subsection{Waveforms characterization} \label{Subsec: Waveforms characterization}
\begin{figure*}[t!]
  	\includegraphics[width=1.\linewidth]{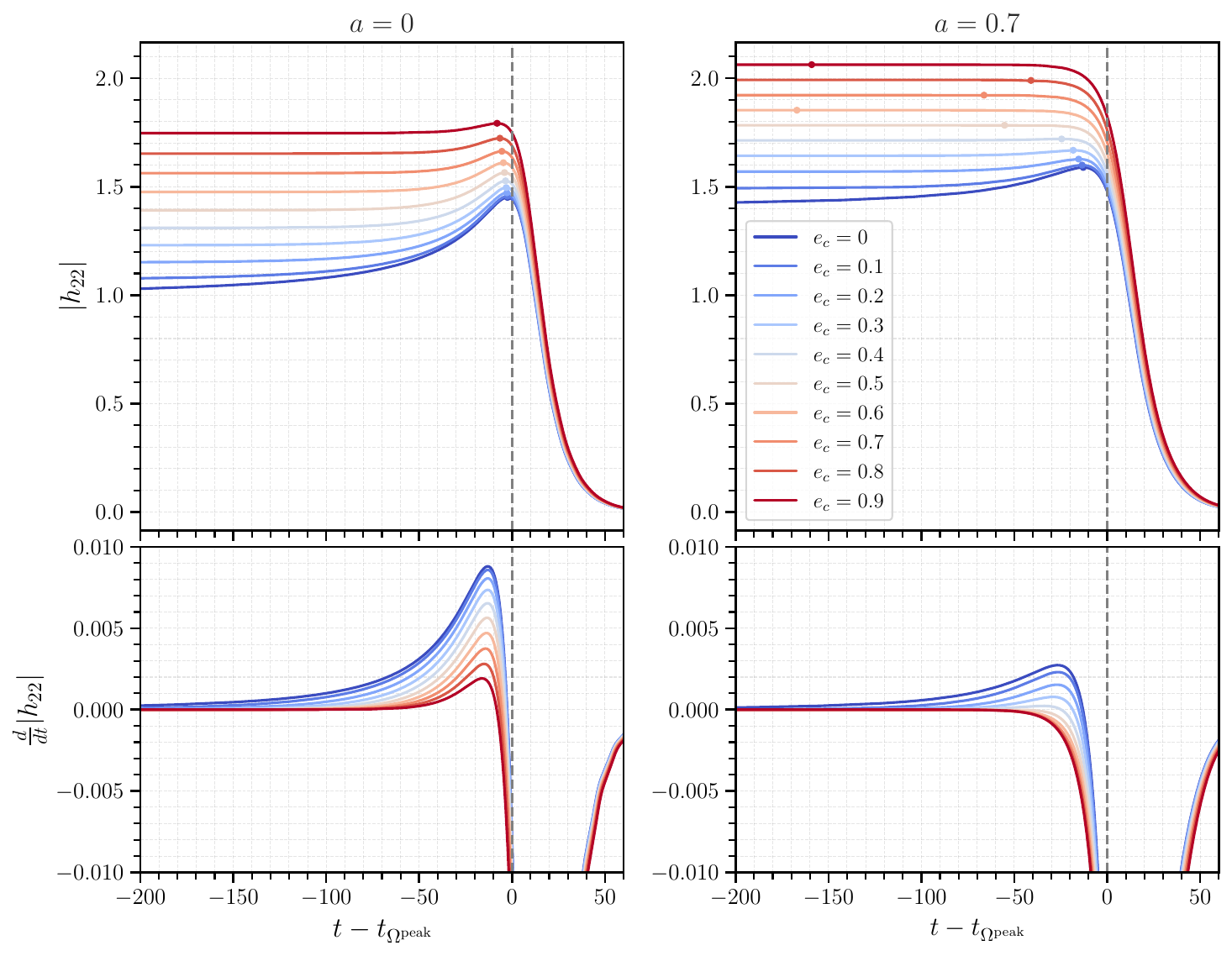}
  	\caption{In this figure we show the gravitational-waveform amplitudes of the $h_{22}$ mode generated by equatorial critical plunge  geodesics of a Kerr BH. The left column specializes to the non-spinning case ($a = 0$), while the right column to the case with spin $a = 0.7$. The upper panels show the amplitude of the mode while the bottom panels show its first time derivative. The different colors represent different critical eccentricities $e_c$, ranging from $e_c = 0$ to $e_c = 0.9$. The peak amplitude $|h_{22}^{\rm peak}|$ and the time of peak amplitude $t_{22}^{\rm peak}$ exhibit a dependence on $e_c$. We find that in the case $a=0.7$ the peak disappears starting from a certain threshold eccentricity $e^{\rm thr}_c \approx 0.5$.}
  	\label{hlm_amplitudes_for_two_spins_compared}
\end{figure*}
\begin{figure}[t!]
  	\includegraphics[width=1.\linewidth]{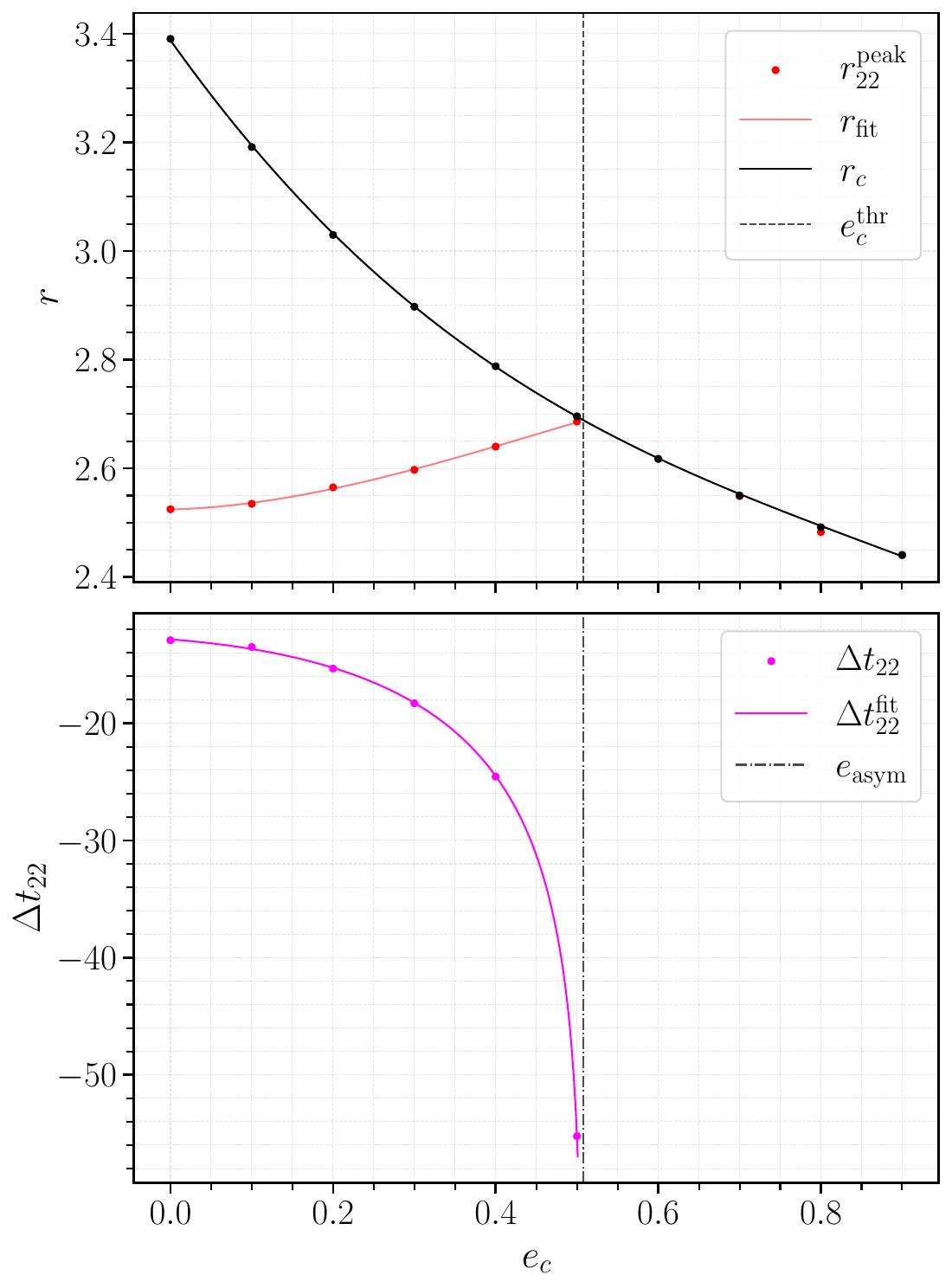}
  	\caption{This figure summarizes the two different methodologies employed to extract $e^{\rm thr}_c$ and $e_{\rm asym}$, which are the threshold eccentricities from which the peak of the amplitude of the $h_{22}$ mode disappears. In the figure we consider configurations with spin $a = 0.7$. The upper panel shows the Method 1, mentioned in Sec.~\ref{Subsec: Waveforms characterization}: the red curve is the fit of $r_{22}^{\rm peak}$ as function of $e_c$, while the black curve is $r_c$, the radius of the UCO. The threshold eccentricity $e^{\rm thr}_c$ corresponds to the eccentricity for which the two curves intersect. We highlight the value of $e^{\rm thr}_c$ with a dashed vertical black line. The bottom panel shows the Method 2: the magenta dots represent the quantity $\Delta t_{22}$, which exhibits a logarithmic divergence at $e_{\rm asym}$. After performing a fit of $\Delta t_{22}$ (magenta curve), we extract the value of the asymptote $e_{\rm asym}$ (dot-dashed vertical black line) and we compare it with $e^{\rm thr}_c$.}	
  	\label{ecc_threshold_values_extraction_methods_compared}
\end{figure}
In Fig.~\ref{hlm_amplitudes_for_two_spins_compared}, we show the amplitudes of the $(2,2)$ waveform mode generated by the TD Teukolsky code sourced by critical plunge geodesics with critical eccentricity in the interval $e_c=[ 0, \ 0.9 ]$. The upper panels show the mode's amplitude, while the bottom panels show their first time derivatives. We show two different spin configurations of the central BH: in the left column, we consider the Schwarzschild case $a=0$, while in the right column, we consider the spinning case with $a=0.7$. We align the waveforms at the time of the peak of the orbital frequency~$t_{\rm \Omega^{\rm peak}}$. 

The amplitude of the $h_{22}$ mode shows a clear dependence on $e_c$. Specifically, we observe that the value of the peak of the amplitude $|h_{22}^{\rm peak}|$ increases monotonically, while, as we will also show in Fig~\ref{h22_delta_tlm_vs_e0}, the time~$t_{22}^{\rm peak}$ of the peak amplitude decreases monotonically with $e_c$. This behavior was previously identified in Ref.~\cite{Albanesi:2023bgi} for the Schwarzschild case, and we observe the same trend here in the more general spinning scenario.

A further feature we identify is the dependence of the prominence of the peak of the amplitude with respect to $e_c$. In the QC case ($e_c=0$), the prominence of the $h_{22}$ amplitude peak depends on the spin $a$ of the Kerr BH, as shown in Ref.~\cite{Taracchini:2014zpa}. In this work, we find that it also depends on $e_c$. In fact, already in the non-spinning case, it is possible to notice that as $e_c$ increases, the peak of the amplitude is less and less prominent as depicted in the upper-left panel of Fig.~\ref{hlm_amplitudes_for_two_spins_compared}.
When considering the prograde spinning case we observe that up to a certain threshold value of the critical eccentricity, $e_c=e_{c}^{\rm thr}$, it is possible to identify a peak of the amplitude, however for values of $e_c \ge e_{c}^{\rm thr}$, there is no discernable peak anymore. 

This transition---from regimes where it is possible to identify a peak of the amplitude to regimes where it is not possible anymore---is made more explicit in the bottom panels of Fig.~\ref{hlm_amplitudes_for_two_spins_compared}, where we show the time-derivative of the amplitude of the $h_{22}$ mode. The quantity $d|h_{22}|/dt$ vanishes in the limit $t \rightarrow - \infty $ because of the fact that critical plunge geodesics asymptotically start back in time $t$ from the UCO, as expressed in Eq.~\eqref{Eq.: asymptotically rc}. When the peak of the amplitude is present, $d|h_{22}|/dt \ge 0$ back in the past and it increases until it reaches a maximum. From that moment, it starts to decrease, and it vanishes at the time of the peak $t_{22}^{\rm peak}$. As $e_c$ increases, the value of the maximum of $d|h_{22}|/dt$ decreases, eventually becoming equal to zero when $e_c=e_{c}^{\rm thr}$. From this particular configuration, it is impossible to identify the peak of the amplitude of the $h_{22}$ mode as it is no longer present. 

Characterizing the peak amplitude of the $h_{22}$ gravitational-waveform mode in the parameter space allows for precise modeling of the final stages of the waveform. Specifically, quantities such as the time at which the amplitude peaks $t_{22}^{\rm peak}$, the peak of the amplitude $|h_{22}^{\rm peak}|$, its first $d |h_{22}|/dt|_{\rm peak}$ and second time derivatives $d^2 |h_{22}|/dt^2|_{\rm peak}$ at the peak time, the mode frequency at the peak $\omega_{22}^{\rm peak}$, and the frequency's time derivative at the peak $\dot{\omega}_{22}^{\rm peak}$ are key inputs to accurately model the waveforms near the merger and to model the merger-ringdown~\cite{Cotesta:2018fcv, Nagar:2020pcj, Riemenschneider:2021ppj, Pompili:2023tna, Barausse:2011kb, Taracchini:2014zpa}. This characterization relies on the existence of a peak of the mode, however, as we summarized above, there are parts of the parameter space where no peak exists. As a first step it is necessary to determine the part of the parameter space in which the amplitude of the $h_{22}$ mode exhibits a peak, hence we are interested in studying $e_c^{\rm thr}$ for each spin $a$ of the Kerr BH.

\input{tab/different_e_thresholds.tex}

To determine the value of eccentricity $e_c^{\rm thr}(a)$ from which the peak of the amplitude of the $h_{22}$ mode vanishes, and its dependence on $a$, we employ the following method:
\begin{enumerate}
\item For every spin configuration we first extract the quantity $t_{22}^{\rm peak}$ that corresponds to the time of the peak of the amplitude of the $h_{22}$ mode.\footnote{A priori, we identify $t_{22}^{\rm peak}$  for every waveform, even when the peak of the mode amplitude is not present and the peak finder algorithm fails, as it is shown in the upper right panel of Fig.~\ref{ecc_threshold_values_extraction_methods_compared} for $e_c>0.4$. In these regimes, the algorithm cannot find the peak, and the nominal $t_{22}^{\rm peak}$ that is found does not exhibit the same regular pattern $t_{22}^{\rm peak}$ has when the peak exists.}
\item At this time $t_{22}^{\rm peak}$, we extract the value of the radial coordinate $r_{22}^{\rm peak} = r(t_{22}^{\rm peak})$ from the trajectory which sources the Teukolsky code employed to generate the waveform.
\item Then, we compare $r_{22}^{\rm peak}$ with the value of the critical radius $r_{c}$ introduced in Sec.~\ref{Sec.:critical plunge geodesics}. This quantity decreases monotonically with $e_c$, as depicted in Eq.~\eqref{Eq.: e_LSO vs rc_c} and as shown in the upper panel of Fig.~\ref{ecc_threshold_values_extraction_methods_compared}, while we find that $r_{22}^{\rm peak}$ increases monotonically up to a certain value of eccentricity for which the value of $r_{22}^{\rm peak}$ reaches the value of the critical radius $r_c$, i.e., $r_{22}^{\rm peak} = r_c$. When this happens, the amplitude of the $h_{22}$ mode reaches its peak at the UCO. We identify the value of eccentricity for which $r_{22}^{\rm peak} = r_c$ as the threshold critical eccentricity $e_c^{\rm thr}$ introduced above. This threshold value separates the regimes where the peak of the $h_{22}$ mode exists to the regimes where it does not.
\item To compute the value of $e_c^{\rm thr}$, we perform a fit of $r_{22}^{\rm peak}$ with respect to $e_c$ in the regime for which it is increasing monotonic. By denoting the fit as $r_{\rm fit}(e_c)$, we determine $e_c^{\rm thr}$ by computing the intersection of the fit with the curve of the critical radius
\begin{equation} \label{Eq: e_thr extraction}
r_{\rm fit}(e_c^{\rm thr}) = r_c(e_c^{\rm thr}) \ .
\end{equation}
\end{enumerate}

The top panel of Fig.~\ref{ecc_threshold_values_extraction_methods_compared} summarizes this procedure: the red curve corresponds to  the fit $r_{\rm fit}$, while the black curve corresponds to $r_c$. The red dashed line is the extrapolation of $r_{\rm fit}$ performed in order to find the intersection with the $r_c$ curve and to compute the threshold eccentricity as mentioned in Eq.~\eqref{Eq: e_thr extraction}. The dashed grey line of Fig.~\ref{ecc_threshold_values_extraction_methods_compared} corresponds to the threshold eccentricity $e_c^{\rm thr}$.

To assess the robustness of the threshold critical eccentricity extraction method (henceforth referred to as Method~1), we extract $e_c^{\rm thr}$ using an alternative approach (henceforth referred to as Method~2), as shown in the bottom panel of Fig.~\ref{ecc_threshold_values_extraction_methods_compared}. In this different method, we identify, for each value of $a$, the eccentric configurations of the trajectories for which a distinct peak in the amplitude of $h_{22}$ exists. This is achieved by focusing on the $t_{22}^{\rm peak}$ values that exhibit a coherent pattern, (i.e. a monotonic decreasing trend with respect to $e_c$) avoiding the values that show a scattered pattern due to the challenges the peak-finder algorithm faces when no peak is present. This scattered pattern is explicit in the upper right panel of Fig.~\ref{hlm_amplitudes_for_two_spins_compared} for eccentricities $e_c \ge 0.5$. We then introduce the quantity $\Delta t_{22}$ to reference the $t_{22}^{\rm peak}$ values to a reference time:
\begin{equation} \label{Eq: Delta_t_22 definition}
\Delta t_{22} = t_{22}^{\rm peak} - t_{\Omega^{\rm peak}} \ 
\end{equation}
Here, $t_{\Omega^{\rm peak}}$ corresponds to the time at which the orbital frequency $\Omega$  of the plunge geodesic, defined in Eq.~\eqref{Eq.: orbitalfreq}, peaks.
What we phenomenologically observe is that the $\Delta t_{22}$ quantity exhibits a divergent-like behaviour at a certain value of $e_c$.

This divergent trend can be understood and interpreted by recognizing that when \( r_{22}^{\rm peak} = r_c \), corresponding to \( e_c = e_c^{\rm thr} \) as introduced earlier, the amplitude of the \( h_{22} \) mode reaches its maximum at the UCO. However, the condition \( r = r_c \) represents an asymptotic configuration for the class of critical plunge geodesics. As discussed in Sec.~\ref{Sec.:critical plunge geodesics} and made explicit in Eq.~\eqref{Eq.: asymptotically rc}, these geodesics asymptotically approach the UCO as \( t \to -\infty \), (i.e., \( r \to r_c \) for \( t \to -\infty \)). Therefore, when the amplitude peaks at the UCO, the corresponding time is \( t_{22}^{\rm peak} = t_c \to -\infty \).

To find the value of the asymptote, we perform a fit using the functional form 
\begin{equation}
\Delta t^{\rm fit}_{22}(e_c) = c_0 + c_1 e_c + c_2 e^2_c + \log{( e_{\rm asym} - e_c)} \ ,
\end{equation}
where $\{c_0, c_1, c_2, e_{\rm asym} \}$ are the free parameters of the fit and $e_{\rm asym}$ is the value of the eccentricity for which $\Delta t_{22}$ exhibits the asymptote.
By repeating this procedure for every spin configuration of the central BH we can extract a set of eccentricity values for the asymptotes that depend on the spin, $e_{\rm asym} = e_{\rm asym}(a)$. In Table~\ref{Table_different_e_thresholds} we show the values of $e_c^{\rm thr}$, computed with Method 1, and of $e_{\rm asym}$, computed with Method 2, for different values of $a$. The absolute differences between these two quantities is $\approx 10^{-3}$ and provides a measure of the error in the extrapolation of $e_c^{\rm thr}$ .

\section{Results} \label{Sec: results}
\subsection{$\ell = 2$ $m = 2$ mode characterization} \label{Subsec: ell = 2 m = 2 mode characterization}
\subsubsection{Amplitude peak characterization} \label{Amplitude peak characterization}
\begin{figure}[t]
  	\includegraphics[width=1.\linewidth]{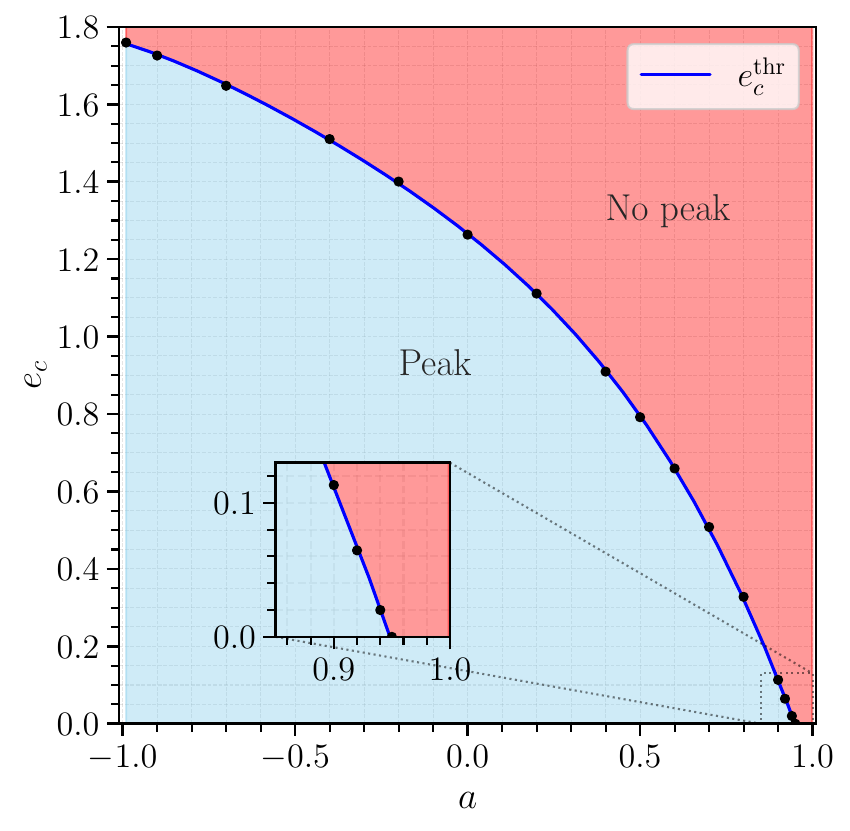}
  	\caption{Characterization over the parameter space $(a, e_c)$ of the peak existence of the $h_{22}$ mode amplitude. The blue area is the part of parameter space for which the peak exists while the red area is the part for which the peak does not exist. The black dots are the values of $e_c^{\rm thr}(a)$ computed with Method 1 as explained in Sec.~\ref{Subsec: Waveforms characterization}. The blue curve is a quadratic polynomial fit of $e_c^{\rm thr}$ as function of $a$. The inset shows a zoom of $e_c^{\rm thr}(a)$ for values of the spin $0.85 \le a \le 1$ and critical eccentricity $0 \le e_c \le 1.2$. We find that for $a \ge 0.95$, it is impossible to find the peak of the amplitude of the $h_{22}$ mode even when $e_c = 0$.
  	}
  	\label{h22_peaks_existence_curve}
\end{figure}
\begin{figure}[t]
  	\includegraphics[width=1.\linewidth]{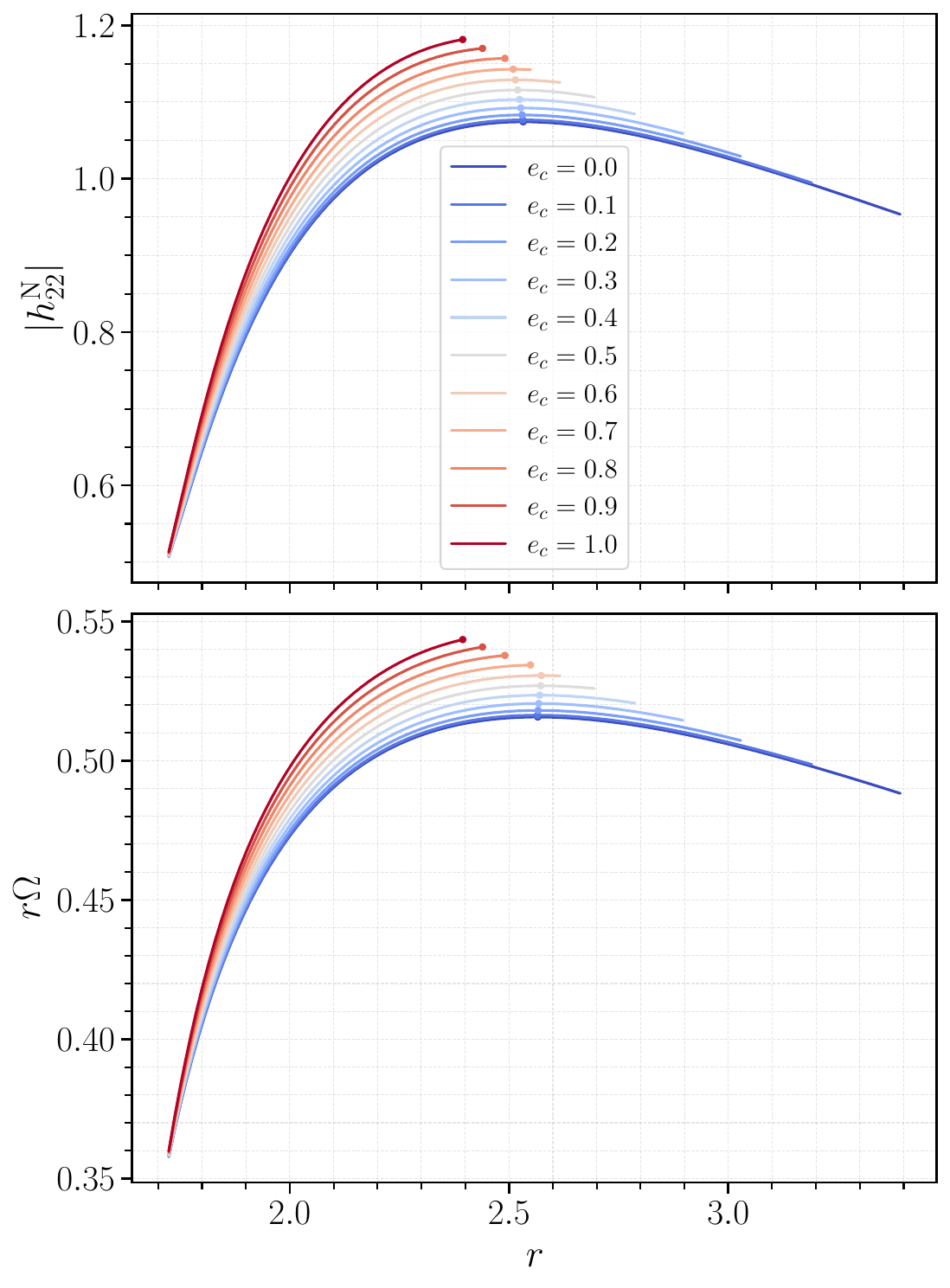}
  	\caption{In the top panel of this figure we show the amplitude of the $h_{22}^{\rm N}$ mode defined in Eq.~\eqref{eq.:h22N expr}, which corresponds to the Newtonian order of the $h_{22}$ mode. We compute the quantity $|h_{22}^{\rm N}|$ on critical plunge geodesics with spin $a = 0.7$ and values of critical eccentricity $e_c \le 1$, and we plot it as a function of the radial coordinate $r$. The bottom panel shows the quantity $r \Omega$ of the considered critical plunge geodesics as a function of $r$. The different curves start at different radii because of the fact that the critical plunge geodesics asymptotically start at the critical radius $r_c$, which is a function of the quantities $a$ and $e_c$, as mentioned in Sec.~\ref{Sec.:critical plunge geodesics}.
  	}
  	\label{Fig.: h22_Newtonian}
\end{figure}
By repeating the procedure of Method~1 for different values of $a$ in the interval $-0.99 \le a \le 0.99$, we extract the threshold values of the critical eccentricity $e_c^{\rm thr}$ from which the $h_{22}$ mode does not exhibit a peak in its amplitude. In doing so, we consider values of the critical eccentricity $0\le e_c < 3$, i.e., we also consider critical plunge geodesics with value $e_c>1$. As we described in Sec.~\ref{Sec.:critical plunge geodesics}, these configurations are associated with homoclinic orbits outside the UCO that are scattering orbits with the same energy $\mathcal{E}_c$ (see Eq.~\ref{eq:energy critical}) of the critical plunge geodesics.

In Fig.~\ref{h22_peaks_existence_curve} we characterize the regions of the $(a, e_c)$ parameter space for which the $|h_{22}^{\rm peak}|$ exist. The blue region corresponds to the part of parameter space where $|h_{22}^{\rm peak}|$ exists. In this region, it is possible to identify and model all the quantities evaluated at the peak and use them as input for merger-ringdown models and respective attachment procedures to the inspiral part of the waveform~\cite{Barausse:2011kb, Cotesta:2018fcv, Riemenschneider:2021ppj, Pompili:2023tna}. The black dots are the values of the threshold eccentricities $e_c^{\rm thr}(a)$ for different values of $a$ derived with Method 1 introduced in Sec.~\ref{Subsec: Waveforms characterization}. The blue curve is a quadratic polynomial fit of $e_c^{\rm thr}$ as a function of $a$. This curve separates the blue region from the part of the parameter space for which it is impossible to identify a peak in the amplitude of the $h_{22}$ mode. In Fig.~\ref{h22_peaks_existence_curve}, we highlight this portion of the parameter space with red color. 

In the inset of Fig.~\ref{h22_peaks_existence_curve}, we show that for the particular value of the spin $a=0.95$, the peak of the amplitude of the $h_{22}$ mode disappears in the QC configuration, $e_c = 0$. For $a \ge 0.95$ values, the $h_{22}$ mode never peaks on the $(a, e_c)$ parameter space. This is similar to the result reported in Ref.~\cite{Taracchini:2014zpa} for equatorial QC trajectories of a Kerr BH.
However the results do not precisely match, since in that work for $a=0.95$ the authors identify a clear peak with $\Delta t_{22} = -103$, while we do not find any peak. We believe this discrepancy comes from the fact that we are using different trajectories. In particular, the work in Ref.~\cite{Taracchini:2014zpa} evolved the QC trajectories with numerical fluxes up to the light-ring of the Kerr BH for a TM with mass-ratio $\mu/M = 0.001$, while the critical geodesics studied in the current work represent the $\mu/M=0$ limit. These differences might be the cause of the discrepancy we observe. As we will show in Sec.~\ref{Subsubsec: Mode features at the peak}, the discrepancy in the quantity $\Delta t_{22}$ appears for spin values $a \ge 0.85$.

As we show in Fig.~\ref{ecc_threshold_values_extraction_methods_compared}, the value of the radial coordinate $r_{22}^{\rm peak}$ for which the amplitude of the $h_{22}$ mode peaks is a monotonic increasing function of $e_c$. At the same time, the critical radius $r_c$ is a monotonic decreasing function of $e_c$. When introducing the Method 1 in Sec.~\ref{Subsec: Waveforms characterization}, we mentioned that when $r_{22}^{\rm peak}(e_c^{\rm thr}) = r_c(e_c^{\rm thr})$, it happens that $|h_{22}|$ peaks at the critical radius $r_c$ of the UCO and we identify a threshold critical eccentricity $e_c^{\rm thr}$. For $e_c \ge e_c^{\rm thr}$, $|h_{22}|$ does not peak anymore. This provides an operational and intuitive explanation for why the peak disappears.
However, it shifts the question to why the peak of the waveform occurs at a relatively large radius compared to the peak of the orbital frequency.
To better understand the behavior of the peak of the amplitude of the $h_{22}$ mode (an $\ell = m$ modes) we consider the Newtonian order of the $h_{22}$ and $h_{33}$ modes
\begin{subequations}
\begin{align}
h^{\rm N}_{22} & \propto \dot{r}^2  + r \ddot{r}  - 2 r^2 \Omega^2 - i(4 \dot{r} r \Omega + \dot{\Omega} r^2) , \label{eq.:h22N expr}  \\
h^{\rm N}_{33} & \propto 6 \dot{r}^3 - 54 i \dot{r}^2 r \Omega + 9 \dot{r} r (2 \ddot{r} - 3 i \dot{\Omega} r - 9 r \Omega^2) \nonumber \\
& + 3 r^2 (\dddot{r} - i (\ddot{\Omega} r + 9 \ddot{r} \Omega - 9 i \dot{\Omega} r \Omega - 9 r \Omega^3)) . \label{eq.:h33N expr}
\end{align}
\end{subequations}
These expressions are derived by computing the coordinate-time derivatives of the Newtonian mass and current multipoles. Specifically, if we introduce the notation $h_{\ell m}^{\rm N} = h^{(\rm N,\epsilon)}_{\ell m}$, where $\epsilon$ is the parity of $\ell + m$, we have $h^{(\rm N,0)}_{\ell m} \propto e^{im \varphi}I^{(\ell)}_{\ell m}$ and $h^{(\rm N,1)}_{\ell m} \propto e^{im\varphi}S^{(\ell)}_{\ell m}$, with $(\ell)$ being the $\ell$-th time-derivative and $I_{\ell m} = r^{\ell}e^{-im\varphi}$ and $S_{\ell m} = r^{\ell + 1} \Omega e^{-im\varphi}$ the Newtonian mass and current multipoles (see, e.g., Refs.~\cite{Thorne:1980ru, Blanchet:1998in, Chiaramello:2020ehz}).
These expressions show how the Newtonian modes depend on the trajectories variables $r$, $\Omega$, and their (Boyer-Lindquist) time derivatives. Using the geodesic equations~\eqref{eq:geoeom} and Eqs.~\eqref{eq:energy critical},\eqref{eq:angular momentum critical}, we can express  the $h_{\ell m}$ modes at Newtonian order on the critical plunges as function of $r$ (plus $a$ and $e_c$).

In the top panel of Fig.~\ref{Fig.: h22_Newtonian}, we show the amplitude of the Newtonian mode $h_{22}^{\rm N}$ of Eq.~\eqref{eq.:h22N expr} as a function of the radial coordinate $r$. We consider critical plunge geodesics with spin $a = 0.7$ and critical eccentricities $e_c \le 1$. In the figure, the curves representing $|h_{22}^{\rm N}|$ start at different radii: this is because the critical plunge geodesics asymptotically start from the critical radius $r_c$, as depicted in Eq.~\eqref{Eq.: asymptotically rc}.  We find that as $e_c$ increases, $|h_{22}^{\rm N}|$ peaks closer to the critical radius $r_c$ (i.e., the value of $r$ where the curves start). This is in agreement with what we observe for the numerical waveform, specifically with the fact that the curves of the radial coordinate $r_{22}^{\rm peak}$ for which the amplitude of the $h_{22}$ mode peaks and the curve of the critical radius $r_c$ intersect for a specific value of critical eccentricity $e_c^{\rm thr}$, as mentioned above and summarized in the top panel of Fig.~\ref{ecc_threshold_values_extraction_methods_compared}.

By studying the expressions in Eq.~\eqref{eq.:h22N expr}, we can connect the peak disappearance of the mode with the features of the dynamical quantities of the critical plunge geodesics. In particular, we find that the quantity $r \Omega$, which appears in Eqs.~\eqref{eq.:h22N expr} -~\eqref{eq.:h33N expr}, also manifests a disappearing peak as $e_c$ increases as we show in the bottom panel of Fig.~\ref{Fig.: h22_Newtonian}. The disappearance of the peak of $r \Omega$ is straightforwardly understood. As noted in Sec.~\ref{Sec.:critical plunge geodesics}, $\Omega$ is a function of $a$, $b$, and $r$ that is universal for all equatorial Kerr geodesics. Consequently, location of its peak does not depend on whether the geodesic is critical or not, depending (weakly) only on the value of $b_c$. By extension the same is true for $r\Omega$, but with a peak at a larger value of $r$, allowing the possibility that a critical plunge starts at a radius $r_c$ smaller than the position of the peak of $r\Omega$.

Through an explicit computation, we find that the quantity $r \Omega$ dominates the different contributions in the Newtonian modes $h_{22}^{\rm N}$ and $h_{33}^{\rm N}$ expressions in Eqs.~\eqref{eq.:h22N expr} and \eqref{eq.:h33N expr}. In particular, we note that as  $e_c$ increases the radial evolution, $\dot{r}$, becomes smaller over the entire plunge (see  Appendix~\ref{Sec.: circularization behaviour} for more detail). Consequently, the dominance of  $r \Omega$ tends to increase with increasing critical eccentricity. Thus, the behavior of the peak  of  $r \Omega$ also explains the disappearance of the peak of  the $h_{22}$ mode.

\subsubsection{Mode features at the peak} \label{Subsubsec: Mode features at the peak}
\begin{figure}[t]
  	\includegraphics[width=1.\linewidth]{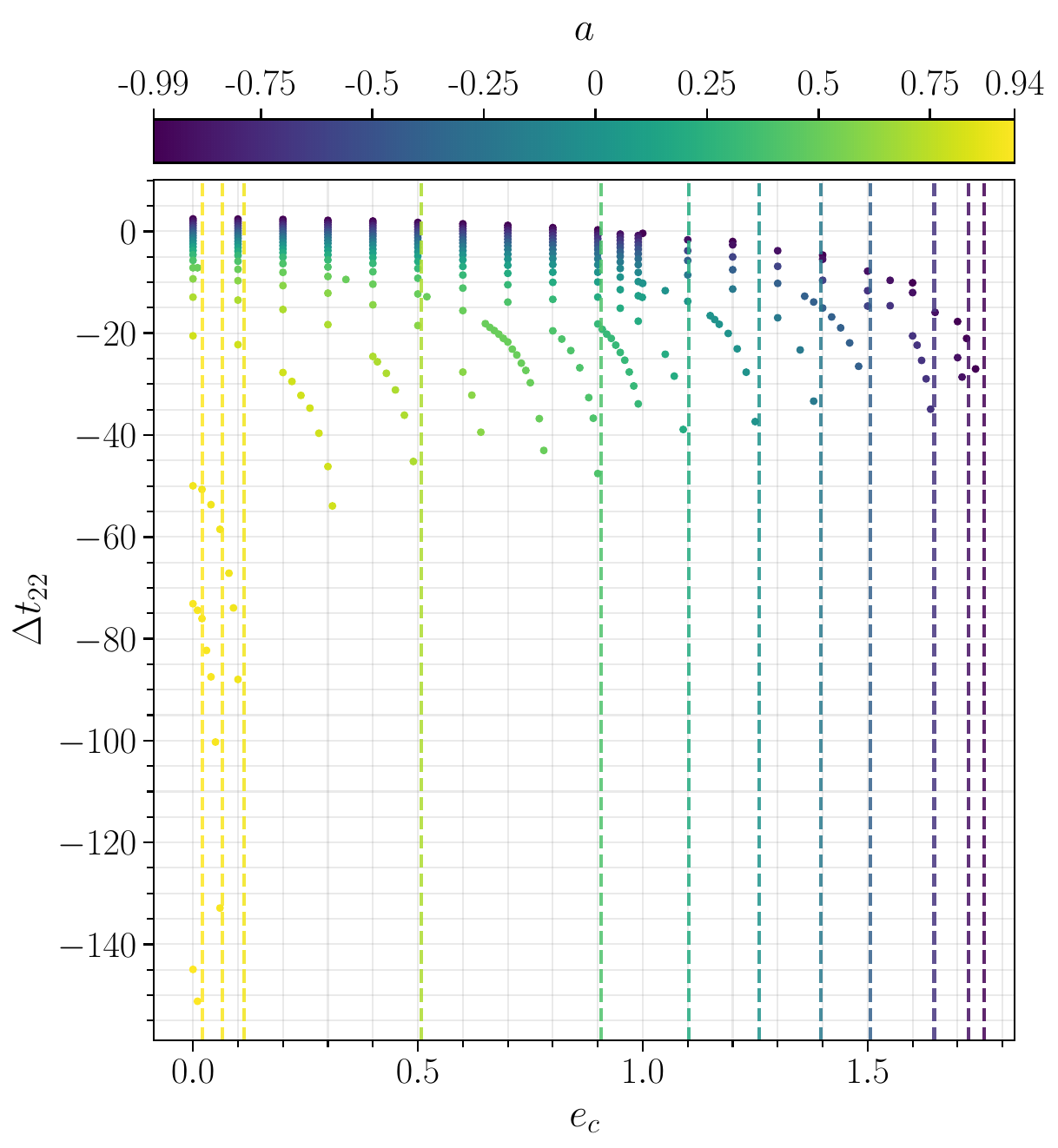}
  	\caption{This figure shows the dependence of the quantity $\Delta t_{22}$ as a function of $e_c$. The color scale represents the spin $a$ of the Kerr BH. In the figure, we consider $-0.99 \le a < 0.95$. The vertical dashed lines correspond to the asymptote values $e_{\rm asym}$ in Table.~\ref{Table_different_e_thresholds}; the colors of the lines correspond to the color of the spin configuration for which the asymptotes occur. For spin values $a \ge 0.95$, we can no longer find a peak even when $e_c = 0$, as shown in Fig.~\ref{h22_peaks_existence_curve}. For these spins, the quantity $\Delta t_{22}$ is not defined, and not shown in the figure.}
  	\label{h22_delta_tlm_vs_e0}
\end{figure}
\begin{figure*}[t]
	\includegraphics[width=1.\linewidth]{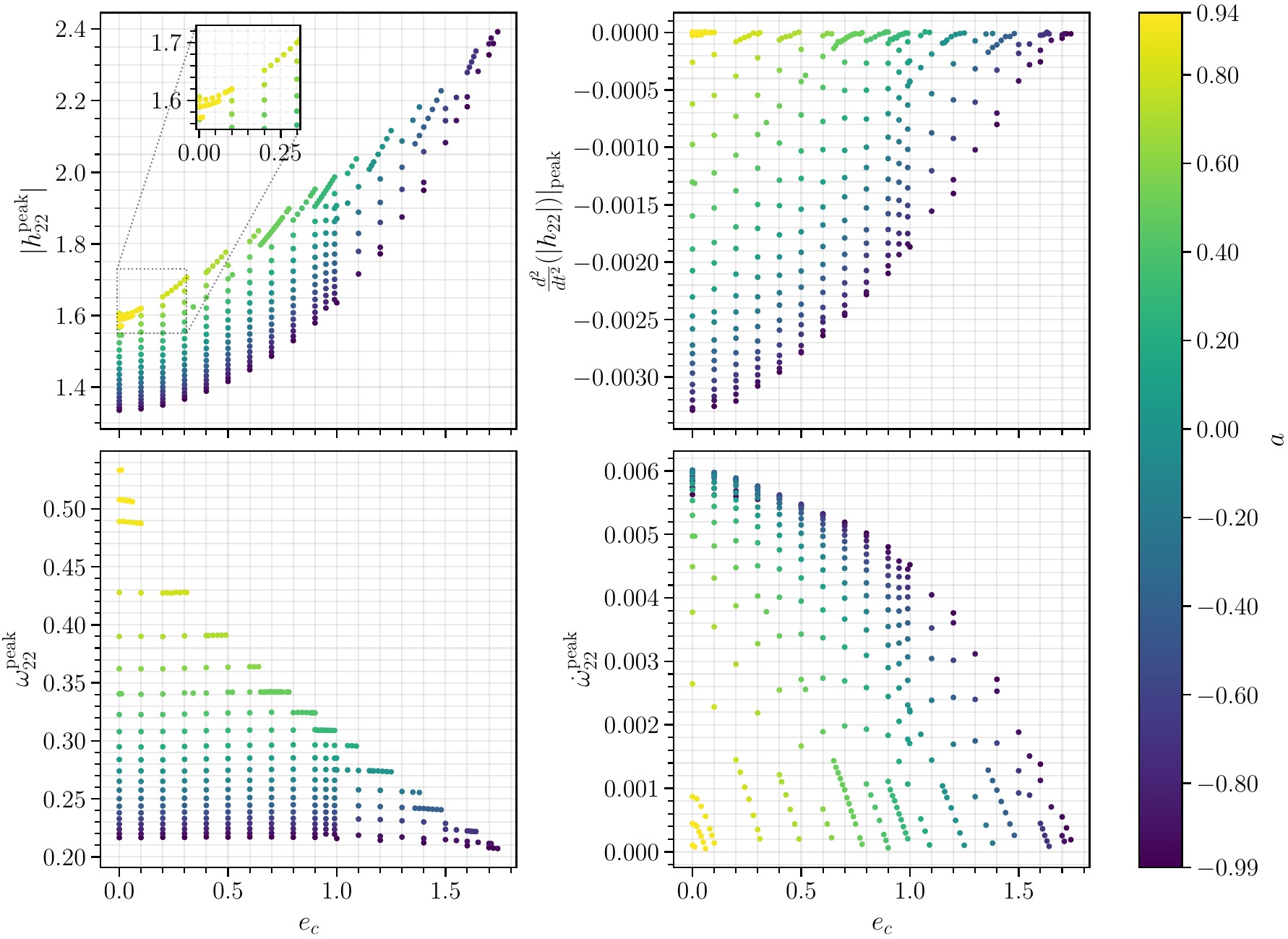}
	\caption{In this figure we show the dependence of the amplitude \( |h_{22}^{\rm peak}| \), second-order time derivative of the amplitude $ d^2|h_{22}|/dt^2|_{\rm peak} $, frequency \( \omega_{22}^{\rm peak} \), and first-order time-derivative of the frequency \( \dot{\omega}_{22}^{\rm peak} \) of the $h_{22}$ mode evaluated at the peak of its amplitude as function of the critical eccentricity $e_c$. We span different values of the spin $-0.99 \le a \le 0.99$ of the central Kerr BH, represented by different colors. We consider configurations for which the peak of the $h_{22}$ exists, as illustrated in the blue area of the parameter space in Fig.~\ref{h22_peaks_existence_curve}. For spin values $a \ge 0.95$, we can no longer find a peak even when $e_c = 0$, as shown in Fig.~\ref{h22_peaks_existence_curve}. For these spins, it is not possible to define quantities at the peak, and they are not shown in the figure.
	}
	\label{h22_NQC_quantities_vs_e0_only_IVs}
\end{figure*}
In GWs modeling, the time of the mode amplitude peak $t_{\ell m}^{\rm peak}$ is a feature inferred from NR waveforms and employed as a calibration parameter to correctly attach ring-down models to the end of the inspiral-plunge waveform. In EOB models for example, the values of $t_{\ell m}^{\rm peak}$ are expressed in terms of time intervals $\Delta t_{\ell m}$ with respect to a reference time associated with particular configurations of the dynamics of the binary, like the peak of the orbital frequency $\Omega$ or the ISCO crossing~\cite{Barausse:2011kb, Cotesta:2018fcv, Riemenschneider:2021ppj, Pompili:2023tna}.

In our work we adopt the same scheme and we refer $t_{\ell m}^{\rm peak}$ with respect to $t_{\Omega^{\rm peak}}$, i.e. we consider
\begin{equation}
	\Delta t_{\ell m} = t_{\ell m}^{\rm peak} - t_{\Omega^{\rm peak}},
\end{equation}
which is a generalization of Eq.~\eqref{Eq: Delta_t_22 definition} in the case of the $h_{22}$ mode. In Fig.~\ref{h22_delta_tlm_vs_e0} we show the quantity $\Delta t_{22}$ as a function of the eccentricity $e_c$. The different colors represent different spin $a$ configurations of the Kerr BH. In the figure, we focus on the section of parameter space for which the peak of the $h_{22}$ mode exists, as depicted in Fig.~\ref{h22_peaks_existence_curve}. 

The time interval $\Delta t_{22}$ exhibits a monotonic decreasing trend with respect to the eccentricity for all the spins. Moreover, for every fixed $e_c$ configuration it also shows a monotonic decreasing behaviour with respect to the spin. As already pointed out in Sec.~\ref{Subsec: Waveforms characterization}, at a certain value of $e_c$ (which depends on the spin) the $\Delta t_{22}$ quantity exhibits a divergent trend because the peak has to move to the infinite past before disappearing. We highlight the presence of these divergences in Fig.~\ref{h22_delta_tlm_vs_e0} by plotting the asymptote values $e_{\rm asym}$ introduced in Sec.~\ref{Subsec: Waveforms characterization} with vertical dashed lines; the colors of these lines correspond to the color of the spin configuration for which the asymptotes occur at the value $e_{\rm asym}(a)$. In the $\Delta t_{22}$ study, we also compared with previous results in Ref.~\cite{Taracchini:2014zpa} for the QC case. We find a good agreement of $\Delta t_{22}$ for the $e_c = 0$ case with the values shown in Fig.~13 of Ref.~\cite{Taracchini:2014zpa} up to spin values $a \sim 0.85$. However, for $a = 0.9$ and $a = 0.95$, we find a discrepancy between the values of $\Delta t_{22}$ we computed from the waveforms produced in our work and the values of $\Delta t_{22}$ in Ref.~\cite{Taracchini:2014zpa}. As we mentioned in Sec.~\ref{Amplitude peak characterization}, we believe this discrepancy is because the trajectories from Ref.~\cite{Taracchini:2014zpa} are considered at finite mass-ratio. These differences are particularly evident when considering $\Delta t_{22}$ for high spins. In Appendix~\ref{Sec.: Comparison of Delta t_22 with previous works} we provide a direct comparison of the $\Delta t_{22}$ quantity extracted considering waveforms produced with critical plunge geodesics in the QC case (i.e., with $e_c = 0$) with the same quantity extracted in Ref.~\cite{Taracchini:2014zpa}.

In Fig.~\ref{h22_NQC_quantities_vs_e0_only_IVs}, we show the quantities employed in typical attachment procedures and in the ringdown fits of the GWs models based on the EOB formalism. These quantities correspond to the amplitude $ |h_{\ell m}^{\rm peak}| $, second-order time derivative $ d^2|h_{\ell m}|/dt^2|_{\rm peak} $, frequency $ \omega_{\ell m}^{\rm peak} $, and first-order time-derivative $ \dot{\omega}_{\ell m}^{\rm peak} $ of the $h_{\ell m}$ modes evaluated at the peak~\cite{Barausse:2011kb, Cotesta:2018fcv, Riemenschneider:2021ppj, Pompili:2023tna}. We specialize our analysis to the dominant $h_{22}$ mode, and we parametrize the quantities with respect to the $e_c$ of the plunging geodesic. 
We restrict our analysis to the section of parameter space for which the peak of the $h_{22}$ mode exists.

In the top-left panel of Fig.~\ref{h22_NQC_quantities_vs_e0_only_IVs}, we show that the peak of the amplitude increases with respect to $e_c$ and $a$, showing an overall monotonic pattern. The exception to this trend is given by low values of critical eccentricity, $e_c \le 0.08$, for the highest prograde spin scenario (i.e. $a=[ 0.9, 0.92, 0.94 ]$). In these cases the peak of the amplitudes is smaller than the  $e_c = 0$ case with spin $a=0.8$, as shown in the inset of the top-left panel of Fig.~\ref{h22_NQC_quantities_vs_e0_only_IVs}. We mention that this result was already found in Ref.~\cite{Taracchini:2014zpa} and is compatible with their value within the numerical error. Moreover, we find that for $a \ge 0.9$ the values of the amplitude of the $h_{22}$ at the peak for fixed values of $e_c$ is not an increasing monotonic function of $a$, e.g., the peak amplitude for $a=0.94$ at $e_c = 0$ is smaller than the peak amplitude for $a=0.92$ at the same critical eccentricity, as shown in the inset of the top-left panel of Fig.~\ref{h22_NQC_quantities_vs_e0_only_IVs}. To quantify how much the eccentricity changes the value of the amplitude at the peak with respect to the QC case, we mention the fact that for $e_c \le 0.5$, we find that the fractional differences of the amplitude at the peak of the eccentric $h_{22}$ mode (i.e. with $e_c \neq 0$) compared to the amplitudes at the peak of the QC modes (i.e. with $e_c = 0$) is $\lessapprox 10 \% $.

The second order time-derivative of the amplitude at the peak $ d^2|h_{22}|/dt^2|_{\rm peak} $ also manifests a monotonic trend with respect to both the spin $a$ and the critical eccentricity $e_c$. We highlight that, when $e_c$ approaches $e_c^{\rm thr}$ it happens $ d^2|h_{22}|/dt^2|_{\rm peak} $ starts to vanish. This is connected to the fact that the mode's amplitude is more and more flattened in this regime.
Regarding the mode frequency $\omega_{22}$, we find that it has an almost flat trend with respect to eccentricity, for all the spin values of the central BH, showing an overall independence from $e_c$. However, we find a very mild dependence on $e_c$ for values of $e_c \ge 1.1$ for the high retrograde spin configurations.
Finally, for the time derivative of the mode frequency $\dot{\omega}_{22}$ we find that this quantity is always positive and that it is characterized by an overall monotonic trend, going to zero as $e_c$ approaches $e_c^{\rm thr}$.
However, we also notice that for small values of the eccentricity, $\dot{\omega}_{22}$ is not monotonic for some negative values of the spin. This result was already found for the QC case (i.e. $e_c = 0$) in Fig.~15 of Ref.~\cite{Taracchini:2014zpa}, and we find this holds also for the eccentric scenarios. As we show in Fig.~\ref{Fig.: h22_h33_h21_IVs_compared} in Appendix~\ref{Sec.: ell = m higher order modes similar behaviour} this small dependence for high values of eccentricity is also present for the time derivative of the frequency $\dot{\omega}_{\ell m}$ of the $\ell = m$ higher order modes.
In Appendix~\ref{Sec.: l=2 m=2 merger values} we provide the values for a subset of the data produced and shown in Fig.~\ref{h22_NQC_quantities_vs_e0_only_IVs}.

\subsection{Higher order modes characterization} \label{Subsec: higher order modes characterization}
\subsubsection{Modes hierarchy and eccentricity} \label{Subsec: hierarchy of the modes}
\begin{figure*}[ht!]
  	\includegraphics[width=1.\linewidth]{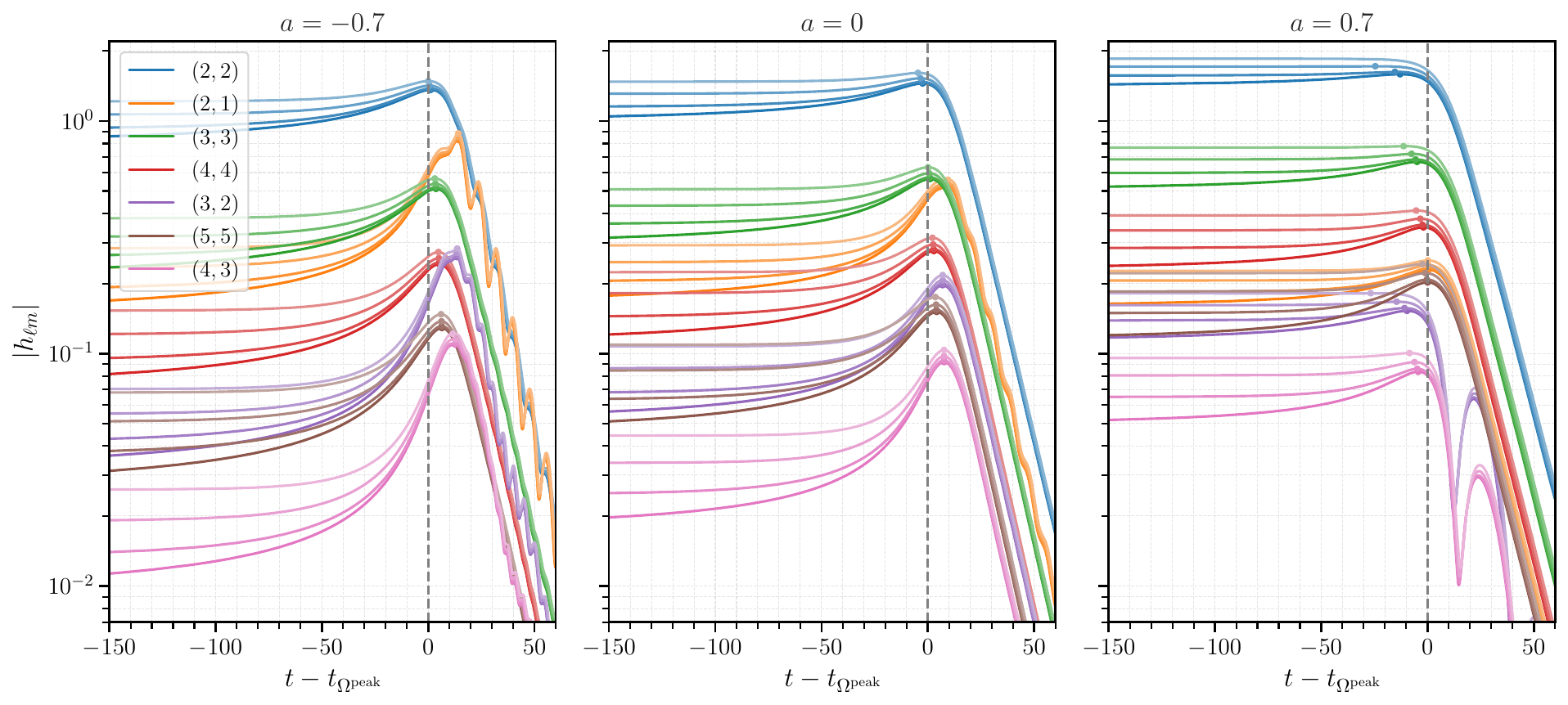}
  	\caption{Amplitude of the modes $h_{22}$ (in blue), $h_{21}$ (in orange), $h_{33}$ (in green), $h_{32}$ (in purple), $h_{44}$ (in red), $h_{43}$ (in pink) and $h_{55}$ (in brown). The first panel on the left shows the case with spin $a=-0.7$, the second panel shows the case with $a=0.0$ and the third panel the case with $a=0.7$. The different shades of the colors represent different values of $e_c$ with $e_c = [0.0, 0.2, 0.4, 0.6]$. Darker shades of the colors are linked to lower values of eccentricity. We align the waveform at the time of the peak of the orbital frequency $t_{\Omega^{\rm peak}}$.}
  	\label{Fig.: hlm hierarchy}
\end{figure*}
As a matter of fact, the hierarchy of the different $h_{\ell m}$ modes, which matters when all modes are summed in Eq.~\eqref{Eq.: strain spherical harmonic decomposition} to provide the total strain $h$, depends on the spin of the central BH~\cite{Barausse:2011kb, Taracchini:2014zpa}. In this section we aim to study the dependence of the hierarchy with respect to $e_c$. In Fig.~\ref{Fig.: hlm hierarchy} we show the different modes amplitudes $|h_{\ell m}|$ for three configurations of spin: in the first panel we consider the case with spin $a = -0.7$, in the second panel the case with $a = 0.0$ and in the third panel we consider the case with $a = 0.7$. In each panel we consider the modes $h_{22}$ (in blue), $h_{21}$ (in orange), $h_{33}$ (in green), $h_{32}$ (in purple), $h_{44}$ (in red), $h_{43}$ (in pink) and $h_{55}$ (in brown). The different shades of the colors, for each mode, represent different values of $e_c$. We consider eccentricity values $e_c = [0.0, 0.2, 0.4, 0.6]$, and, in Fig.~\ref{Fig.: hlm hierarchy}, darker shades of the colors correspond to lower values of eccentricity. As in Fig.~\ref{hlm_amplitudes_for_two_spins_compared} we align the waveforms at the time of the peak of the orbital frequency $t_{\Omega^{\rm peak}}$ of the critical plunge geodesics used to produce the waveforms.

From Fig.~\ref{Fig.: hlm hierarchy}, we notice that, regardless of the value of $a$, the eccentricity has more impact on the value of the amplitude of the modes during the pre-merger part of the waveform (i.e., when the critical plunge geodesics evolve on the UCO) rather than on the value of the amplitude in the merger part of the waveforms (i.e., close to the peak, just before the beginning of the ringdown part).
From Fig.~\ref{Fig.: hlm hierarchy} we also notice that the hierarchy of the modes mostly depends on the spin $a$ of the Kerr BH, while the eccentricity seems not to play a significant role. In fact, by inspecting the three panels of the figure, it turns out that waveforms connected to the same eccentricity (same shade of the different colors) do not exhibit significant changes in the hierarchy (i.e., in the relative magnitude of their amplitudes). However, in Fig.~\ref{Fig.: hlm hierarchy} there are some configurations for which the independence of the hierarchy from eccentricity does not always hold. In particular, we observe that for all the three spin scenarios the $h_{32}$ (purple) and the $h_{55}$ (brown) modes become more and more comparable for times $t \le 20$ when the eccentricity increases at the point that when $e_c > 0.2$ their amplitudes are almost the same. However, we remark that this effect is not present close to merger.
In the $a=0.7$ case (third panel of Fig.~\ref{Fig.: hlm hierarchy}) we observe that different eccentricities affect the hierarchy of the $h_{21}$ (orange) and the $h_{55}$ (brown) modes. In fact for low eccentricities (i.e., $e_c \le 0.2$) the $h_{21}$ is slightly more dominant, but as $e_c$ increases, the two modes become more and more comparable.
However, in all these cases, the change in mode hierarchy from different eccentricities is smaller than the much larger impact due to the change in the spin.

\subsubsection{$\ell = m$ modes}
\begin{figure*}[ht!]
  	\includegraphics[width=1.\linewidth]{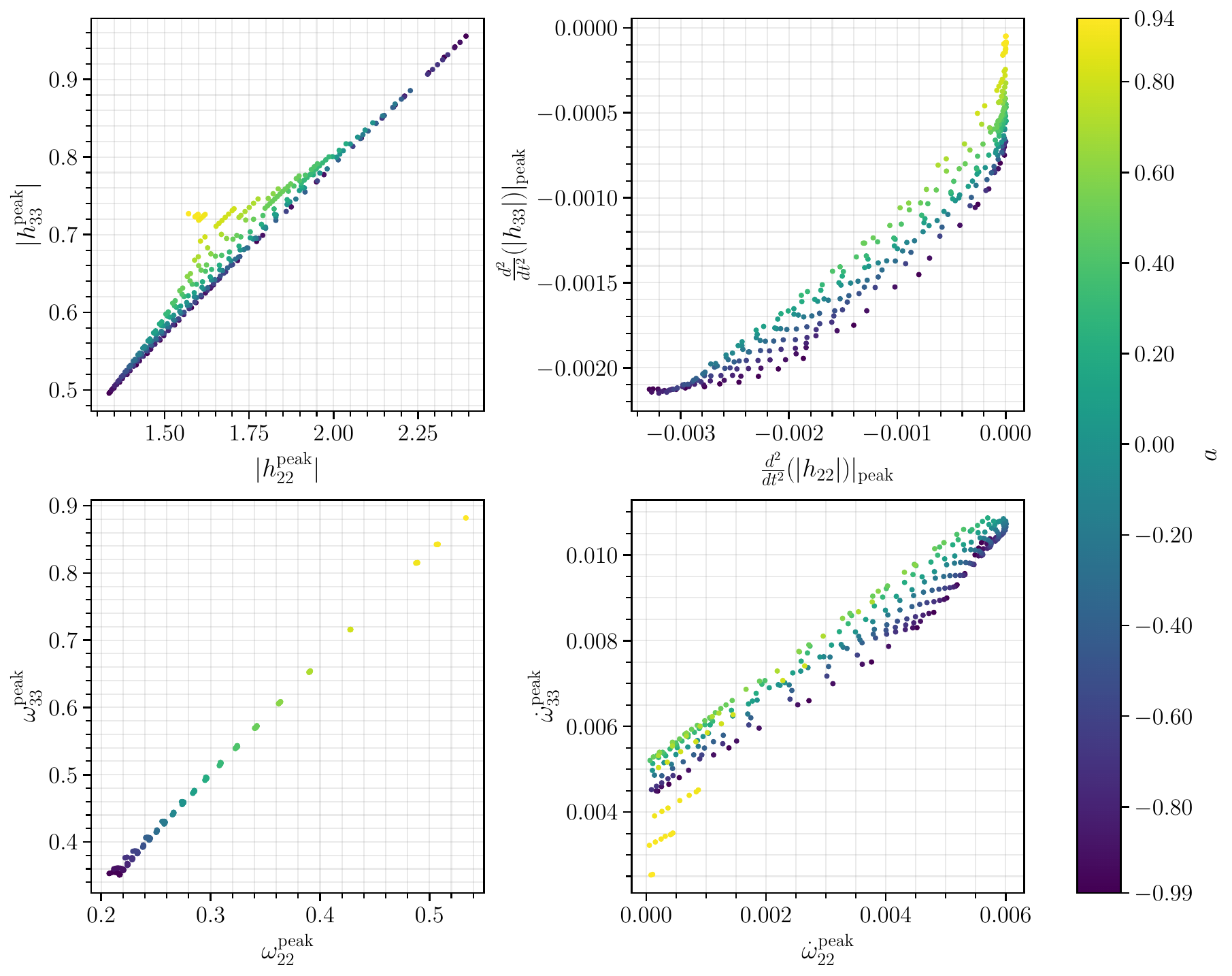}
  	\caption{In this figure we show the dependence of the amplitude \( |h_{33}^{\rm peak}| \), second-order time derivative of the amplitude $ d^2 |h_{33}| / dt^2|_{\rm peak} $, frequency \( \omega_{33}^{\rm peak} \), and first-order time-derivative of the frequency \( \dot{\omega}_{33}^{\rm peak} \) of the $h_{33}$ mode evaluated at the peak of its amplitude as a function of the same corresponding quantities for the $h_{22}$ mode. We span different values of the spin $-0.99 \le a \le 0.99$ of the central Kerr BH, represented by different colors. We consider configurations for which the peak of the $h_{22}$ exists, as illustrated in the blue area of the parameter space in Fig.~\ref{h22_peaks_existence_curve}. For spin values $a \ge 0.95$, we can no longer find a peak even when $e_c = 0$, as shown in Fig.~\ref{h22_peaks_existence_curve}. For these spins, it is not possible to define quantities at the peak, and they are not shown in the figure.}
  	\label{h33_NQC_quantities_vs_h22_quantities_only_IVs}
\end{figure*}

For the $\ell = m$ higher-order modes, similar to the dominant $h_{22}$ mode, we also observe that the amplitude peak disappears from a certain threshold value of $e_c$, as can be noticed from the third panel of Fig.~\ref{Fig.: hlm hierarchy}. However what we notice is that this transition occurs at larger $e_c$ compared to the $h_{22}$ mode case. Phenomenologically we find that when the $h_{22}$ mode has a peak also the amplitude of the $\ell = m$ higher-order modes we consider, i.e. $h_{33}$, $h_{44}$ and $h_{55}$, have peaks. By inspecting the Newtonian terms in Eqs.~\eqref{eq.:h22N expr} -~\eqref{eq.:h33N expr} on the critical plunge geodesic, we find that the threshold eccentricity for $h_{33}^{\rm N}$ is higher than that for $h_{22}^{\rm N}$. This shows that the hierarchy of the different threshold critical eccentricities among the $\ell = m$ modes is evident already at the Newtonian level and agrees with what we find with the numerical waveforms.

Moreover, we observe that the features and patterns we identified for the $h_{22}$ mode in Fig.~\ref{h22_NQC_quantities_vs_e0_only_IVs} are similar also for the higher-order modes. In this section we limit to show the results for the quantities related to the $h_{33}$ mode and we refer the reader to Appendix~\ref{Sec.: ell = m higher order modes similar behaviour} for the other modes.

In Fig.~\ref{h33_NQC_quantities_vs_h22_quantities_only_IVs} we show $ |h_{33}^{\rm peak}| $, $ d^2 (|h_{33}|) /dt^2|_{\rm peak} $, $ \omega_{33}^{\rm peak} $, and $ \dot{\omega}_{33}^{\rm peak} $ of the $h_{33}$ mode.  Differently from Sec.~\ref{Subsec: ell = 2 m = 2 mode characterization}, where we show these quantities for the $h_{22}$ mode as function of $e_c$,  we choose to show $h_{33}$ quantities with respect to the corresponding quantities of the $h_{22}$ mode. In doing so we consider the region of the parameter space for which the peak of the $h_{22}$ exists, as depicted in Fig.~\ref{h22_peaks_existence_curve}. 

In the top-left panel of Fig.~\ref{h33_NQC_quantities_vs_h22_quantities_only_IVs}, we observe that, for each spin configuration, the $|h_{33}^{\rm peak}|$ shows a linear dependence with respect to the $|h_{22}^{\rm peak}|$ quantity. This suggests the value of $|h_{33}^{\rm peak}|$ is a (spin-dependent) multiple of  $|h_{22}^{\rm peak}|$.

Concerning $ d^2 (|h_{33}|) /dt^2|_{\rm peak} $ (top-right panel of Fig.~\ref{h33_NQC_quantities_vs_h22_quantities_only_IVs}), we find a monotonic increasing trend with respect to the corresponding $h_{22}$ quantity and with respect to the spin of the central BH. We also highlight the fact that for configurations close to the threshold eccentricity $e^{\rm thr}_c(a)$ (for which the peak of the amplitude of the $h_{22}$ mode is not anymore manifest), when $ d^2 (|h_{22}|) / dt^2|_{\rm peak} = 0 $, the corresponding $h_{33}$ quantity $ d^2 |h_{33}| / dt^2|_{\rm peak}  < 0 $. This confirms what we mention at the beginning of this section, i.e., that whenever the peak of the amplitude of the $h_{22}$ mode exists also the peak of the amplitude of the higher order modes exists, being its second-time derivative of the amplitude non vanishing.

From the bottom-left panel of Fig.~\ref{h33_NQC_quantities_vs_h22_quantities_only_IVs} we observe that the relationship between the frequencies $\omega_{22}^{\rm peak}$ and $\omega_{33}^{\rm peak}$ does not depend on $e_c$, except a small dependence for high retrograde spins, which is inherited from the small dependence of $\omega_{22}^{\rm peak}$ with respect to $e_c$ for values of eccentricity $e_c \ge 1.1$ (bottom left panel of Fig.~\ref{h22_NQC_quantities_vs_e0_only_IVs}), as we already commented in Sec.~\ref{Subsec: ell = 2 m = 2 mode characterization}. This reflects the fact that, also for the higher-order modes, there is no dependence of the frequency on $e_c$, as shown for the $h_{22}$ mode in Sec.~\ref{Subsec: ell = 2 m = 2 mode characterization}. 

Finally, regarding $ \dot{\omega}_{33}^{\rm peak} $ (bottom right panel of Fig.~\ref{h33_NQC_quantities_vs_h22_quantities_only_IVs}) we observe an increasing monotonic behaviour with respect to the relative quantity of the $h_{22}$ mode but, as for the $h_{22}$ mode, the dependency on the spin of the central BH is not monotonic. 

\subsubsection{$\ell \neq m$ modes}
In our work we also consider higher order modes with $\ell \neq m$, in particular we consider the off-diagonal $h_{21}$, $h_{32}$ and $h_{43}$ modes. As for the $\ell = m$ case, we first qualitatively inspect the behaviour of the peak of the different modes on the parameter space. As in the $\ell = m$ case, we find that whenever the peak of the amplitude of the $h_{22}$ mode exists also the peak of these higher order modes is present, as can be observed in Fig.~\ref{Fig.: hlm hierarchy} for a subset of waveforms. From the figure it is also possible to notice how the $\ell \neq m$ modes are more affected by the mode-mixing, also when considering eccentric systems. Without going into more details, we only mention that for retrograde spins the mixing is mostly with the mirror fundamental quasi-normal-mode (see e.g. Ref.~\cite{Krivan:1997hc}). While for prograde spins the mixing comes primarily from the lower-$\ell$ harmonics with the same $m$. 
Many works have studied the nature of this mixing, and since its analysis is not among the scope of this article we refer the reader to Ref.~\cite{Taracchini:2014zpa} for more details, noting that the inclusion of eccentricity does not seem to have a large impact on the phenomenology relative to the QC case. 

In this work, we perform the same analysis we do for the $\ell = m$ case for the considered $\ell \neq m$ modes, and we find that in general the $h_{21}$, $h_{32}$ and $h_{43}$ modes have a merger structure that is different from the structure of the merger of the $\ell = m$ modes. However, we find that the $h_{21}$, $h_{32}$ and $h_{43}$ have a similar merger structure among them. Furthermore, we also find that the merger structure of the negative spin scenarios presents differences with respect to the merger structure of the positive spin scenarios. We link these different behaviours to the presence of the mode-mixing mentioned above and we provide more details about this analysis and characterization in Appendix~\ref{Sec.: ell = m higher order modes similar behaviour}. 

\subsection{Bondi news and $\psi_4$ peaks characterization} \label{Sec.: Bondi news and psi_4 peaks characterization}
\begin{figure*}[t!]
	\includegraphics[width=1.\linewidth]{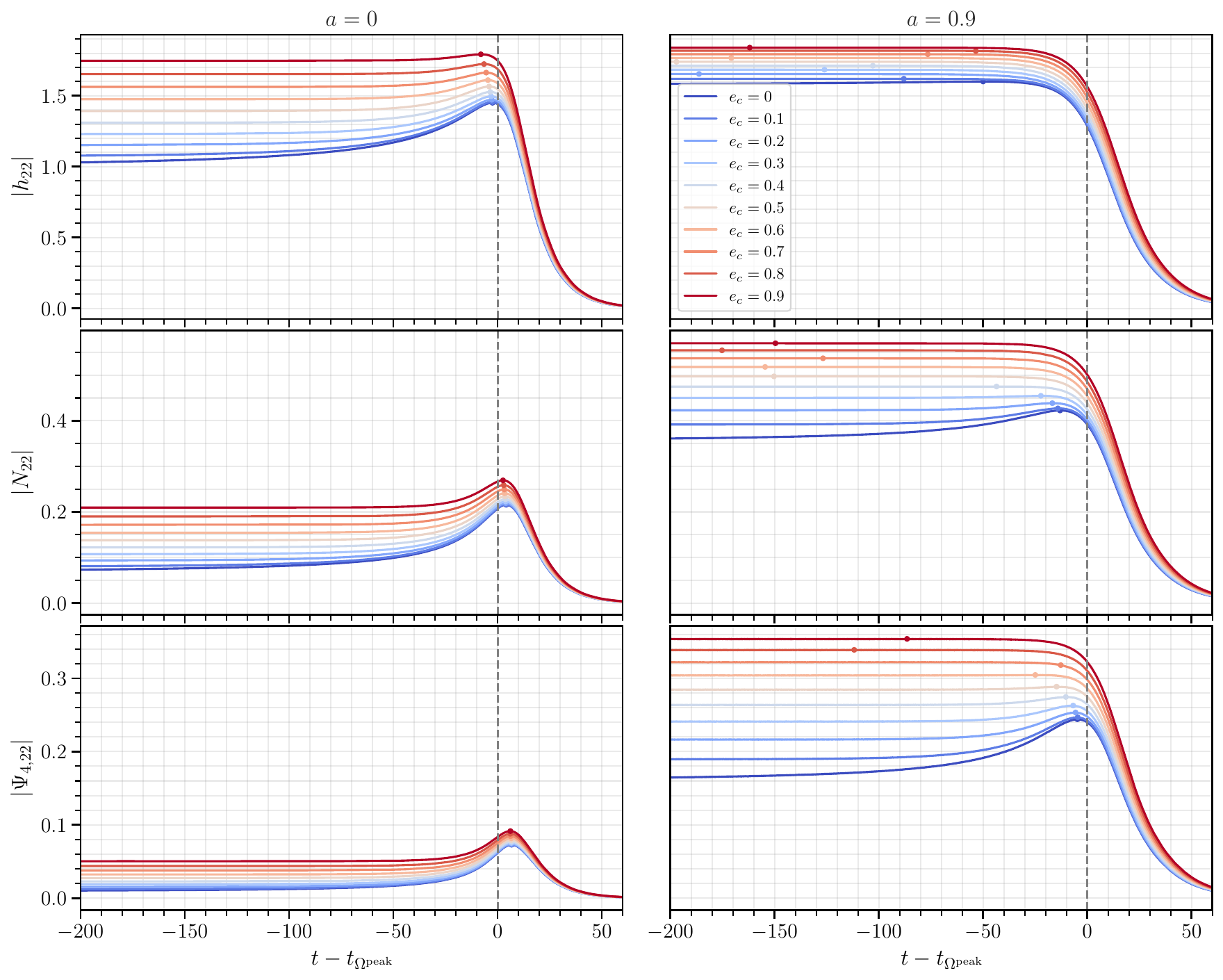}
	\caption{In this plot we show the amplitude peak behaviour of the $(2,2)$ spin-weighted spherical harmonic mode of the strain, $h_{22}$ (first row), Bondi news $N_{22}$ (second row) and Weyl scalar $\psi_{4, 22}$ (third row). In the left column we consider spin $a = 0$ configurations, while the plots in the right column correspond to spin $a = 0.9$ configurations. The different colors represent different critical eccentricities, ranging from $e_c = 0$ to $e_c = 0.9$. The peak amplitudes of all the quantities and the time of the peak amplitude exhibit a dependence on the eccentricity, as already mentioned in Fig.~\ref{hlm_amplitudes_for_two_spins_compared}. We notice that both the amplitudes of $N_{22}$ and of $\psi_{4, 22}$ are characterized by regimes for which their peaks disappear, as for the strain $h_{22}$. However the threshold eccentricities for which they disappear are different from the threshold eccentricity $e^{\rm thr}_c(a)$ we identified for the $h_{22}$ mode in Sec.~\ref{Subsec: Waveforms characterization}.}
	\label{Bondi_news_and_Psi_4_peaks}
\end{figure*}
In Sec.~\ref{Subsec: Waveforms characterization} we characterized the existence of the peak of the amplitude of the $h_{22}$ mode in the parameter space. For the sake of completeness we aim to understand how the amplitude peaks behave for two other quantities related to the waveform modes, the Weyl scalar $\psi_4$ defined in Eq.~\eqref{Eq.: psi_4 definition} and the Bondi news $N$, defined by
\begin{equation} \label{Eq.: Bondi news definition}
    N = \frac{1}{2}\frac{\partial h}{\partial t} = \frac{1}{2} \left( \frac{\partial h_{+}}{\partial t} - i\frac{\partial h_{\times}}{\partial t} \right) .
  \end{equation}
In particular, as for the strain $h$, we are interested in the 
$(2,2)$ spherical harmonic mode of these two quantities, $N_{22}$ and $\psi_{4,22}$.
In the second and third rows of Fig.~\ref{Bondi_news_and_Psi_4_peaks} we show the amplitudes of these quantities $|N_{22}|$ and $|\psi_{4,22}|$ generated by trajectories with critical eccentricities $e_c = [0, 0.1, 0.2, 0.3, 0.4, 0.5, 0.6, 0.7, 0.8, 0.9]$ and for two spin configurations of the central BH: on the left column we consider $a = 0$ while on the right column we consider $a= 0.9$. The first row of Fig.~\ref{Bondi_news_and_Psi_4_peaks} shows the amplitude of the $h_{22}$ mode: for $a=0$ the values of $e_c$ are such that the peak of the amplitude is always present. For $a=0.9$, this is not the case, since we consider values of critical eccentricity that are greater than the threshold value $e^{\rm thr}_c$ we identified in our characterization performed in Sec.~\ref{Subsec: Waveforms characterization} and illustrated in Fig.~\ref{h22_peaks_existence_curve}. Note that this result was already found for the QC case (i.e. $e_c = 0$) for values of spin $a>0.95$ in Ref.~\cite{Taracchini:2014zpa}.

When analyzing the $N_{22}$ and $\psi_{4,22}$, we find that, similar to the $h_{22}$ mode, their amplitude peak ceases to exist beyond certain threshold values of $e_c$. However, we observe a different pattern when considering the $N_{22}$ and $\psi_{4,22}$, which we show in the second and third rows of Fig.~\ref{Bondi_news_and_Psi_4_peaks}, respectively. We find that the peaks' existence phenomenology of these quantities is different. In fact, by focusing on the $a=0.9$ case, we observe that the threshold values of the critical eccentricity, from which the peak of the amplitudes disappears, for the Bondi news $e^{\rm thr}_{N_{22}}$ and for the Weyl scalar $e^{\rm thr}_{\psi_{4,22}}$ are different compared to the $h_{22}$ mode, and they appear to be greater, i.e., $e^{\rm thr}_c<e^{\rm thr}_{N_{22}}<e^{\rm thr}_{\psi_{4,22}}$. For example, for $a=0.9$ and $e_c = 0.4$, the $h_{22}$ does not show any peak. However, both the Bondi news and the Weyl scalar manifest a peak at this value of the critical eccentricity. A possible explanation of this behaviour is linked to the fact that $N_{22} \sim \dot{h}_{22} \approx \omega_{22} h_{22}$ and $\psi_{4,22} \sim \ddot{h}_{22} \approx (\omega_{22})^2 h_{22}$, where $\omega_{22}$ is the frequency of the $h_{22}$ mode. Hence, the presence of the peak of the amplitudes of $N_{22}$ and $\psi_{4,22}$ changes on the parameter space $(a, e_c)$ due to the fact that $|N_{22}|$ and $|\psi_{4,22}|$ inherit the features of the factor $\omega_{22}$.

\section{Conclusions} \label{Sec: conclusions}
In this work, we characterized the properties at merger of the GW modes for equatorial eccentric configurations of the Kerr metric in the TM limit. This is done by considering a particular class of critical plunge geodesics parametrized by the spin $a$ of the central Kerr BH and by the critical eccentricity $e_c$ introduced in Sec.~\ref{Sec.:critical plunge geodesics}. We employed a TD Teukolsky code to produce the GW signals generated by this class of plunge geodesics.

By inspecting different configurations of the spin $a$ and different $e_c$, we found that the amplitude of the $h_{22}$ mode exhibits a peak in a particular part of the parameter space, while this peak disappears in a disjoint part of the parameter space. In Sec.~\ref{Sec.: methodology} we introduced a methodology that allows to characterize the existence of the peak of the amplitude of the $h_{22}$ mode and we identified the region of the parameter space for which its peak exists, as we show in Fig.~\ref{h22_peaks_existence_curve}. This behavior is qualitatively understood by studying the analytic expression for $h_{22}^{\rm N}$, the Newtonian order part of the $h_{22}$, which in turn is dominated by the quantity $r \Omega$. The location of the peak of $r \Omega$ is at a fixed value of $r$, and the disappearance of the peak for larger values of $e_c$ is explained by the critical plunge geodesics starting at a radius $r_c$ that is smaller than this value.

We restricted the merger characterization of the waveform modes to the scenarios for which the peak of the amplitude of the $h_{22}$ mode exists. On the part of parameter space $(a, e_c)$ for which the peak of the $h_{22}$ mode exists, we described the quantities usually employed in the modeling of the merger and post-merger parts of gravitational waveforms . These quantities correspond to the amplitude $ |h_{\ell m}^{\rm peak}| $, second-order time derivative of the amplitude $ d^2 |h_{\ell m}| /dt^2|_{\rm peak} $, frequency $ \omega_{\ell m}^{\rm peak} $, and first-order time-derivative $ \dot{\omega}_{\ell m}^{\rm peak} $ of the $h_{\ell m}$ waveforms modes evaluated at the peak of the amplitude of the modes~\cite{Barausse:2011kb, Cotesta:2018fcv, Riemenschneider:2021ppj, Pompili:2023tna}. 
In our work we also inspected the higher-order $h_{33}$, $h_{44}$, $h_{55}$, $h_{21}$, $h_{32}$ and $h_{43}$ modes. For the $\ell = m$ modes, we found that the peak of the amplitude also disappears at a certain value of $e_c$. However, this threshold value is larger than the value, we found for the $h_{22}$ mode. Consequently, when the peak of the $h_{22}$ mode exists the peaks of the other $\ell = m$ modes also exist. 
For these higher-order modes, we found that the quantities at the peak exhibit a similar dependence on $e_c$ and $a$ as for the case of the $h_{22}$ mode, as shown in Fig.~\ref{Fig.: h22_h33_h21_IVs_compared}, and in Fig.~\ref{h33_NQC_quantities_vs_h22_quantities_only_IVs} we shown a different way of parametrizing the quantities at the peak. This different parametrization describes the quantities $ |h_{\ell m}^{\rm peak}| $, $ d^2 |h_{\ell m}| /dt^2|_{\rm peak} $, $ \omega_{\ell m}^{\rm peak} $, and $ \dot{\omega}_{\ell m}^{\rm peak} $ of the $\ell = m > 2$ modes as function of the same quantities of the $h_{22}$ mode. In Fig.~\ref{h33_NQC_quantities_vs_h22_quantities_only_IVs} we specialized to the case $\ell = m = 3$.

Finally, we also explored what happens to the peak of the amplitude of the $(2,2)$ mode of the Bondi news $N$ and of the Weyl scalar $\psi_4$. We observed that also for these quantities the peak of the amplitude of the dominant spin-weighted spherical harmonic mode disappears at a certain value of $e_c$. This threshold value is different from the threshold value, we found and characterized for the $h_{22}$ mode, and appears to follow the general rule that $e_{\rm thr}<e_{\rm thr}^{N_{22}}<e_{\rm thr}^{\psi_{4,22}}$. Although, we have not exhaustively probed the parameter space to confirm this.

The methodology we employed in this work unveils the properties of the merger and post-merger parts of the gravitational waveforms in the TM limit of the eccentric equatorial Kerr scenario. This work can be extended in multiple directions: to study the merger properties of the waveforms in the more general case of non-equatorial eccentric critical plunge geodesics of Kerr and to study the impact of eccentricity in the ringdown part of the waveforms. Furthermore, it can be possible to extend current phenomenological EOB QC ringdown models in the TM limit~\cite{Barausse:2011kb, Taracchini:2014zpa, Pompili:2023tna} to take into account eccentricity and spin.

We believe all these directions will provide important insights about the nature of the merger and post-merger of the waveforms emitted by a TM plunging and merging in a Kerr BH and they will be essential in informing the current~\cite{Gamboa:2024hli, Nagar:2024dzj} and future gravitational-waveform models with TM limit information.

\section*{Acknowledgments}
The authors are grateful to Hector Estelles, Aldo Gamboa, Marcus Haberland, Elisa Maggio, Alessandro Nagar, Nami Nishimura, Lorenzo Pompili, Stefano Savastano and Sebastian Völkel for the useful comments and discussions.
MvdM acknowledges financial support by 
the VILLUM Foundation (grant no. VIL37766),
the DNRF Chair program (grant no. DNRF162) by the Danish National Research Foundation and the MPI for Gravitational Physics,
and the European Union’s Horizon ERC Synergy Grant “Making Sense of the Unexpected in the Gravitational-Wave Sky” grant agreement no.\ GWSky–101167314. 
G.K. acknowledges support from US National Science Foundation grants No. DMS-2309609 and PHY-2307236. Simulations were performed on the UMass-URI UNITY HPC/AI  supercomputer supported by the Massachusetts Green High Performance Computing Center (MGHPCC). 

This work makes use of the Black Hole Perturbation Toolkit~\cite{BHPToolkit}.

%\clearpage % Ensure bibliography starts on a new page
\appendix
\section{Why critical plunge geodesics?} \label{Sec.: appendix why critical plunges?}
\begin{figure}
  	\includegraphics[width=1.\linewidth]{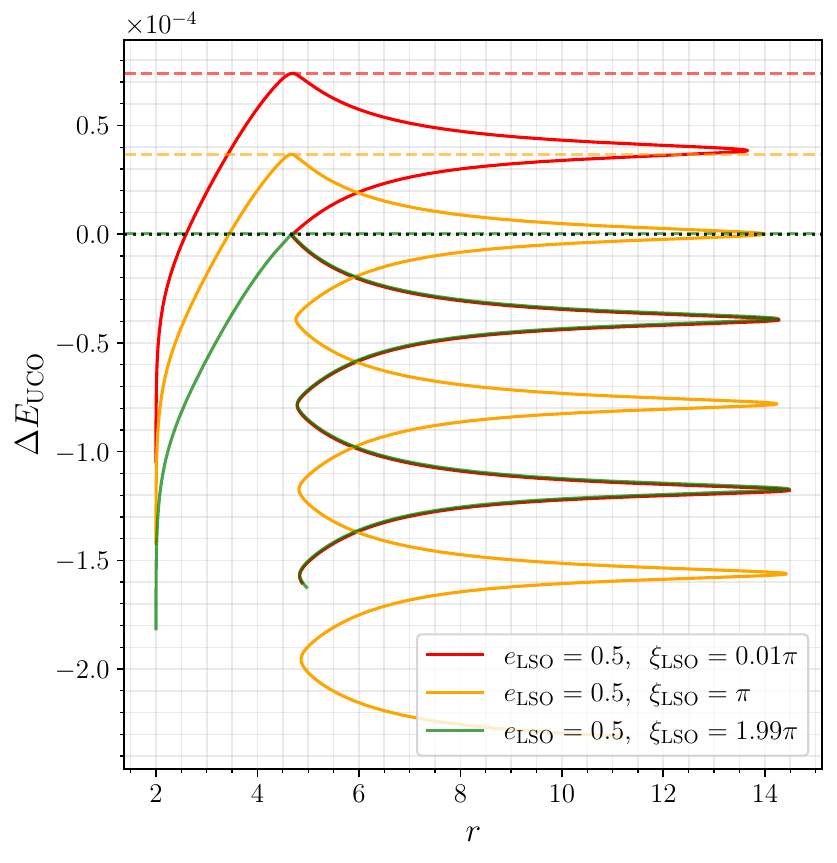}
  	\caption{Evolution of the quantity $\Delta E_{\rm UCO}$ as a function of the radial coordinate $r$ for three trajectories characterized by $\nu = 10^{-3}$, $e_{\rm LSO} = 0.5$, and three distinct values of $\xi_{\rm LSO} = [0.01\pi, \pi, 1.99\pi]$ corresponding respectively to the scenarios \textit{ii} (red curve), \textit{iii} (orange curve), \textit{i} (green curve) mentioned in the main text. The quantity $\Delta E_{\rm UCO}$ exhibits an increasing monotonic trend until it reaches a maximum (after the LSO crossing) and starts to decrease monotonically during the plunge. The horizontal dashed lines highlight the maximum value $\Delta E_{\rm UCO}^{\rm onset}$ which defines the onset of the plunge. }
  	\label{evolution E_c vs r}
\end{figure}
\begin{figure}
	\includegraphics[width=1.\linewidth]{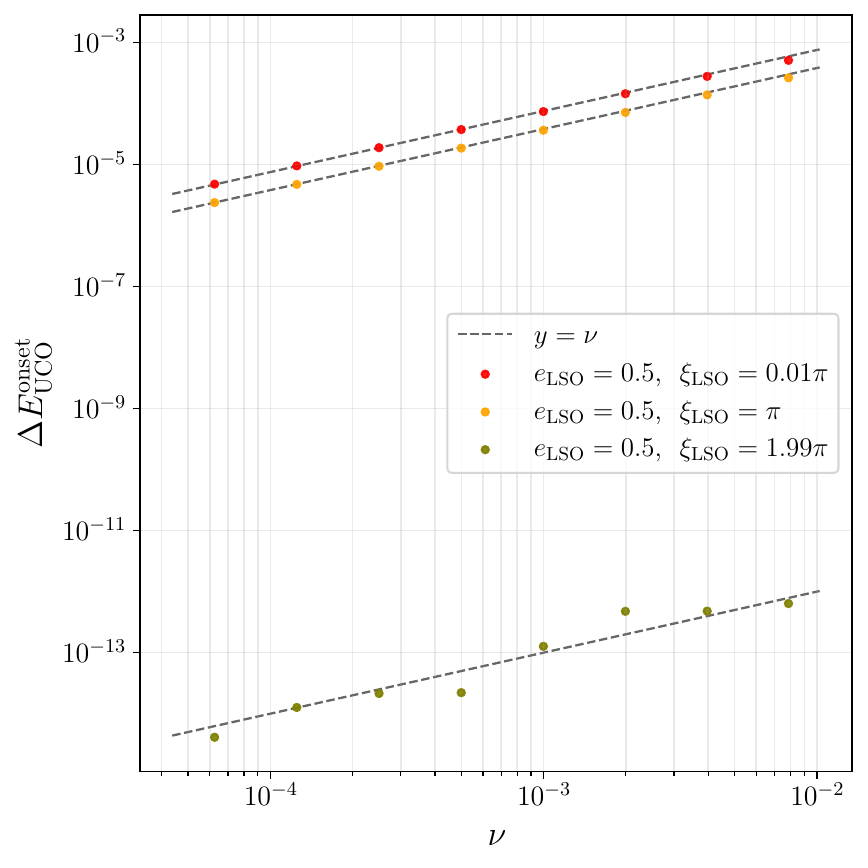}  	
	\caption{Scaling of $\Delta E_{\rm UCO}^{\rm onset}$ with respect to the symmetric mass ratio $\nu$. The data is shown for fixed eccentricity $e_{\rm LSO} = 0.5$ and three configurations of the radial phase at the LSO, $\xi_{\rm LSO} = [ 0.01\pi, \ \pi , \ 1.99\pi ]$, respectively represented with red, orange and green dots. These configurations correspond respectively to the scenarios \textit{ii}, \textit{iii}, \textit{i} mentioned in the main text of Appendix~\ref{Sec.: appendix why critical plunges?}. We observe a scaling $\Delta E_{\rm UCO}^{\rm onset} \sim \nu$ for all the different $\xi_{\mathrm{LSO}}$. }
	\label{Delta_E_USO_scaling}
\end{figure}

The class of critical plunge geodesics, for which analytical closed form expressions can be found in Refs.~\cite{Mummery:2023hlo, Dyson:2023fws} and that we reviewed in Sec.~\ref{Sec.:critical plunge geodesics}, describes the final part of the trajectories of a TM in Kerr spacetime. In this limit, when the symmetric mass ratio $\nu = \mu/M \to 0$, the dynamics of the plunge becomes geodesic~\cite{Buonanno:2000ef, Ori:2000zn}, and it converges to a critical plunge geodesic, see~\cite{Lhost:2024jmw} for an analytical treatment in Schwarzschild.

To have better insight into this, it is instructive to inspect the phenomenology of the eccentric plunge transition of a small mass $\mu$ orbiting the equatorial plane of the Kerr spacetime, and then consider the TM limit. 
During the inspiral phase, the bound system evolves adiabatically and the orbital motion of the small mass is represented as a sequence of eccentric geodesics, which at any given instant can be parametrized by the eccentricity $e$, the semilatus rectum $p$, and the relativistic anomaly $\xi$ expressed in Eq.~\eqref{eq.: e, p, xi def}.

In the eccentric case, as the system approaches the LSO, the plunge dynamics that follow, exhibit richer phenomenology compared to the QC case, which, by contrast, follows a universal behavior~\cite{Buonanno:2000ef, Ori:2000zn}. Notably, the features of the eccentric plunge depend sensitively on the value of the relativistic anomaly $\xi_{\rm LSO}$ at the LSO~\cite{Becker:2024xdi, Nee:2025zdy}.
Combining the expressions in Eq.~\eqref{eq.: e, p, xi def} of $e$ and $p$ evaluated at the LSO (i.e., with $r_2 = r_3 = r_c$, as summarized in Sec.~\ref{Sec.:critical plunge geodesics}) in the definition of the relativistic anomaly $\xi$, it reads
\begin{equation} \label{eq.:rel anomlay at LSO}
\cos \xi_{\rm LSO} = \frac{2 r_1 r_c - r_{\rm LSO} (r_1+r_c) }{r_{\rm LSO} (r_1 - r_c)},
\end{equation}
where, $r_c$ is the critical radius introduced in Sec.~\ref{Sec.:critical plunge geodesics}, $r_{\rm LSO}$ corresponds to the value of the radial coordinate $r$ when the TM is at the LSO and where $r_1$ is the apocenter radial coordinate at the LSO.  We remark that Eq.~\eqref{eq.:rel anomlay at LSO} typically has two solutions between $0$ and $2\pi$. We solve this degeneracy by considering the time derivative of the radial coordinate $\dot{r}$ at the LSO.

Depending on the value of $\xi_{\rm LSO}$, it is possible to identify three possible scenarios that characterize the eccentric plunge transition: 
\begin{enumerate}
	\item[i)] 	when $\xi_{\rm LSO}= 0^{-}$
		\footnote{We are considering $\xi_{\rm LSO}$ mod $2\pi$. }
		\footnote{The critical case $\xi_{\rm LSO}= 0$ produces the same critical phenomenon for critical scattering as studied (in Schwarzschild) in Ref.~\cite{Gundlach:2012aj}, but reaching the UCO from a bound orbit rather than a scattering orbit; the TM continues orbiting the UCO, which radius $r_c$ grows under the loss of energy, until it reaches a transition regime when it the reaches the ISCO. }, 
		equations~\eqref{eq.:rel anomlay at LSO} and~\eqref{eq:radialeom} imply $r_{\rm LSO} = r_c$ and $(dr/d\lambda)_{\rm LSO} = 0$, the TM orbits the UCO and then plunges into the BH,
	
	\item[ii)]  when $\xi_{\rm LSO}= 0^{+}$ the TM orbits the UCO but then performs a last radial cycle before plunging,
	
	\item[iii)]  when $0 < \xi_{\rm LSO} <  2\pi$, the LSO occurs when $r_c<r_{\rm LSO}<r_1$ and the TM completes a last fraction of radial cycle before plunging.
\end{enumerate} 
To better understand how the plunges of these scenarios converge to the critical plunge geodesic in the TM limit, it is instructive to consider the difference between the energy of the system $E$ and the energy $\mathcal{E}_{\rm UCO}$ of the UCO with the same angular momentum $L$,
\begin{equation}
\Delta E_{\rm UCO} = E - \mathcal{E}_{\rm UCO}(L) \ .
\end{equation}
As long as $ E_{\rm UCO}$ is negative the centrifugal barrier keeps the particle in an eccentric orbit outside the UCO. However, when $ E_{\rm UCO}>0$ the particle is free to pass across the radius of the UCO and start its plunge.
We remark on the fact that while the system evolves both $E$ and $\mathcal{E}_{\rm UCO}$ decrease because of the gravitational radiation (in particular $\mathcal{E}_{\rm UCO}$ evolves through the evolution in time of $L$), however they generally evolve with different rates.

We numerically study the behaviour of the quantity $\Delta E_{\rm UCO}$ by considering trajectories generated by evolving the Hamiltonian system described in Eqs. (5a)-(5d) of Ref.~\cite{Faggioli:2024ugn}. The system incorporates the dissipative effects due to GWs emission through a radiation-reaction (RR) force that includes resummed PN eccentric corrections~\cite{Khalil:2021txt, Faggioli:2024ugn, Gamboa:2024imd}.
In Fig.~\ref{evolution E_c vs r} we show the evolution of the quantity $\Delta E_{\rm UCO}$ with respect to the radial coordinate $r$ for three trajectories with $\nu = 10^{-3}$, $e_{\rm LSO}=0.5$ and $\xi_{\rm LSO} = [ 0.01\pi, \ \pi, \ 1.99\pi ]$ orbiting Schwarzschild. These three values of $\xi_{\rm LSO}$ yield, respectively, to the scenarios \textit{ii} (red curve), \textit{iii} (orange curve), \textit{i} (green curve) described above. Figure~\ref{evolution E_c vs r} shows that during the inspiral $\Delta E_{\rm UCO}<0$ and monotonically increases. When the system reaches the LSO configuration by definition
\begin{equation}\label{Eq. E_LSO condition}
E = E_{\rm LSO} = \mathcal{E}_{\rm UCO} = \mathcal{E}_c \ ,
\end{equation}
and $\Delta E_{\rm UCO}=0$. Note that the three curves in Fig.~\ref{evolution E_c vs r} reach this configuration having different values of the radial coordinate corresponding to the different radial phase $\xi_{\rm LSO}$ values.
 
After the LSO, the three scenarios evolve differently before plunging, but for all of them the plunge begins when $r \approx r_c$. 
In this configuration $\Delta E_{\rm UCO}$ reaches its maximum value and it starts to decrease monotonically until the end of the plunge. We define the moment for which $\Delta E_{\rm UCO}$ is in its maximum as the \textit{onset} of the plunge. 
In Fig.~\ref{evolution E_c vs r} we highlight this configuration with three horizontal dashed lines of the same color of the related $\Delta E_{\rm UCO}$ curves. 
Since the onset of the plunge happens after the LSO crossing and when $r \approx r_c$ we have $\Delta E_{\rm UCO}^{\rm onset} \geq 0$. 
As we motivate in the following, if $\Delta E_{\rm UCO}^{\rm onset} \rightarrow 0$ in the TM limit, then the different plunges of scenarios \textit{i},\textit{ii} and \textit{iii} converge to the critical plunge geodesic.
In fact, as $\nu \rightarrow 0$ the plunge becomes geodesic~\cite{Buonanno:2000ef, Ori:2000zn} and the value of the energy of the plunge geodesic freezes at the value it has at its onset, i.e. $E^{\rm plunge} = E^{\rm onset}$. 
Hence, by showing that $\Delta E_{\rm UCO}^{\rm onset} \rightarrow 0$ in the TM limit, then, from Eq.~\eqref{Eq. E_LSO condition} we get $E^{\rm plunge} = E^{\rm onset}\rightarrow \mathcal{E}_c$ .

To study the behaviour of $\Delta E_{\rm UCO}^{\rm onset}$ when $\nu \rightarrow 0$, it is insightful to consider the total time derivative of $\Delta E_{\rm UCO}$
\begin{align} \label{Eq: derivative delta E UCO}
\Delta \dot{E}_{\rm UCO} & = \dot{E} - \dot{\mathcal{E}}_{\rm UCO} \nonumber \\
& = \dot{E} - \frac{d\mathcal{E}_{\rm UCO}}{dL} \dot{L} \nonumber \\ 
& = \dot{E} - \Omega_{\rm UCO}\dot{L},
\end{align}
where 
$\Omega_{\rm UCO}$ corresponds to the orbital frequency of the UCO. In the last passage of Eq.~\eqref{Eq: derivative delta E UCO} we employ the fact that $L(t)=\mathcal{L}_{\rm UCO}(t)$ and that $d\mathcal{E}_{\rm UCO}/d\mathcal{L}_{\rm UCO} = \Omega_{\rm UCO}$~\cite{Bardeen:1972fi}.
Since both $\dot{E}$ and $\dot{L}$ scale with $\nu$, also $\Delta \dot{E}_{\rm UCO} \sim \nu$. Accordingly, we introduce the notation  $\Delta \dot{E}_{\rm UCO} = \nu \mathcal{F}_{\Delta E_{\rm UCO}}$.
Furthermore, we highlight the fact that the quantity $\mathcal{F}_{\Delta E_{\rm UCO}}$ vanishes 
at the UCO, since on circular orbits $\dot{E} = \Omega \dot{L}$~\cite{Bardeen:1972fi}. 

In general, the value of $\Delta E^{\rm onset}_{\rm UCO} \ge 0$ and it is bounded by the value of $\Delta E^{\rm onset, \textit{ii}}_{\rm UCO}$ of the the scenario \textit{ii}, which represents the configuration for which $\Delta E^{\rm onset}_{\rm UCO}$ reaches its maximum among all the possible scenarios, as depicted in Fig.~\ref{evolution E_c vs r}:
\begin{align} \label{Eq: bound delta E onset UCO}
\Delta E^{\rm onset}_{\rm UCO} \le \Delta E^{\rm onset, \textit{ii}}_{\rm UCO} & = \nu\int_{r_c}^{r_{\rm la}}\mathcal{F}_{\Delta E_{\rm UCO}}\frac{dt}{dr}dr \nonumber \\ 
& +\nu\int_{r_{\rm la}}^{r_{\rm on}}\mathcal{F}_{\Delta E_{\rm UCO}}\frac{dt}{dr}dr,
\end{align}
where $r_c$ is the critical radius introduced in Sec.~\ref{Sec.:critical plunge geodesics}, $r_{\rm la}$ corresponds to the radial position of the \textit{last apocenter} passage of the scenario \textit{ii} and $r_{\rm on}$ is the value of the radial coordinate of the TM when it is at the onset of the plunge. While the second integral (computed between $r_{\rm la}$ and $r_{\rm on}$) is always finite, this is not as obvious for the first integral, since at $r=r_c$ the quantity $dt/dr$ diverges. However, as mentioned above, at the UCO $\mathcal{F}_{\Delta E_{\rm UCO}}$ vanishes, and we find empirically that this happens quick enough to render the integral finite, and conjecture that this is true in general, and remains bounded in the $\nu\to 0$ limit.
Equation~\eqref{Eq: bound delta E onset UCO} shows that $\Delta E^{\rm onset, \textit{ii}}_{\rm UCO} \sim \nu$ and since this quantity bounds $\Delta E^{\rm onset}_{\rm UCO}$, it holds
\begin{equation} \label{Eq.: Delta Eonset UCO TP limit}
\Delta E^{\rm onset}_{\rm UCO} \rightarrow 0 \ , \ \text{when} \ \nu \rightarrow 0.
\end{equation}

As introduced above, when considering the TM limit, the plunge becomes geodesic, and its energy corresponds to the value $E^{\rm plunge} = E^{\rm onset}$ the system has at the onset of the plunge. Since in the TM  limit $\Delta E^{\rm onset}_{\rm UCO} \rightarrow 0$, it follows from Eq.~\eqref{Eq. E_LSO condition} that the plunge becomes geodesic with energy $E^{\rm plunge} \to \mathcal{E}_c$, converging to a critical plunge geodesic with energy $\mathcal{E}_c$.

We numerically test the limit in Eq.~\eqref{Eq.: Delta Eonset UCO TP limit} and study the scaling of $\Delta E^{\rm onset}_{\rm UCO}$ by considering a set of eccentric trajectories like in Fig.~\ref{evolution E_c vs r} but with different symmetric mass ratios. The trajectories we consider have residual eccentricity at the LSO $e_{\rm LSO} = 0.5$, different symmetric mass ratios 
$\nu = [0.06, \ 0.13, \ 0.25, \ 0.50, \ 1.00, \ 2.00, \ 4.00, \ 8.00]\times 10^{-3}$
and we examine the same three families of trajectories characterized by values of the relativistic anomaly at LSO, 
$\xi_{\rm LSO} = [ 0.01\pi, \ \pi, \ 1.99\pi ]$  
that correspond, respectively, to the scenarios \textit{ii}, \textit{iii}, \textit{i}, as in Fig.~\ref{evolution E_c vs r}.

\begin{figure*}
	\includegraphics[width=1.\linewidth]{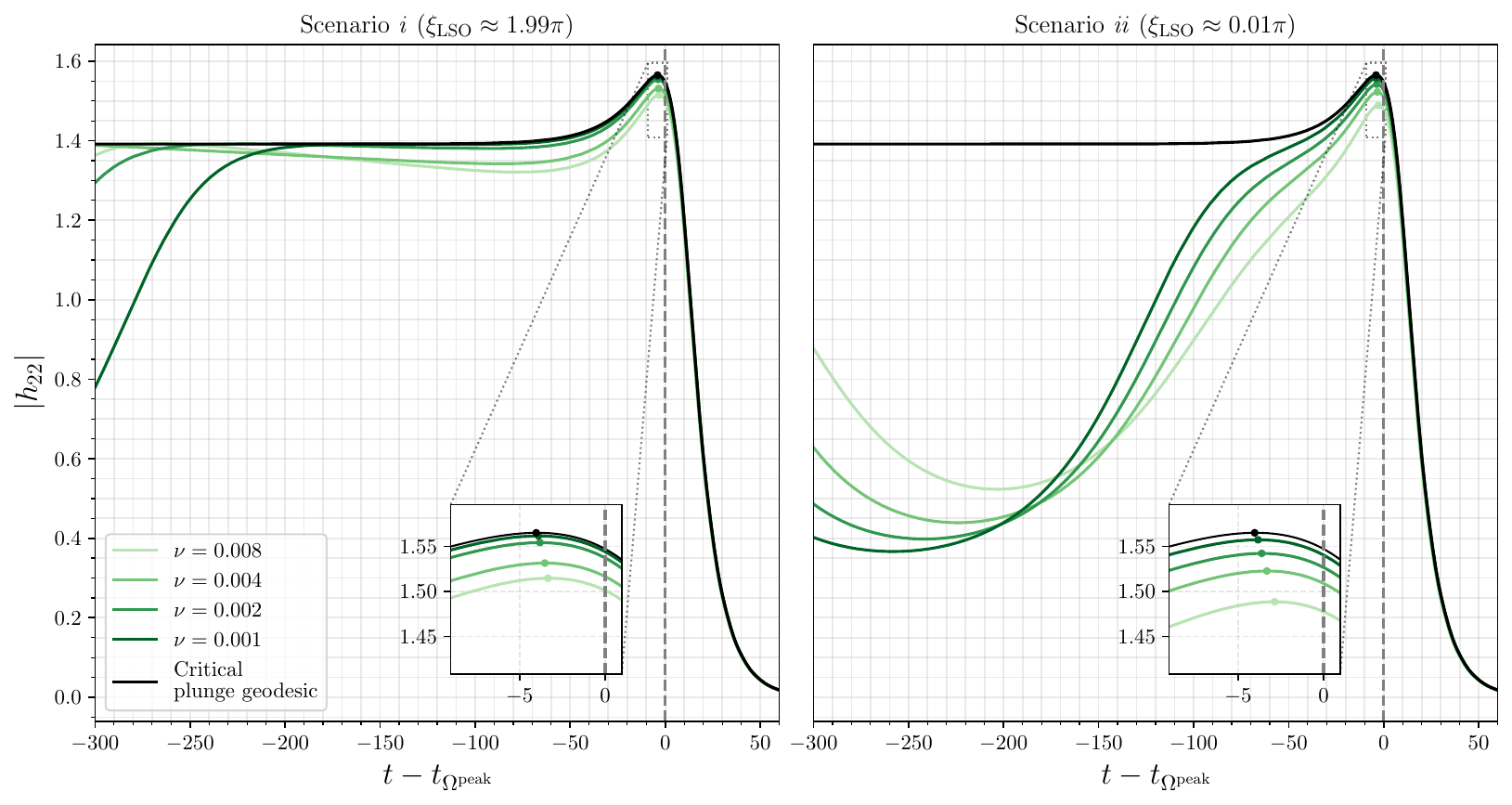}
	\caption{Comparison of the amplitude of the $h_{22}$ mode generated considering different trajectories with different mass ratios and same eccentricity $e_{\rm LSO} = 0.5$ at the LSO. The different green shades represent the $|h_{22}|$ generated considering a subset of the trajectories evolved with the RR force of Ref.~\cite{Faggioli:2024ugn} introduced in Appendix~\ref{Sec.: appendix why critical plunges?}. 
		We consider the subset with different symmetric mass ratios $\nu = [0.008, \ 0.004, \ 0.002, \ 0.001]$ and $\xi_{\rm LSO} = [1.99\pi, \ 0.01\pi]$. The black curve corresponds to the amplitude of the $h_{22}$ mode of the critical plunge geodesic having the same properties of the RR evolved trajectories at the LSO , i.e. having  $e_c = e_{\rm LSO} = 0.5$. In the left panel we show the case with $\xi_{\rm LSO} = 1.99\pi$, corresponding to the scenario \textit{i} mentioned in Appendix~\ref{Sec.: appendix why critical plunges?}, while in the right panel we show the case with $\xi_{\rm LSO} = 0.01\pi$ corresponding to the scenario \textit{ii}. With this test, we phenomenologically corroborate the fact that as the symmetric mass ratio $\nu$ decreases, the merger properties of the mode  converge to the merger properties of the mode generated employing critical plunge geodesics.}
	\label{different mass ratios merger convergence}
\end{figure*}

For each of these trajectories, we compute the quantity $\Delta E^{\rm onset}_{\rm UCO}$ at the onset of the plunge, when $r \approx r_c$. In Fig.~\ref{Delta_E_USO_scaling}, we show the scaling of $\Delta E^{\rm onset}_{\rm UCO}$ with respect to the the symmetric mass ratio $\nu$ for the three families of trajectories we considered. Red dots are the values connected to the scenario \textit{ii}, orange dots to the scenario \textit{ii} and green dots to the scenario \textit{i}. Figure~\ref{Delta_E_USO_scaling} shows that when $\nu \rightarrow 0$ the excess $\Delta E^{\rm onset}_{\rm UCO}$ scales as a power of $\nu$ which does not depend on the value of the radial phase at the LSO, since the slope of the lines is the same for every value of $\xi_{\rm LSO}$. This scaling corroborates what we described above, that in the TM limit, the plunge converges to the class of critical plunge geodesics discussed in Sec.~\ref{Sec.:critical plunge geodesics}. Figure~\ref{Delta_E_USO_scaling} also strongly suggests that $\Delta E^{\rm onset}_{\rm UCO}$ scales linearly with $\nu$, i.e., $\Delta E^{\rm onset}_{\rm UCO} \sim \nu$.

\section{Test-particle limit convergence of eccentric waveform merger properties} \label{Sec.: convergence with mass ratio}
In order to illustrate and corroborate the convergence of the merger properties of the waveforms to the merger properties of the critical plunge geodesic waveforms, we consider a subset of the trajectories mentioned in Appendix~\ref{Sec.: appendix why critical plunges?}, which are evolved through the use of the RR force of Ref.~\cite{Faggioli:2024ugn} and we generate the waveforms produced by these trajectories. In particular, we consider the subset of trajectories with symmetric mass ratio $\nu = [0.008, \ 0.004, \ 0.002, \ 0.001]$ of the two families having relativistic anomaly at LSO $\xi_{\rm LSO} = [0.01\pi, \ 1.99\pi]$, which correspond respectively to the scenarios \textit{ii} and \textit{i} introduced in Appendix~\ref{Sec.: appendix why critical plunges?}. In Fig.~\ref{different mass ratios merger convergence} we represent in different shades of green the amplitude of the $h_{22}$ mode, generated considering the different symmetric mass ratios. These waveforms have the same features at LSO, i.e. the same values of eccentricity at LSO $e_{\rm LSO} = 0.5$. The black curve represents the $h_{22}$ mode generated considering a critical plunge geodesic with with $e_c = e_{\rm LSO} = 0.5$, i.e., with critical eccentricity equal to the same value of eccentricity at LSO of the trajectories evolved with the radiation-reaction force of Ref.~\cite{Faggioli:2024ugn}. The left panel of Fig.~\ref{different mass ratios merger convergence} illustrates the configurations in the scenario \textit{i} , i.e. with $\xi_{\rm LSO} = 1.99\pi$, while the right panel considers the scenario \textit{ii} with $\xi_{\rm LSO}\ = 0.01\pi$. From Fig.~\ref{different mass ratios merger convergence} we observe that as the symmetric mass ratio $\nu$ decreases (curves with darker shades of green), the merger properties of the waveforms generated through the RR force evolved trajectories approach the merger properties of the waveform generated employing critical plunge geodesic, as highlighted by the inset plots in the figure.

From the figure it is also possible to observe the impact of the value at LSO of the relativistic anomaly $\xi_{\rm LSO}$ on the merger features of the waveforms. In particular, we mention the effect of the evolution of the small mass on the UCO, when $\xi_{\rm LSO} = 1.99\pi$. From the left panel of Fig.~\ref{different mass ratios merger convergence} we observe that the coordinate time interval $\Delta t_{\rm UCO}$ the small mass spends on the UCO ranges in between $\Delta t_{\rm UCO} \sim 200$ when $\nu = 0.001$ and $\Delta t_{\rm UCO} \sim 300$ for the higher mass ratio case with $\nu = 0.008$. For the cases characterized by higher values of $\nu$, the amplitude shows a clear decreasing trend. This trend is due to the fact that when the small mass evolution is stuck on the UCO, its orbital frequency $\Omega$ decreases and the radial coordinate $r$ (which corresponds to the radius $r_{\rm UCO}(t)$ of the UCO) increases with time, because of the evolution in time of the UCO configuration due to the radiated angular-momentum in the emission of GWs. The decreasing trend of $\Omega$ influences the waveform properties, and in particular it manifests with a decreasing behaviour of the amplitude of the modes (and of their frequencies), as can be observed in the left panel of Fig.~\ref{different mass ratios merger convergence} for the cases with higher values of $\nu$ (curves with lighter shades of green). The reason why this decreasing behaviour is more manifest for higher values of $\nu$ is connected to the fact that the amount of radiated angular-momentum scales with $\nu$, consequently, these effects are more evident with higher values of $\nu$.

Finally, another feature that we mention here and that is explored deeper in Refs.~\cite{Becker:2024xdi, Nee:2025zdy}, is the fact that the merger properties of the waveforms depend on the value of $\xi_{\rm LSO}$. By comparing the inset plots in the two panels of Fig.~\ref{different mass ratios merger convergence} it is possible to notice, for example, how the values of the peak of the amplitudes of the $h_{22}$ mode differ for waveforms that have the same LSO configuration but not the same $\xi_{\rm LSO}$. However, in the TM limit, as $\nu \rightarrow 0$, these differences vanish and the waveform's merger converges to the critical plunge geodesic merger, corroborating the fact that in this limit the eccentric aligned geodesic plunges converge to this particular class of geodesics.

\section{Comparison of $\Delta t_{22}$ for $e_c = 0$ with previous works} \label{Sec.: Comparison of Delta t_22 with previous works}
\begin{figure}[tp!]
  	\includegraphics[width=1.\linewidth]{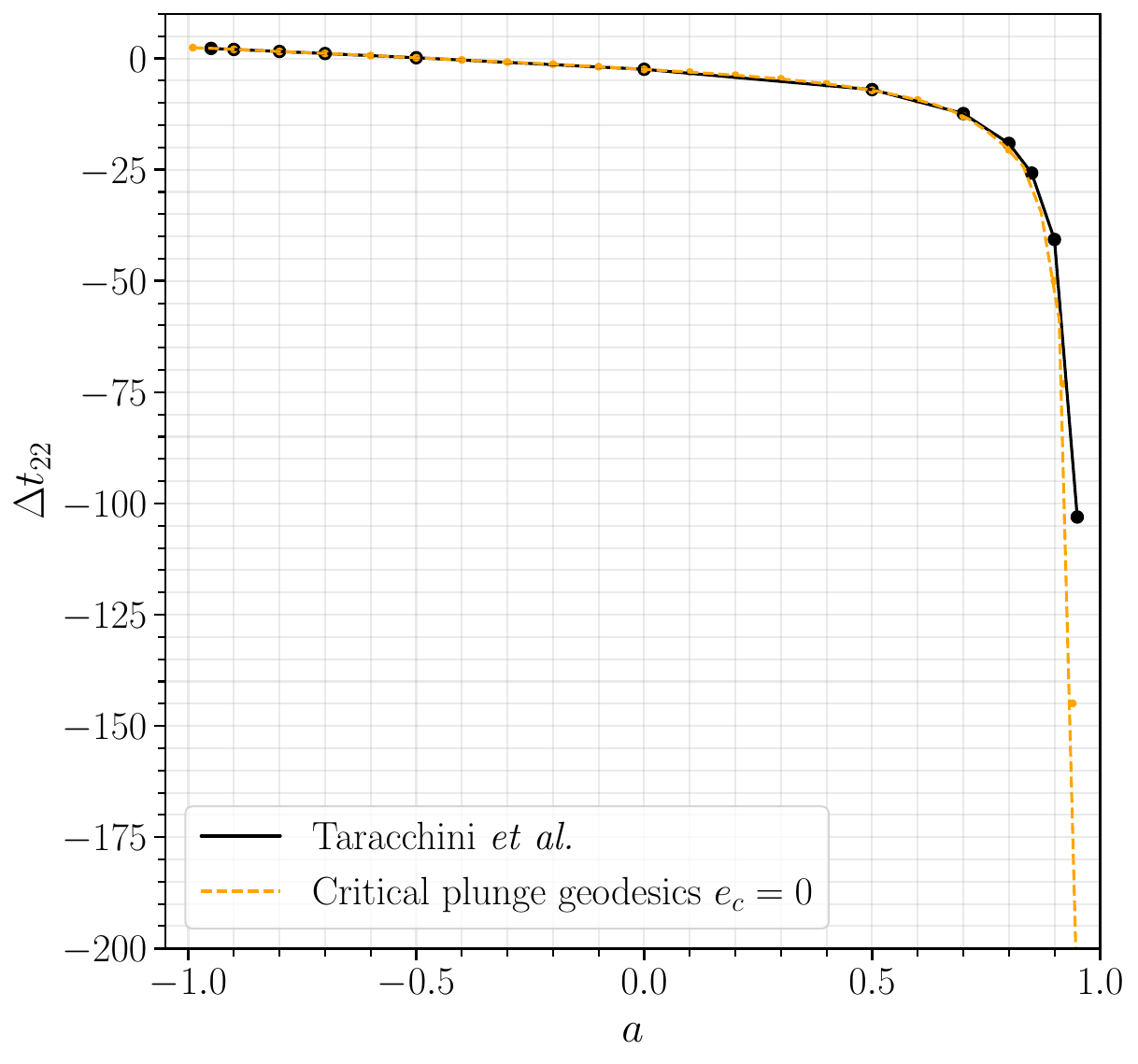}
\caption{In this figure we show the values of quantity $\Delta t_{22}$ defined in Eq.~\eqref{Eq: Delta_t_22 definition}. In orange we show $\Delta t_{22}$ values extracted from the $h_{22}$ mode generated using critical plunge geodesics with $e_c = 0$. In black we plot the same quantity computed in Ref.~\cite{Taracchini:2014zpa}, referred as \textit{Taracchini et al.}.}
\label{Fig.: comparison of Delta t_22 with previous works}
\end{figure}
In order to assess the consistency of our computation of the quantity $\Delta t_{22}$ defined in Eq.~\eqref{Eq: Delta_t_22 definition} with previous results in the literature, we perform a direct comparison with the values of $\Delta t_{22}$ obtained in Ref.~\cite{Taracchini:2014zpa}, hereafter referred to as \textit{Taracchini et al.}. In that work, the authors generated waveforms for the Kerr equatorial QC case, using the same methodology employed in the present analysis, but considering trajectories evolved through numerical fluxes computed via a frequency-domain Teukolsky code~\cite{Taracchini:2013rva, Taracchini:2013wfa}. The specific scenario considered in \textit{Taracchini et al.} corresponds to a system with $\nu = 0.001$.

In Fig.~\ref{Fig.: comparison of Delta t_22 with previous works} we show the comparison between the $\Delta t_{22}$ values extracted in \textit{Taracchini et al.} and the values extracted in our work considering critical plunge geodesics in the QC case, i.e. with $e_c = 0$.
In Fig.~\ref{Fig.: comparison of Delta t_22 with previous works} the black curve represents $\Delta t_{22}$ computed in \textit{Taracchini et al.} while the dashed orange curve represents $\Delta t_{22}$ extracted considering the geodesics. We observe that for spins up to $a \sim 0.85$, the values of $\Delta t_{22}$ extracted considering waveforms produced with the critical plunge geodesics are in agreement with the corresponding values computed by \textit{Taracchini et al.}.

However, for higher spin values $a > 0.85$, a discrepancy begins to emerge between the two works. We interpret this deviation as a consequence of the fact that the trajectories of the two works are different. Specifically, while \textit{Taracchini et al.} consider trajectories at finite $\nu = 0.001$, our computation strictly corresponds to the $\nu = 0$ limit. For lower spin values ($a \lesssim 0.85$), the influence of the difference in the symmetric mass-ratio on $\Delta t_{22}$ remains negligible, resulting in a good agreement. As the spin increases, however, these corrections become increasingly significant, since $\Delta t_{22}$ diverges for geodesics while remaining finite for finite values of $\nu$.

\section{More eccentric plunges behave more circular} \label{Sec.: circularization behaviour}
\begin{figure*}[htbp]
  	\includegraphics[width=1.\linewidth]{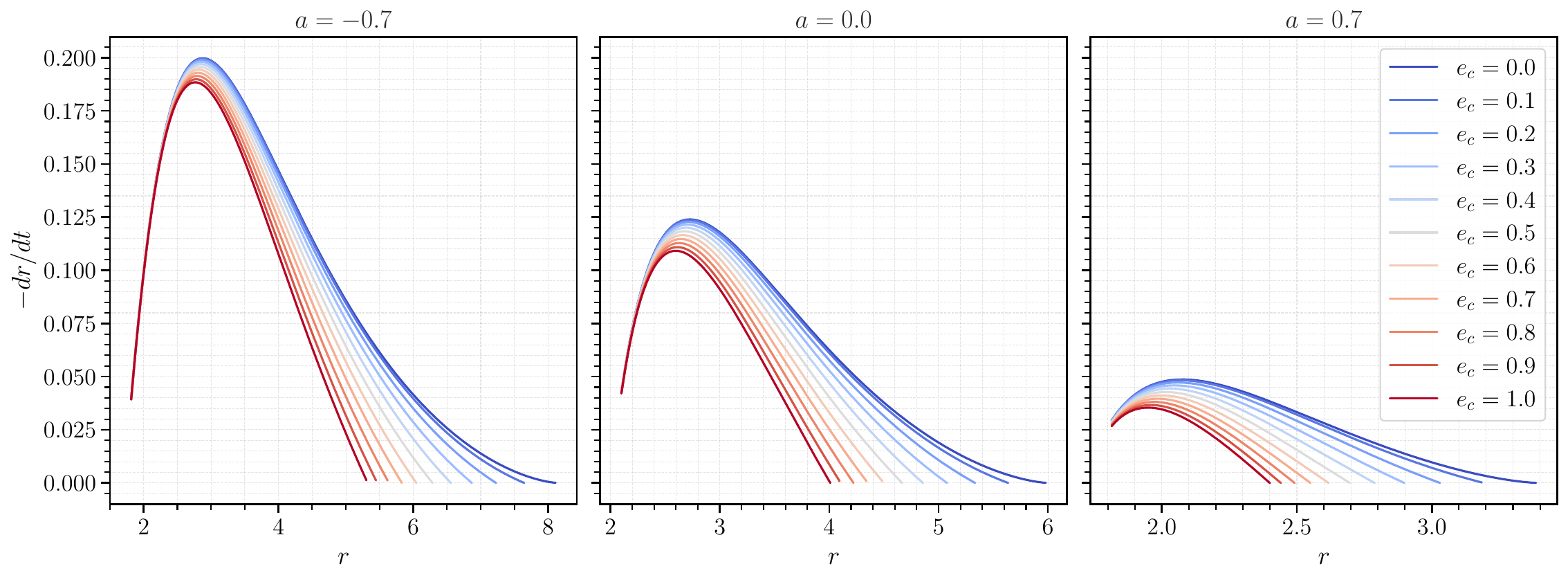}
\caption{In this figure we show the time-derivative of the radial coordinate $dr/dt$ of the critical plunge geodesics as function of $r$. We consider spin values $a=-0.7$ (first column), $a=0$ (second column) and $a=0.7$ (third column). For every spin case we consider critical eccentricities $e_c \le 1$.}
\label{Fig.: quasi circularity study}
\end{figure*}
We study the behaviour of the time-derivative of the radial coordinate, i.e. the quantity $dr/dt$ of the critical plunge geodesics as a function of $r$. In Fig.~\ref{Fig.: quasi circularity study} we show $dr/dt$ for spin values $a=-0.7$ (first column), $a=0$ (second column) and $a=0.7$ (third column). For every spin case we consider critical eccentricities $e_c \le 1$. As in Ref.~\cite{Taracchini:2014zpa} for the QC case ($e_c = 0$), we find that the radial velocity $dr/dt$ reaches maximum values that decrease with $a$, meaning that for large spins the plunge becomes more circular. Interestingly, we find that, when keeping the spin at a fix value, the maximum values of $dr/dt$ decrease with $e_c$. We remark that this result can be found in Ref.~\cite{Albanesi:2023bgi} for the case $a = 0$. We extended the result to the more general Kerr case. Hence, paradoxically, by increasing the critical eccentricity the plunge becomes more circular.

\section{Higher order modes merger features} \label{Sec.: ell = m higher order modes similar behaviour}
\begin{figure*}[htbp]
  	\includegraphics[width=1.\linewidth]{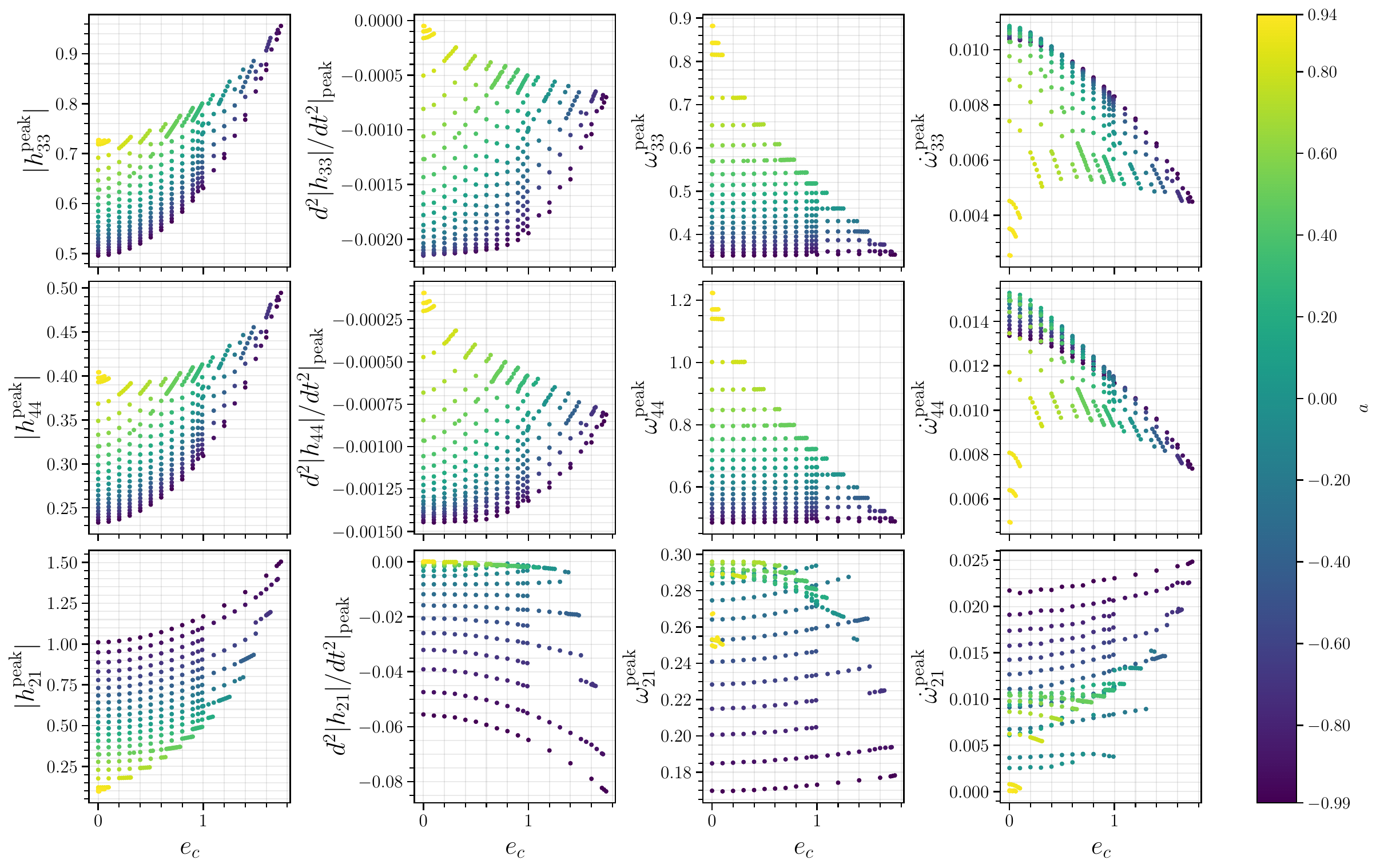}
\caption{In this figure we show the dependence of the amplitude \( |h_{\ell m}^{\rm peak}| \), second-order time derivative of the amplitude $d^2 |h_{\ell m}| / dt^2|_{\rm peak}$, frequency \( \omega_{\ell m}^{\rm peak} \), and first-order time derivative of the frequency \( \dot{\omega}_{\ell m}^{\rm peak} \) of the $h_{33}$, $h_{44}$ and $h_{21}$ modes evaluated at the peak of their respective amplitudes as function of the critical eccentricity $e_c$. We span different values of the spin $-0.99 \le a \le 0.99$ of the central Kerr BH, represented by different colors. We consider the section of parameter space for which the amplitude of the $h_{22}$ mode has a peak, i.e., the blue area of Fig.~\ref{h22_peaks_existence_curve}. For spin values $a \ge 0.95$, we can no longer find a peak of the amplitude of the $h_{22}$ mode even when $e_c = 0$, as shown in Fig.~\ref{h22_peaks_existence_curve}. These configurations are not shown in the figure.}
\label{Fig.: h22_h33_h21_IVs_compared}
\end{figure*}

\input{tab/h22_ampl_values.tex}
\input{tab/ddt_h22_peak_values.tex}
\input{tab/omega22_h22_peak.tex}
\input{tab/dt_omega22_h22_peak.tex}
In this appendix, we assess what was mentioned in Sec.~\ref{Subsec: ell = 2 m = 2 mode characterization} regarding the fact that the higher-order $\ell = m$ modes are characterized by a merger which has the same structure and similar trend of the $h_{22}$ mode case. Specifically, we investigate the dependence of the waveform quantities at merger like the mode amplitude $|h_{\ell m}^{\rm peak}|$, the second-order time derivative of the amplitude $d^2 |h_{\ell m}| / dt^2|_{\rm peak}$, the mode frequency $\omega_{\ell m}^{\rm peak}$, and its time derivative $\dot{\omega}_{\ell m}^{\rm peak}$ with respect to the critical eccentricity $e_c$, for the range of spin values considered in our analysis.

In the first two rows of Fig.~\ref{Fig.: h22_h33_h21_IVs_compared} we plot these quantities for the $h_{33}$ and $h_{44}$ modes as function of $e_c$. As in Fig.~\ref{h22_NQC_quantities_vs_e0_only_IVs}, we restrict to the region of the parameter space for which the peak of the amplitude of the $h_{22}$ mode exists, as depicted in the blue region of Fig.~\ref{h22_peaks_existence_curve}. By comparing the results in Fig.~\ref{Fig.: h22_h33_h21_IVs_compared} with the results in Fig.~\ref{h22_NQC_quantities_vs_e0_only_IVs} of Sec.~\ref{Sec: results}, we confirm that the higher-order $\ell = m$ modes merger structure is similar to the structure of the $h_{22}$ mode merger. More precisely, the values of the plotted quantities $|h_{\ell m}^{\rm peak}|$, $d^2 |h_{\ell m}| / dt^2|_{\rm peak}$, $\omega_{\ell m}^{\rm peak}$ and $\dot{\omega}_{\ell m}^{\rm peak}$ have a trend similar to the same quantities of the $h_{22}$ mode. In particular, we find that the peak of the amplitude $|h_{\ell m}^{\rm peak}|$ exhibits a monotonic increase with $e_c$ and spin $a$. The second-order time derivative of the amplitude, $d^2 |h_{\ell m}| / dt^2|_{\rm peak}$, increases as $e_c$ approaches the threshold eccentricity $e_c^{\rm thr}$, indicating a flattening of the modes amplitude at merger when $e_c$ increases.
The modes frequency $\omega_{\ell m}^{\rm peak}$ remains largely insensitive to variations in $e_c$, showing a nearly flat trend across all spin configurations.
The time derivative of the mode frequency, $\dot{\omega}_{\ell m}^{\rm peak}$, generally follows a monotonic decreasing pattern with respect to $e_c$. As in the $h_{22}$ mode case, we observe deviations from monotonicity concerning the spin dependence, and we find that these deviations exhibit the same structure for all the $\ell = m$ modes, as can be noticed in the first two panels of the third column of Fig.~\ref{Fig.: h22_h33_h21_IVs_compared}. The fact that for fixed eccentricity $\dot{\omega}_{\ell m}^{\rm peak}$ is not monotonic (especially for lower values of $e_c$) for all the $\ell = m$ modes strongly suggests this behaviour is a physical feature of the waveform modes close to merger.

The third row of Fig.~\ref{Fig.: h22_h33_h21_IVs_compared} shows the same quantities mentioned above evaluated at merger for the $h_{21}$ mode. From the figure it is possible to notice how the merger structure presents differences with respect to the cases where $\ell = m$. In fact, for the amplitude of the mode at the peak $|h_{\ell m}^{\rm peak}|$ we observe a decreasing monotonic trend with respect to the spin for all the values of $e_c$, contrary to the increasing monotonic trend observed in the $\ell = m$ cases. Concerning the dependence on the eccentricity, we still observe that $|h_{\ell m}^{\rm peak}|$ increases when $e_c$ increases. 
We also notice a different structure regarding the quantity $d^2 |h_{\ell m}| / dt^2|_{\rm peak}$, which has significant smaller values for the retrograde spin cases compared to the prograde ones. This is connected to the fact that in the retrograde spin configurations the mixing of the mode with its mirror quasi-normal-mode with complex frequency $-\omega_{\ell m}^*$ is significant close to merger. This mixing is due to the frame-dragging which characterizes the dynamics of retrograde spin trajectories in the regimes of the Kerr ergo-sphere. As a matter of fact, the resulting inversion of the azimuthal motion of the trajectories due to the dragging significantly excites the mirror quasi-normal-modes. For this reason, in the retrograde scenarios the mixing with the mirror modes generates a peak of the $h_{21}$ mode which always happens at times $t > t_{\Omega^{\rm peak}}$, as can be observed in the first panel of Fig.~\ref{Fig.: hlm hierarchy}. 
This mode-mixing also impacts the structure of the frequency $\omega_{\ell m}^{\rm peak}$ of the mode since it has a different pattern from the frequencies of the $\ell = m$ cases, and this holds also for the quantity $\dot{\omega}_{\ell m}^{\rm peak}$. Furthermore, examining both the quantities $\omega_{\ell m}^{\rm peak}$ and $\dot{\omega}_{\ell m}^{\rm peak}$ in Fig.~\ref{Fig.: h22_h33_h21_IVs_compared} it is possible to notice how the dependence on eccentricity of the merger of the $h_{21}$ mode is different between the retrograde and prograde spin scenarios. In fact $\omega_{\ell m}^{\rm peak}$ and $\dot{\omega}_{\ell m}^{\rm peak}$ increase monotonically with respect to $e_c$ for the retrograde spin cases, while they decrease with $e_c$ for the prograde spin scenarios.
As we mentioned above, we believe this different structure between retrograde and prograde orbits, and with respect to the $\ell = m$ cases, is due to the mode-mixing that affects the $h_{21}$ mode merger, and which makes its merger structure less trivial than the $\ell = m$ case. However, we found (without showing in Fig.~\ref{Fig.: h22_h33_h21_IVs_compared}) that the $h_{32}$ and $h_{43}$ modes have a similar merger structure as the $h_{21}$ mode. 

\section{$\ell = 2$ $m = 2$ merger values} \label{Sec.: l=2 m=2 merger values}
In this section, we present a subset of the data for the quantities of the $h_{22}$ mode studied in Sec.~\ref{Subsec: ell = 2 m = 2 mode characterization}, a part of which is shown in Fig.~\ref{h22_NQC_quantities_vs_e0_only_IVs}. The subset is selected considering values of the critical eccentricity $e_c \leq 0.9$ and values of the spin $a=[-0.9, -0.7, -0.5, -0.3, -0.1, 0.0, 0.1, 0.3, 0.5, 0.7, 0.9]$.
In Table~\ref{Tab: h22 amplitude peak} we show the values of the peak of the amplitude $|h^{\rm peak}_{22}|$ of the $h_{22}$ mode in the section of parameter space we consider. In Table~\ref{Tab: ddth22 amplitude peak}, Table~\ref{Tab: omega22 amplitude peak} and Table~\ref{Tab: dtomega22 amplitude peak} we show, respectively, the values for the quantities $d^2 |h_{22}| / dt^2|_{\rm peak}$, $\omega^{\rm peak}_{22}$ and $\dot{\omega}^{\rm peak}_{22}$.
In the tables, the entries containing the ``-'' symbol represents regions of the parameter space where the peak of the amplitude of the $h_{22}$ mode is not present. 

%\clearpage % Ensure bibliography starts on a new page
\bibliography{references.bib}

\end{document}

%% file: tab/different_e_thresholds.tex
\begin{table}
    \setlength\tabcolsep{18pt}
    \centering
    \begin{ruledtabular}
    \begin{tabular}{rlll}
%        \toprule
        \multicolumn{1}{c}{$a$} & \multicolumn{1}{c}{$e_{c}^{\rm thr}$} & \multicolumn{1}{c}{$e_{\rm asym}$} & \multicolumn{1}{c}{$\delta$} \\
        \midrule
        -0.99 & 1.760 & 1.756 & $3.2  \times 10^{-3}$ \\
        -0.90 & 1.724 & 1.732 & $7.85 \times 10^{-3}$ \\
        -0.70 & 1.648 & 1.650 & $1.60 \times 10^{-3}$ \\
        -0.40 & 1.506 & 1.507 & $1.04 \times 10^{-3}$ \\
        -0.20 & 1.396 & 1.393 & $3.17 \times 10^{-3}$ \\
        0.00 & 1.2590 & 1.2587 & $0.3 \times 10^{-3}$ \\
        0.20 & 1.1020 & 1.1010 & $1.0 \times 10^{-3}$ \\
        0.40 & 0.9072 & 0.9051 & $2.1 \times 10^{-3}$ \\
        0.70 & 0.5082 & 0.5075 & $0.7  \times 10^{-3}$ \\
        0.90 & 0.1133 & 0.1115 & $1.8  \times 10^{-3}$ \\
        0.92 & 0.0645 & 0.0619 & $2.7  \times 10^{-3}$ \\
        0.94 & 0.0201 & 0.0211 & $1.0  \times 10^{-3}$ \\
%        \bottomrule
    \end{tabular}
    \end{ruledtabular}
    \caption{Table of values $e_{c}^{\rm thr}$ and $e_{\rm asym}$ for different spins $a$, of the central BH. We also compute the absolute difference $\delta$ between the two quantities.}
    \label{Table_different_e_thresholds}
\end{table}

%% file: tab/h22_ampl_values.tex
\begin{table*}[htbp]
\setlength \tabcolsep{10pt}
\centering
\begin{ruledtabular}
\begin{tabular}{cccccccccccc}
\diagbox{e}{a} & -0.9 & -0.7 & -0.5 & -0.3 & -0.1 & 0.0 & 0.1 & 0.3 & 0.5 & 0.7 & 0.9 \\
\midrule
0.0 & 1.343 & 1.361 & 1.383 & 1.407 & 1.435 & 1.451 & 1.467 & 1.504 & 1.545 & 1.589 & 1.601 \\
0.1 & 1.347 & 1.366 & 1.387 & 1.412 & 1.441 & 1.456 & 1.473 & 1.511 & 1.553 & 1.599 & 1.620 \\
0.2 & 1.358 & 1.377 & 1.400 & 1.426 & 1.455 & 1.472 & 1.490 & 1.529 & 1.575 & 1.627 & - \\
0.3 & 1.375 & 1.395 & 1.419 & 1.447 & 1.478 & 1.496 & 1.515 & 1.557 & 1.608 & 1.668 & - \\
0.4 & 1.397 & 1.419 & 1.445 & 1.474 & 1.508 & 1.527 & 1.547 & 1.594 & 1.650 & 1.720 & - \\
0.5 & 1.425 & 1.449 & 1.476 & 1.508 & 1.545 & 1.565 & 1.588 & 1.639 & 1.702 & 1.783 & - \\
0.6 & 1.459 & 1.484 & 1.514 & 1.548 & 1.588 & 1.610 & 1.635 & 1.692 & 1.763 & - & - \\
0.7 & 1.497 & 1.525 & 1.558 & 1.595 & 1.638 & 1.663 & 1.690 & 1.753 & 1.834 & - & - \\
0.8 & 1.542 & 1.572 & 1.608 & 1.648 & 1.696 & 1.723 & 1.753 & 1.824 & - & - & - \\
0.9 & 1.593 & 1.626 & 1.665 & 1.709 & 1.762 & 1.792 & 1.826 & 1.905 & - & - & - \\
\end{tabular}
\end{ruledtabular}
\caption{A subset of the values of the peak of the amplitude $|h^{\rm peak}_{22}|$ of the $h_{22}$ mode. 
%In red we highlight the values which are extracted from the waveforms in the region of the parameter space for which the peak of the amplitude of the $h_{22}$ mode is not present.}
}
\label{Tab: h22 amplitude peak}
\end{table*}

%% file: tab/ddt_h22_peak_values.tex
\begin{table*}[htbp]
\setlength \tabcolsep{5pt}
\centering
\begin{ruledtabular}
\begin{tabular}{cccccccccccc}
\diagbox{e}{a} & -0.9 & -0.7 & -0.5 & -0.3 & -0.1 & 0.0 & 0.1 & 0.3 & 0.5 & 0.7 & 0.9 \\
\midrule
0.0 & -0.00327 & -0.00313 & -0.00297 & -0.00280 & -0.00258 & -0.00247 & -0.00230 & -0.00189 & -0.00130 & -0.00062 & -0.00003 \\
0.1 & -0.00320 & -0.00311 & -0.00293 & -0.00278 & -0.00252 & -0.00243 & -0.00223 & -0.00183 & -0.00122 & -0.00055 & -0.00000 \\
0.2 & -0.00315 & -0.00303 & -0.00289 & -0.00266 & -0.00243 & -0.00229 & -0.00210 & -0.00169 & -0.00107 & -0.00038 & - \\
0.3 & -0.00304 & -0.00292 & -0.00278 & -0.00256 & -0.00230 & -0.00213 & -0.00196 & -0.00150 & -0.00089 & -0.00022 & - \\
0.4 & -0.00292 & -0.00281 & -0.00259 & -0.00237 & -0.00211 & -0.00195 & -0.00176 & -0.00129 & -0.00064 & -0.00006 & - \\
0.5 & -0.00279 & -0.00263 & -0.00246 & -0.00222 & -0.00191 & -0.00174 & -0.00154 & -0.00103 & -0.00043 & -0.00000 & - \\
0.6 & -0.00260 & -0.00248 & -0.00226 & -0.00202 & -0.00168 & -0.00152 & -0.00130 & -0.00077 & -0.00021 & - & - \\
0.7 & -0.00244 & -0.00229 & -0.00205 & -0.00179 & -0.00143 & -0.00126 & -0.00103 & -0.00051 & -0.00006 & - & - \\
0.8 & -0.00226 & -0.00207 & -0.00184 & -0.00153 & -0.00118 & -0.00100 & -0.00075 & -0.00027 & - & - & - \\
0.9 & -0.00203 & -0.00183 & -0.00158 & -0.00128 & -0.00091 & -0.00071 & -0.00049 & -0.00010 & - & - & - \\
\end{tabular}
\end{ruledtabular}
\caption{A subset of the values of the second-order time derivative of the amplitude $d^2 |h_{22}| / dt^2|_{\rm peak}$ of the $h_{22}$ mode evaluated at its amplitude peak. 
%In red we highlight the values which are extracted from the waveforms in the region of the parameter space for which the peak of the amplitude of the $h_{22}$ mode is not present.}
}
\label{Tab: ddth22 amplitude peak}
\end{table*}

%% file: tab/omega22_h22_peak.tex
\begin{table*}[htbp]
\setlength \tabcolsep{10pt}
\centering
\begin{ruledtabular}
\begin{tabular}{cccccccccccc}
\diagbox{e}{a} & -0.9 & -0.7 & -0.5 & -0.3 & -0.1 & 0.0 & 0.1 & 0.3 & 0.5 & 0.7 & 0.9 \\
\midrule
0.0 & 0.220 & 0.228 & 0.238 & 0.250 & 0.265 & 0.274 & 0.284 & 0.308 & 0.341 & 0.390 & 0.489 \\
0.1 & 0.220 & 0.228 & 0.238 & 0.250 & 0.265 & 0.274 & 0.284 & 0.308 & 0.340 & 0.390 & 0.487 \\
0.2 & 0.220 & 0.228 & 0.238 & 0.250 & 0.265 & 0.274 & 0.284 & 0.308 & 0.340 & 0.390 & - \\
0.3 & 0.220 & 0.228 & 0.239 & 0.251 & 0.266 & 0.274 & 0.284 & 0.308 & 0.341 & 0.391 & - \\
0.4 & 0.221 & 0.229 & 0.238 & 0.251 & 0.266 & 0.274 & 0.284 & 0.309 & 0.341 & 0.391 & - \\
0.5 & 0.221 & 0.229 & 0.239 & 0.251 & 0.266 & 0.275 & 0.285 & 0.309 & 0.342 & 0.391 & - \\
0.6 & 0.220 & 0.229 & 0.239 & 0.251 & 0.266 & 0.275 & 0.285 & 0.309 & 0.342 & - & - \\
0.7 & 0.221 & 0.229 & 0.239 & 0.251 & 0.266 & 0.275 & 0.285 & 0.309 & 0.342 & - & - \\
0.8 & 0.221 & 0.229 & 0.239 & 0.251 & 0.266 & 0.276 & 0.285 & 0.309 & - & - & - \\
0.9 & 0.220 & 0.228 & 0.239 & 0.251 & 0.266 & 0.275 & 0.285 & 0.310 & - & - & - \\
\end{tabular}

\end{ruledtabular}
\caption{A subset of the values of the frequency $\omega^{\rm peak}_{22}$ of the $h_{22}$ mode evaluated at its amplitude peak. 
%In red we highlight the values which are extracted from the waveforms in the region of the parameter space for which the peak of the amplitude of the $h_{22}$ mode is not present.}
}
\label{Tab: omega22 amplitude peak}
\end{table*}

%% file: tab/dt_omega22_h22_peak.tex
\begin{table*}[t!]
\setlength \tabcolsep{5pt}
%\centering
\begin{ruledtabular}
\begin{tabular}{cccccccccccc}
\diagbox{e}{a} & -0.9 & -0.7 & -0.5 & -0.3 & -0.1 & 0.0 & 0.1 & 0.3 & 0.5 & 0.7 & 0.9 \\
\midrule
0.0 & 0.00574 & 0.00591 & 0.00599 & 0.00601 & 0.00596 & 0.00588 & 0.00580 & 0.00553 & 0.00497 & 0.00377 & 0.00087 \\
0.1 & 0.00573 & 0.00589 & 0.00596 & 0.00597 & 0.00590 & 0.00582 & 0.00572 & 0.00544 & 0.00482 & 0.00354 & 0.00014 \\
0.2 & 0.00569 & 0.00583 & 0.00589 & 0.00586 & 0.00577 & 0.00565 & 0.00554 & 0.00521 & 0.00448 & 0.00296 & - \\
0.3 & 0.00563 & 0.00574 & 0.00576 & 0.00571 & 0.00557 & 0.00543 & 0.00530 & 0.00487 & 0.00403 & 0.00219 & - \\
0.4 & 0.00555 & 0.00562 & 0.00559 & 0.00549 & 0.00531 & 0.00515 & 0.00498 & 0.00445 & 0.00340 & 0.00121 & - \\
0.5 & 0.00545 & 0.00546 & 0.00540 & 0.00525 & 0.00502 & 0.00481 & 0.00461 & 0.00395 & 0.00271 & 0.00009 & - \\
0.6 & 0.00532 & 0.00529 & 0.00516 & 0.00497 & 0.00464 & 0.00444 & 0.00418 & 0.00337 & 0.00189 & - & - \\
0.7 & 0.00517 & 0.00507 & 0.00489 & 0.00463 & 0.00424 & 0.00399 & 0.00366 & 0.00269 & 0.00098 & - & - \\
0.8 & 0.00498 & 0.00480 & 0.00457 & 0.00424 & 0.00378 & 0.00349 & 0.00309 & 0.00193 & - & - & - \\
0.9 & 0.00472 & 0.00449 & 0.00419 & 0.00381 & 0.00326 & 0.00290 & 0.00245 & 0.00111 & - & - & - \\
\end{tabular}

\end{ruledtabular}
\caption{A subset of the values of the first order time derivative of the frequency $\dot{\omega}^{\rm peak}_{22}$ of the $h_{22}$ mode evaluated at its amplitude peak. 
%In red we highlight the values which are extracted from the waveforms in the region of the parameter space for which the peak of the amplitude of the $h_{22}$ mode is not present.}
}
\label{Tab: dtomega22 amplitude peak}
\end{table*}

%% file: geo_spin_ecc_plunge_WF.bbl
\begin{thebibliography}{142}%
\makeatletter
\providecommand \@ifxundefined [1]{%
 \@ifx{#1\undefined}
}%
\providecommand \@ifnum [1]{%
 \ifnum #1\expandafter \@firstoftwo
 \else \expandafter \@secondoftwo
 \fi
}%
\providecommand \@ifx [1]{%
 \ifx #1\expandafter \@firstoftwo
 \else \expandafter \@secondoftwo
 \fi
}%
\providecommand \natexlab [1]{#1}%
\providecommand \enquote  [1]{``#1''}%
\providecommand \bibnamefont  [1]{#1}%
\providecommand \bibfnamefont [1]{#1}%
\providecommand \citenamefont [1]{#1}%
\providecommand \href@noop [0]{\@secondoftwo}%
\providecommand \href [0]{\begingroup \@sanitize@url \@href}%
\providecommand \@href[1]{\@@startlink{#1}\@@href}%
\providecommand \@@href[1]{\endgroup#1\@@endlink}%
\providecommand \@sanitize@url [0]{\catcode `\\12\catcode `\$12\catcode
  `\&12\catcode `\#12\catcode `\^12\catcode `\_12\catcode `\%12\relax}%
\providecommand \@@startlink[1]{}%
\providecommand \@@endlink[0]{}%
\providecommand \url  [0]{\begingroup\@sanitize@url \@url }%
\providecommand \@url [1]{\endgroup\@href {#1}{\urlprefix }}%
\providecommand \urlprefix  [0]{URL }%
\providecommand \Eprint [0]{\href }%
\providecommand \doibase [0]{https://doi.org/}%
\providecommand \selectlanguage [0]{\@gobble}%
\providecommand \bibinfo  [0]{\@secondoftwo}%
\providecommand \bibfield  [0]{\@secondoftwo}%
\providecommand \translation [1]{[#1]}%
\providecommand \BibitemOpen [0]{}%
\providecommand \bibitemStop [0]{}%
\providecommand \bibitemNoStop [0]{.\EOS\space}%
\providecommand \EOS [0]{\spacefactor3000\relax}%
\providecommand \BibitemShut  [1]{\csname bibitem#1\endcsname}%
\let\auto@bib@innerbib\@empty
%</preamble>
\bibitem [{\citenamefont {Abbott}\ \emph {et~al.}(2016)\citenamefont {Abbott}
  \emph {et~al.}}]{LIGOScientific:2016aoc}%
  \BibitemOpen
  \bibfield  {author} {\bibinfo {author} {\bibfnamefont {B.~P.}\ \bibnamefont
  {Abbott}} \emph {et~al.} (\bibinfo {collaboration} {LIGO Scientific,
  Virgo}),\ }\bibfield  {title} {\bibinfo {title} {{Observation of
  Gravitational Waves from a Binary Black Hole Merger}},\ }\href
  {https://doi.org/10.1103/PhysRevLett.116.061102} {\bibfield  {journal}
  {\bibinfo  {journal} {Phys. Rev. Lett.}\ }\textbf {\bibinfo {volume} {116}},\
  \bibinfo {pages} {061102} (\bibinfo {year} {2016})},\ \Eprint
  {https://arxiv.org/abs/1602.03837} {arXiv:1602.03837 [gr-qc]} \BibitemShut
  {NoStop}%
\bibitem [{\citenamefont {Abbott}\ \emph {et~al.}(2019)\citenamefont {Abbott}
  \emph {et~al.}}]{LIGOScientific:2018mvr}%
  \BibitemOpen
  \bibfield  {author} {\bibinfo {author} {\bibfnamefont {B.~P.}\ \bibnamefont
  {Abbott}} \emph {et~al.} (\bibinfo {collaboration} {LIGO Scientific,
  Virgo}),\ }\bibfield  {title} {\bibinfo {title} {{GWTC-1: A
  Gravitational-Wave Transient Catalog of Compact Binary Mergers Observed by
  LIGO and Virgo during the First and Second Observing Runs}},\ }\href
  {https://doi.org/10.1103/PhysRevX.9.031040} {\bibfield  {journal} {\bibinfo
  {journal} {Phys. Rev. X}\ }\textbf {\bibinfo {volume} {9}},\ \bibinfo {pages}
  {031040} (\bibinfo {year} {2019})},\ \Eprint
  {https://arxiv.org/abs/1811.12907} {arXiv:1811.12907 [astro-ph.HE]}
  \BibitemShut {NoStop}%
\bibitem [{\citenamefont {Abbott}\ \emph
  {et~al.}(2021{\natexlab{a}})\citenamefont {Abbott} \emph
  {et~al.}}]{LIGOScientific:2019lzm}%
  \BibitemOpen
  \bibfield  {author} {\bibinfo {author} {\bibfnamefont {R.}~\bibnamefont
  {Abbott}} \emph {et~al.} (\bibinfo {collaboration} {LIGO Scientific,
  Virgo}),\ }\bibfield  {title} {\bibinfo {title} {{Open data from the first
  and second observing runs of Advanced LIGO and Advanced Virgo}},\ }\href
  {https://doi.org/10.1016/j.softx.2021.100658} {\bibfield  {journal} {\bibinfo
   {journal} {SoftwareX}\ }\textbf {\bibinfo {volume} {13}},\ \bibinfo {pages}
  {100658} (\bibinfo {year} {2021}{\natexlab{a}})},\ \Eprint
  {https://arxiv.org/abs/1912.11716} {arXiv:1912.11716 [gr-qc]} \BibitemShut
  {NoStop}%
\bibitem [{\citenamefont {Abbott}\ \emph
  {et~al.}(2021{\natexlab{b}})\citenamefont {Abbott} \emph
  {et~al.}}]{LIGOScientific:2020ibl}%
  \BibitemOpen
  \bibfield  {author} {\bibinfo {author} {\bibfnamefont {R.}~\bibnamefont
  {Abbott}} \emph {et~al.} (\bibinfo {collaboration} {LIGO Scientific,
  Virgo}),\ }\bibfield  {title} {\bibinfo {title} {{GWTC-2: Compact Binary
  Coalescences Observed by LIGO and Virgo During the First Half of the Third
  Observing Run}},\ }\href {https://doi.org/10.1103/PhysRevX.11.021053}
  {\bibfield  {journal} {\bibinfo  {journal} {Phys. Rev. X}\ }\textbf {\bibinfo
  {volume} {11}},\ \bibinfo {pages} {021053} (\bibinfo {year}
  {2021}{\natexlab{b}})},\ \Eprint {https://arxiv.org/abs/2010.14527}
  {arXiv:2010.14527 [gr-qc]} \BibitemShut {NoStop}%
\bibitem [{\citenamefont {Abbott}\ \emph {et~al.}(2024)\citenamefont {Abbott}
  \emph {et~al.}}]{LIGOScientific:2021usb}%
  \BibitemOpen
  \bibfield  {author} {\bibinfo {author} {\bibfnamefont {R.}~\bibnamefont
  {Abbott}} \emph {et~al.} (\bibinfo {collaboration} {LIGO Scientific,
  VIRGO}),\ }\bibfield  {title} {\bibinfo {title} {{GWTC-2.1: Deep extended
  catalog of compact binary coalescences observed by LIGO and Virgo during the
  first half of the third observing run}},\ }\href
  {https://doi.org/10.1103/PhysRevD.109.022001} {\bibfield  {journal} {\bibinfo
   {journal} {Phys. Rev. D}\ }\textbf {\bibinfo {volume} {109}},\ \bibinfo
  {pages} {022001} (\bibinfo {year} {2024})},\ \Eprint
  {https://arxiv.org/abs/2108.01045} {arXiv:2108.01045 [gr-qc]} \BibitemShut
  {NoStop}%
\bibitem [{\citenamefont {Abbott}\ \emph {et~al.}(2018)\citenamefont {Abbott}
  \emph {et~al.}}]{KAGRA:2013rdx}%
  \BibitemOpen
  \bibfield  {author} {\bibinfo {author} {\bibfnamefont {B.~P.}\ \bibnamefont
  {Abbott}} \emph {et~al.} (\bibinfo {collaboration} {KAGRA, LIGO Scientific,
  Virgo, VIRGO}),\ }\bibfield  {title} {\bibinfo {title} {{Prospects for
  observing and localizing gravitational-wave transients with Advanced LIGO,
  Advanced Virgo and KAGRA}},\ }\href
  {https://doi.org/10.1007/s41114-020-00026-9} {\bibfield  {journal} {\bibinfo
  {journal} {Living Rev. Rel.}\ }\textbf {\bibinfo {volume} {21}},\ \bibinfo
  {pages} {3} (\bibinfo {year} {2018})},\ \Eprint
  {https://arxiv.org/abs/1304.0670} {arXiv:1304.0670 [gr-qc]} \BibitemShut
  {NoStop}%
\bibitem [{\citenamefont {Abbott}\ \emph
  {et~al.}(2021{\natexlab{c}})\citenamefont {Abbott} \emph
  {et~al.}}]{LIGOScientific:2021djp}%
  \BibitemOpen
  \bibfield  {author} {\bibinfo {author} {\bibfnamefont {R.}~\bibnamefont
  {Abbott}} \emph {et~al.} (\bibinfo {collaboration} {LIGO Scientific, VIRGO,
  KAGRA}),\ }\bibfield  {title} {\bibinfo {title} {{GWTC-3: Compact Binary
  Coalescences Observed by LIGO and Virgo During the Second Part of the Third
  Observing Run}},\ }\href@noop {} {\bibfield  {journal} {\bibinfo  {journal}
  {Phys. Rev. X"}\ } (\bibinfo {year} {2021}{\natexlab{c}})},\ \Eprint
  {https://arxiv.org/abs/2111.03606} {arXiv:2111.03606 [gr-qc]} \BibitemShut
  {NoStop}%
\bibitem [{\citenamefont {Abbott}\ \emph {et~al.}(2023)\citenamefont {Abbott}
  \emph {et~al.}}]{KAGRA:2021vkt}%
  \BibitemOpen
  \bibfield  {author} {\bibinfo {author} {\bibfnamefont {R.}~\bibnamefont
  {Abbott}} \emph {et~al.} (\bibinfo {collaboration} {KAGRA, VIRGO, LIGO
  Scientific}),\ }\bibfield  {title} {\bibinfo {title} {{GWTC-3: Compact Binary
  Coalescences Observed by LIGO and Virgo during the Second Part of the Third
  Observing Run}},\ }\href {https://doi.org/10.1103/PhysRevX.13.041039}
  {\bibfield  {journal} {\bibinfo  {journal} {Phys. Rev. X}\ }\textbf {\bibinfo
  {volume} {13}},\ \bibinfo {pages} {041039} (\bibinfo {year} {2023})},\
  \Eprint {https://arxiv.org/abs/2111.03606} {arXiv:2111.03606 [gr-qc]}
  \BibitemShut {NoStop}%
\bibitem [{\citenamefont {Punturo}\ \emph {et~al.}(2010)\citenamefont {Punturo}
  \emph {et~al.}}]{Punturo_2010}%
  \BibitemOpen
  \bibfield  {author} {\bibinfo {author} {\bibfnamefont {M.}~\bibnamefont
  {Punturo}} \emph {et~al.},\ }\bibfield  {title} {\bibinfo {title} {The
  einstein telescope: a third-generation gravitational wave observatory},\
  }\href {https://doi.org/10.1088/0264-9381/27/19/194002} {\bibfield  {journal}
  {\bibinfo  {journal} {Classical and Quantum Gravity}\ }\textbf {\bibinfo
  {volume} {27}},\ \bibinfo {pages} {194002} (\bibinfo {year}
  {2010})}\BibitemShut {NoStop}%
\bibitem [{\citenamefont {Abac}\ \emph {et~al.}(2025)\citenamefont {Abac} \emph
  {et~al.}}]{Abac:2025saz}%
  \BibitemOpen
  \bibfield  {author} {\bibinfo {author} {\bibfnamefont {A.}~\bibnamefont
  {Abac}} \emph {et~al.},\ }\bibfield  {title} {\bibinfo {title} {{The Science
  of the Einstein Telescope}},\ }\Eprint {https://arxiv.org/abs/2503.12263}
  {arXiv:2503.12263 [gr-qc]} \BibitemShut {NoStop}%
\bibitem [{\citenamefont {Evans}\ \emph {et~al.}(2021)\citenamefont {Evans}
  \emph {et~al.}}]{Evans:2021gyd}%
  \BibitemOpen
  \bibfield  {author} {\bibinfo {author} {\bibfnamefont {M.}~\bibnamefont
  {Evans}} \emph {et~al.},\ }\bibfield  {title} {\bibinfo {title} {{A Horizon
  Study for Cosmic Explorer: Science, Observatories, and Community}},\
  }\href@noop {} {\  (\bibinfo {year} {2021})},\ \Eprint
  {https://arxiv.org/abs/2109.09882} {arXiv:2109.09882 [astro-ph.IM]}
  \BibitemShut {NoStop}%
\bibitem [{\citenamefont {Amaro-Seoane}\ \emph {et~al.}(2017)\citenamefont
  {Amaro-Seoane} \emph {et~al.}}]{LISA}%
  \BibitemOpen
  \bibfield  {author} {\bibinfo {author} {\bibfnamefont {P.}~\bibnamefont
  {Amaro-Seoane}} \emph {et~al.},\ }\href
  {https://doi.org/10.48550/ARXIV.1702.00786} {\bibinfo {title} {Laser
  interferometer space antenna}} (\bibinfo {year} {2017})\BibitemShut {NoStop}%
\bibitem [{\citenamefont {Colpi}\ \emph {et~al.}(2024)\citenamefont {Colpi}
  \emph {et~al.}}]{LISA:2024hlh}%
  \BibitemOpen
  \bibfield  {author} {\bibinfo {author} {\bibfnamefont {M.}~\bibnamefont
  {Colpi}} \emph {et~al.} (\bibinfo {collaboration} {LISA}),\ }\bibfield
  {title} {\bibinfo {title} {{LISA Definition Study Report}},\ }\Eprint
  {https://arxiv.org/abs/2402.07571} {arXiv:2402.07571 [astro-ph.CO]}
  \BibitemShut {NoStop}%
\bibitem [{\citenamefont {Mandel}\ and\ \citenamefont
  {O'Shaughnessy}(2010)}]{Mandel:2009nx}%
  \BibitemOpen
  \bibfield  {author} {\bibinfo {author} {\bibfnamefont {I.}~\bibnamefont
  {Mandel}}\ and\ \bibinfo {author} {\bibfnamefont {R.}~\bibnamefont
  {O'Shaughnessy}},\ }\bibfield  {title} {\bibinfo {title} {{Compact Binary
  Coalescences in the Band of Ground-based Gravitational-Wave Detectors}},\
  }\href {https://doi.org/10.1088/0264-9381/27/11/114007} {\bibfield  {journal}
  {\bibinfo  {journal} {Class. Quant. Grav.}\ }\textbf {\bibinfo {volume}
  {27}},\ \bibinfo {pages} {114007} (\bibinfo {year} {2010})},\ \Eprint
  {https://arxiv.org/abs/0912.1074} {arXiv:0912.1074 [astro-ph.HE]}
  \BibitemShut {NoStop}%
\bibitem [{\citenamefont {Rodriguez}\ and\ \citenamefont
  {Loeb}(2018)}]{Rodriguez:2018rmd}%
  \BibitemOpen
  \bibfield  {author} {\bibinfo {author} {\bibfnamefont {C.~L.}\ \bibnamefont
  {Rodriguez}}\ and\ \bibinfo {author} {\bibfnamefont {A.}~\bibnamefont
  {Loeb}},\ }\bibfield  {title} {\bibinfo {title} {{Redshift Evolution of the
  Black Hole Merger Rate from Globular Clusters}},\ }\href
  {https://doi.org/10.3847/2041-8213/aae377} {\bibfield  {journal} {\bibinfo
  {journal} {Astrophys. J. Lett.}\ }\textbf {\bibinfo {volume} {866}},\
  \bibinfo {pages} {L5} (\bibinfo {year} {2018})},\ \Eprint
  {https://arxiv.org/abs/1809.01152} {arXiv:1809.01152 [astro-ph.HE]}
  \BibitemShut {NoStop}%
\bibitem [{\citenamefont {Fragione}\ and\ \citenamefont
  {Kocsis}(2018)}]{Fragione:2018vty}%
  \BibitemOpen
  \bibfield  {author} {\bibinfo {author} {\bibfnamefont {G.}~\bibnamefont
  {Fragione}}\ and\ \bibinfo {author} {\bibfnamefont {B.}~\bibnamefont
  {Kocsis}},\ }\bibfield  {title} {\bibinfo {title} {{Black hole mergers from
  an evolving population of globular clusters}},\ }\href
  {https://doi.org/10.1103/PhysRevLett.121.161103} {\bibfield  {journal}
  {\bibinfo  {journal} {Phys. Rev. Lett.}\ }\textbf {\bibinfo {volume} {121}},\
  \bibinfo {pages} {161103} (\bibinfo {year} {2018})},\ \Eprint
  {https://arxiv.org/abs/1806.02351} {arXiv:1806.02351 [astro-ph.GA]}
  \BibitemShut {NoStop}%
\bibitem [{\citenamefont {Zevin}\ \emph {et~al.}(2021)\citenamefont {Zevin},
  \citenamefont {Romero-Shaw}, \citenamefont {Kremer}, \citenamefont {Thrane},\
  and\ \citenamefont {Lasky}}]{Zevin:2021rtf}%
  \BibitemOpen
  \bibfield  {author} {\bibinfo {author} {\bibfnamefont {M.}~\bibnamefont
  {Zevin}}, \bibinfo {author} {\bibfnamefont {I.~M.}\ \bibnamefont
  {Romero-Shaw}}, \bibinfo {author} {\bibfnamefont {K.}~\bibnamefont {Kremer}},
  \bibinfo {author} {\bibfnamefont {E.}~\bibnamefont {Thrane}},\ and\ \bibinfo
  {author} {\bibfnamefont {P.~D.}\ \bibnamefont {Lasky}},\ }\bibfield  {title}
  {\bibinfo {title} {{Implications of Eccentric Observations on Binary Black
  Hole Formation Channels}},\ }\href {https://doi.org/10.3847/2041-8213/ac32dc}
  {\bibfield  {journal} {\bibinfo  {journal} {Astrophys. J. Lett.}\ }\textbf
  {\bibinfo {volume} {921}},\ \bibinfo {pages} {L43} (\bibinfo {year}
  {2021})},\ \Eprint {https://arxiv.org/abs/2106.09042} {arXiv:2106.09042
  [astro-ph.HE]} \BibitemShut {NoStop}%
\bibitem [{\citenamefont {Kozai}(1962)}]{Kozai:1962zz}%
  \BibitemOpen
  \bibfield  {author} {\bibinfo {author} {\bibfnamefont {Y.}~\bibnamefont
  {Kozai}},\ }\bibfield  {title} {\bibinfo {title} {{Secular perturbations of
  asteroids with high inclination and eccentricity}},\ }\href
  {https://doi.org/10.1086/108790} {\bibfield  {journal} {\bibinfo  {journal}
  {Astron. J.}\ }\textbf {\bibinfo {volume} {67}},\ \bibinfo {pages} {591}
  (\bibinfo {year} {1962})}\BibitemShut {NoStop}%
\bibitem [{\citenamefont {Lidov}(1962)}]{LIDOV1962719}%
  \BibitemOpen
  \bibfield  {author} {\bibinfo {author} {\bibfnamefont {M.}~\bibnamefont
  {Lidov}},\ }\bibfield  {title} {\bibinfo {title} {The evolution of orbits of
  artificial satellites of planets under the action of gravitational
  perturbations of external bodies},\ }\href
  {https://doi.org/https://doi.org/10.1016/0032-0633(62)90129-0} {\bibfield
  {journal} {\bibinfo  {journal} {Planetary and Space Science}\ }\textbf
  {\bibinfo {volume} {9}},\ \bibinfo {pages} {719} (\bibinfo {year}
  {1962})}\BibitemShut {NoStop}%
\bibitem [{\citenamefont {Samsing}\ \emph {et~al.}(2014)\citenamefont
  {Samsing}, \citenamefont {MacLeod},\ and\ \citenamefont
  {Ramirez-Ruiz}}]{Samsing:2013kua}%
  \BibitemOpen
  \bibfield  {author} {\bibinfo {author} {\bibfnamefont {J.}~\bibnamefont
  {Samsing}}, \bibinfo {author} {\bibfnamefont {M.}~\bibnamefont {MacLeod}},\
  and\ \bibinfo {author} {\bibfnamefont {E.}~\bibnamefont {Ramirez-Ruiz}},\
  }\bibfield  {title} {\bibinfo {title} {{The Formation of Eccentric Compact
  Binary Inspirals and the Role of Gravitational Wave Emission in Binary-Single
  Stellar Encounters}},\ }\href {https://doi.org/10.1088/0004-637X/784/1/71}
  {\bibfield  {journal} {\bibinfo  {journal} {Astrophys. J.}\ }\textbf
  {\bibinfo {volume} {784}},\ \bibinfo {pages} {71} (\bibinfo {year} {2014})},\
  \Eprint {https://arxiv.org/abs/1308.2964} {arXiv:1308.2964 [astro-ph.HE]}
  \BibitemShut {NoStop}%
\bibitem [{\citenamefont {Zevin}\ \emph {et~al.}(2019)\citenamefont {Zevin},
  \citenamefont {Samsing}, \citenamefont {Rodriguez}, \citenamefont {Haster},\
  and\ \citenamefont {Ramirez-Ruiz}}]{Zevin:2018kzq}%
  \BibitemOpen
  \bibfield  {author} {\bibinfo {author} {\bibfnamefont {M.}~\bibnamefont
  {Zevin}}, \bibinfo {author} {\bibfnamefont {J.}~\bibnamefont {Samsing}},
  \bibinfo {author} {\bibfnamefont {C.}~\bibnamefont {Rodriguez}}, \bibinfo
  {author} {\bibfnamefont {C.-J.}\ \bibnamefont {Haster}},\ and\ \bibinfo
  {author} {\bibfnamefont {E.}~\bibnamefont {Ramirez-Ruiz}},\ }\bibfield
  {title} {\bibinfo {title} {{Eccentric Black Hole Mergers in Dense Star
  Clusters: The Role of Binary\textendash{}Binary Encounters}},\ }\href
  {https://doi.org/10.3847/1538-4357/aaf6ec} {\bibfield  {journal} {\bibinfo
  {journal} {Astrophys. J.}\ }\textbf {\bibinfo {volume} {871}},\ \bibinfo
  {pages} {91} (\bibinfo {year} {2019})},\ \Eprint
  {https://arxiv.org/abs/1810.00901} {arXiv:1810.00901 [astro-ph.HE]}
  \BibitemShut {NoStop}%
\bibitem [{\citenamefont {Portegies~Zwart}\ and\ \citenamefont
  {McMillan}(2000)}]{PortegiesZwart:1999nm}%
  \BibitemOpen
  \bibfield  {author} {\bibinfo {author} {\bibfnamefont {S.~F.}\ \bibnamefont
  {Portegies~Zwart}}\ and\ \bibinfo {author} {\bibfnamefont {S.}~\bibnamefont
  {McMillan}},\ }\bibfield  {title} {\bibinfo {title} {{Black hole mergers in
  the universe}},\ }\href {https://doi.org/10.1086/312422} {\bibfield
  {journal} {\bibinfo  {journal} {Astrophys. J. Lett.}\ }\textbf {\bibinfo
  {volume} {528}},\ \bibinfo {pages} {L17} (\bibinfo {year} {2000})},\ \Eprint
  {https://arxiv.org/abs/astro-ph/9910061} {arXiv:astro-ph/9910061}
  \BibitemShut {NoStop}%
\bibitem [{\citenamefont {Miller}\ and\ \citenamefont
  {Hamilton}(2002)}]{Miller:2001ez}%
  \BibitemOpen
  \bibfield  {author} {\bibinfo {author} {\bibfnamefont {M.~C.}\ \bibnamefont
  {Miller}}\ and\ \bibinfo {author} {\bibfnamefont {D.~P.}\ \bibnamefont
  {Hamilton}},\ }\bibfield  {title} {\bibinfo {title} {{Production of
  intermediate-mass black holes in globular clusters}},\ }\href
  {https://doi.org/10.1046/j.1365-8711.2002.05112.x} {\bibfield  {journal}
  {\bibinfo  {journal} {Mon. Not. Roy. Astron. Soc.}\ }\textbf {\bibinfo
  {volume} {330}},\ \bibinfo {pages} {232} (\bibinfo {year} {2002})},\ \Eprint
  {https://arxiv.org/abs/astro-ph/0106188} {arXiv:astro-ph/0106188}
  \BibitemShut {NoStop}%
\bibitem [{\citenamefont {Divyajyoti}\ \emph {et~al.}(2024)\citenamefont
  {Divyajyoti}, \citenamefont {Kumar}, \citenamefont {Tibrewal}, \citenamefont
  {Romero-Shaw},\ and\ \citenamefont {Mishra}}]{Divyajyoti:2023rht}%
  \BibitemOpen
  \bibfield  {author} {\bibinfo {author} {\bibnamefont {Divyajyoti}}, \bibinfo
  {author} {\bibfnamefont {S.}~\bibnamefont {Kumar}}, \bibinfo {author}
  {\bibfnamefont {S.}~\bibnamefont {Tibrewal}}, \bibinfo {author}
  {\bibfnamefont {I.~M.}\ \bibnamefont {Romero-Shaw}},\ and\ \bibinfo {author}
  {\bibfnamefont {C.~K.}\ \bibnamefont {Mishra}},\ }\bibfield  {title}
  {\bibinfo {title} {{Blind spots and biases: The dangers of ignoring
  eccentricity in gravitational-wave signals from binary black holes}},\ }\href
  {https://doi.org/10.1103/PhysRevD.109.043037} {\bibfield  {journal} {\bibinfo
   {journal} {Phys. Rev. D}\ }\textbf {\bibinfo {volume} {109}},\ \bibinfo
  {pages} {043037} (\bibinfo {year} {2024})},\ \Eprint
  {https://arxiv.org/abs/2309.16638} {arXiv:2309.16638 [gr-qc]} \BibitemShut
  {NoStop}%
\bibitem [{\citenamefont {Gupte}\ \emph {et~al.}(2024)\citenamefont {Gupte}
  \emph {et~al.}}]{Gupte:2024jfe}%
  \BibitemOpen
  \bibfield  {author} {\bibinfo {author} {\bibfnamefont {N.}~\bibnamefont
  {Gupte}} \emph {et~al.},\ }\bibfield  {title} {\bibinfo {title} {{Evidence
  for eccentricity in the population of binary black holes observed by
  LIGO-Virgo-KAGRA}},\ }\Eprint {https://arxiv.org/abs/2404.14286}
  {arXiv:2404.14286 [gr-qc]} \BibitemShut {NoStop}%
\bibitem [{\citenamefont {Favata}(2014)}]{Favata:2013rwa}%
  \BibitemOpen
  \bibfield  {author} {\bibinfo {author} {\bibfnamefont {M.}~\bibnamefont
  {Favata}},\ }\bibfield  {title} {\bibinfo {title} {{Systematic parameter
  errors in inspiraling neutron star binaries}},\ }\href
  {https://doi.org/10.1103/PhysRevLett.112.101101} {\bibfield  {journal}
  {\bibinfo  {journal} {Phys. Rev. Lett.}\ }\textbf {\bibinfo {volume} {112}},\
  \bibinfo {pages} {101101} (\bibinfo {year} {2014})},\ \Eprint
  {https://arxiv.org/abs/1310.8288} {arXiv:1310.8288 [gr-qc]} \BibitemShut
  {NoStop}%
\bibitem [{\citenamefont {Ramos-Buades}\ \emph {et~al.}(2020)\citenamefont
  {Ramos-Buades}, \citenamefont {Husa}, \citenamefont {Pratten}, \citenamefont
  {Estell\'es}, \citenamefont {Garc\'\i{}a-Quir\'os}, \citenamefont
  {Mateu-Lucena}, \citenamefont {Colleoni},\ and\ \citenamefont
  {Jaume}}]{Ramos-Buades:2019uvh}%
  \BibitemOpen
  \bibfield  {author} {\bibinfo {author} {\bibfnamefont {A.}~\bibnamefont
  {Ramos-Buades}}, \bibinfo {author} {\bibfnamefont {S.}~\bibnamefont {Husa}},
  \bibinfo {author} {\bibfnamefont {G.}~\bibnamefont {Pratten}}, \bibinfo
  {author} {\bibfnamefont {H.}~\bibnamefont {Estell\'es}}, \bibinfo {author}
  {\bibfnamefont {C.}~\bibnamefont {Garc\'\i{}a-Quir\'os}}, \bibinfo {author}
  {\bibfnamefont {M.}~\bibnamefont {Mateu-Lucena}}, \bibinfo {author}
  {\bibfnamefont {M.}~\bibnamefont {Colleoni}},\ and\ \bibinfo {author}
  {\bibfnamefont {R.}~\bibnamefont {Jaume}},\ }\bibfield  {title} {\bibinfo
  {title} {{First survey of spinning eccentric black hole mergers: Numerical
  relativity simulations, hybrid waveforms, and parameter estimation}},\ }\href
  {https://doi.org/10.1103/PhysRevD.101.083015} {\bibfield  {journal} {\bibinfo
   {journal} {Phys. Rev. D}\ }\textbf {\bibinfo {volume} {101}},\ \bibinfo
  {pages} {083015} (\bibinfo {year} {2020})},\ \Eprint
  {https://arxiv.org/abs/1909.11011} {arXiv:1909.11011 [gr-qc]} \BibitemShut
  {NoStop}%
\bibitem [{\citenamefont {Cho}(2022)}]{Cho:2022cdy}%
  \BibitemOpen
  \bibfield  {author} {\bibinfo {author} {\bibfnamefont {H.-S.}\ \bibnamefont
  {Cho}},\ }\bibfield  {title} {\bibinfo {title} {{Systematic bias due to
  eccentricity in parameter estimation for merging binary neutron stars}},\
  }\href {https://doi.org/10.1103/PhysRevD.105.124022} {\bibfield  {journal}
  {\bibinfo  {journal} {Phys. Rev. D}\ }\textbf {\bibinfo {volume} {105}},\
  \bibinfo {pages} {124022} (\bibinfo {year} {2022})},\ \Eprint
  {https://arxiv.org/abs/2205.12531} {arXiv:2205.12531 [gr-qc]} \BibitemShut
  {NoStop}%
\bibitem [{\citenamefont {Guo}\ \emph {et~al.}(2022)\citenamefont {Guo},
  \citenamefont {Williams}, \citenamefont {Heng}, \citenamefont {Gabbard},
  \citenamefont {Bae}, \citenamefont {Kang},\ and\ \citenamefont
  {Zhu}}]{Guo:2022ehk}%
  \BibitemOpen
  \bibfield  {author} {\bibinfo {author} {\bibfnamefont {W.}~\bibnamefont
  {Guo}}, \bibinfo {author} {\bibfnamefont {D.}~\bibnamefont {Williams}},
  \bibinfo {author} {\bibfnamefont {I.~S.}\ \bibnamefont {Heng}}, \bibinfo
  {author} {\bibfnamefont {H.}~\bibnamefont {Gabbard}}, \bibinfo {author}
  {\bibfnamefont {Y.-B.}\ \bibnamefont {Bae}}, \bibinfo {author} {\bibfnamefont
  {G.}~\bibnamefont {Kang}},\ and\ \bibinfo {author} {\bibfnamefont {Z.-H.}\
  \bibnamefont {Zhu}},\ }\bibfield  {title} {\bibinfo {title} {{Mimicking
  mergers: mistaking black hole captures as mergers}},\ }\href
  {https://doi.org/10.1093/mnras/stac2385} {\bibfield  {journal} {\bibinfo
  {journal} {Mon. Not. Roy. Astron. Soc.}\ }\textbf {\bibinfo {volume} {516}},\
  \bibinfo {pages} {3847} (\bibinfo {year} {2022})},\ \Eprint
  {https://arxiv.org/abs/2203.06969} {arXiv:2203.06969 [gr-qc]} \BibitemShut
  {NoStop}%
\bibitem [{\citenamefont {Gil~Choi}\ \emph {et~al.}(2024)\citenamefont
  {Gil~Choi}, \citenamefont {Yang},\ and\ \citenamefont
  {Lee}}]{GilChoi:2022waq}%
  \BibitemOpen
  \bibfield  {author} {\bibinfo {author} {\bibfnamefont {H.}~\bibnamefont
  {Gil~Choi}}, \bibinfo {author} {\bibfnamefont {T.}~\bibnamefont {Yang}},\
  and\ \bibinfo {author} {\bibfnamefont {H.~M.}\ \bibnamefont {Lee}},\
  }\bibfield  {title} {\bibinfo {title} {{Importance of eccentricities in
  parameter estimation of compact binary inspirals with decihertz
  gravitational-wave detectors}},\ }\href
  {https://doi.org/10.1103/PhysRevD.110.024025} {\bibfield  {journal} {\bibinfo
   {journal} {Phys. Rev. D}\ }\textbf {\bibinfo {volume} {110}},\ \bibinfo
  {pages} {024025} (\bibinfo {year} {2024})},\ \Eprint
  {https://arxiv.org/abs/2210.09541} {arXiv:2210.09541 [gr-qc]} \BibitemShut
  {NoStop}%
\bibitem [{\citenamefont {Das}\ \emph {et~al.}(2024)\citenamefont {Das},
  \citenamefont {Gayathri}, \citenamefont {Divyajyoti}, \citenamefont {Jose},
  \citenamefont {Bartos}, \citenamefont {Klimenko},\ and\ \citenamefont
  {Mishra}}]{Das:2024zib}%
  \BibitemOpen
  \bibfield  {author} {\bibinfo {author} {\bibfnamefont {R.}~\bibnamefont
  {Das}}, \bibinfo {author} {\bibfnamefont {V.}~\bibnamefont {Gayathri}},
  \bibinfo {author} {\bibnamefont {Divyajyoti}}, \bibinfo {author}
  {\bibfnamefont {S.}~\bibnamefont {Jose}}, \bibinfo {author} {\bibfnamefont
  {I.}~\bibnamefont {Bartos}}, \bibinfo {author} {\bibfnamefont
  {S.}~\bibnamefont {Klimenko}},\ and\ \bibinfo {author} {\bibfnamefont
  {C.~K.}\ \bibnamefont {Mishra}},\ }\bibfield  {title} {\bibinfo {title}
  {{Inferring additional physics through unmodelled signal reconstructions}},\
  }\Eprint {https://arxiv.org/abs/2412.11749} {arXiv:2412.11749 [gr-qc]}
  \BibitemShut {NoStop}%
\bibitem [{\citenamefont {Saini}\ \emph {et~al.}(2022)\citenamefont {Saini},
  \citenamefont {Favata},\ and\ \citenamefont {Arun}}]{Saini:2022igm}%
  \BibitemOpen
  \bibfield  {author} {\bibinfo {author} {\bibfnamefont {P.}~\bibnamefont
  {Saini}}, \bibinfo {author} {\bibfnamefont {M.}~\bibnamefont {Favata}},\ and\
  \bibinfo {author} {\bibfnamefont {K.~G.}\ \bibnamefont {Arun}},\ }\bibfield
  {title} {\bibinfo {title} {{Systematic bias on parametrized tests of general
  relativity due to neglect of orbital eccentricity}},\ }\href
  {https://doi.org/10.1103/PhysRevD.106.084031} {\bibfield  {journal} {\bibinfo
   {journal} {Phys. Rev. D}\ }\textbf {\bibinfo {volume} {106}},\ \bibinfo
  {pages} {084031} (\bibinfo {year} {2022})},\ \Eprint
  {https://arxiv.org/abs/2203.04634} {arXiv:2203.04634 [gr-qc]} \BibitemShut
  {NoStop}%
\bibitem [{\citenamefont {Saini}\ \emph {et~al.}(2024)\citenamefont {Saini},
  \citenamefont {Bhat}, \citenamefont {Favata},\ and\ \citenamefont
  {Arun}}]{Saini:2023rto}%
  \BibitemOpen
  \bibfield  {author} {\bibinfo {author} {\bibfnamefont {P.}~\bibnamefont
  {Saini}}, \bibinfo {author} {\bibfnamefont {S.~A.}\ \bibnamefont {Bhat}},
  \bibinfo {author} {\bibfnamefont {M.}~\bibnamefont {Favata}},\ and\ \bibinfo
  {author} {\bibfnamefont {K.~G.}\ \bibnamefont {Arun}},\ }\bibfield  {title}
  {\bibinfo {title} {{Eccentricity-induced systematic error on parametrized
  tests of general relativity: Hierarchical Bayesian inference applied to a
  binary black hole population}},\ }\href
  {https://doi.org/10.1103/PhysRevD.109.084056} {\bibfield  {journal} {\bibinfo
   {journal} {Phys. Rev. D}\ }\textbf {\bibinfo {volume} {109}},\ \bibinfo
  {pages} {084056} (\bibinfo {year} {2024})},\ \Eprint
  {https://arxiv.org/abs/2311.08033} {arXiv:2311.08033 [gr-qc]} \BibitemShut
  {NoStop}%
\bibitem [{\citenamefont {Narayan}\ \emph {et~al.}(2023)\citenamefont
  {Narayan}, \citenamefont {Johnson-McDaniel},\ and\ \citenamefont
  {Gupta}}]{Narayan:2023vhm}%
  \BibitemOpen
  \bibfield  {author} {\bibinfo {author} {\bibfnamefont {P.}~\bibnamefont
  {Narayan}}, \bibinfo {author} {\bibfnamefont {N.~K.}\ \bibnamefont
  {Johnson-McDaniel}},\ and\ \bibinfo {author} {\bibfnamefont {A.}~\bibnamefont
  {Gupta}},\ }\bibfield  {title} {\bibinfo {title} {{Effect of ignoring
  eccentricity in testing general relativity with gravitational waves}},\
  }\href {https://doi.org/10.1103/PhysRevD.108.064003} {\bibfield  {journal}
  {\bibinfo  {journal} {Phys. Rev. D}\ }\textbf {\bibinfo {volume} {108}},\
  \bibinfo {pages} {064003} (\bibinfo {year} {2023})},\ \Eprint
  {https://arxiv.org/abs/2306.04068} {arXiv:2306.04068 [gr-qc]} \BibitemShut
  {NoStop}%
\bibitem [{\citenamefont {Gupta}\ \emph {et~al.}(2024)\citenamefont {Gupta}
  \emph {et~al.}}]{Gupta:2024gun}%
  \BibitemOpen
  \bibfield  {author} {\bibinfo {author} {\bibfnamefont {A.}~\bibnamefont
  {Gupta}} \emph {et~al.},\ }\bibfield  {title} {\bibinfo {title} {{Possible
  Causes of False General Relativity Violations in Gravitational Wave
  Observations}},\ }\Eprint {https://arxiv.org/abs/2405.02197}
  {arXiv:2405.02197 [gr-qc]} \BibitemShut {NoStop}%
\bibitem [{\citenamefont {Shaikh}\ \emph {et~al.}(2024)\citenamefont {Shaikh},
  \citenamefont {Bhat},\ and\ \citenamefont {Kapadia}}]{Shaikh:2024wyn}%
  \BibitemOpen
  \bibfield  {author} {\bibinfo {author} {\bibfnamefont {M.~A.}\ \bibnamefont
  {Shaikh}}, \bibinfo {author} {\bibfnamefont {S.~A.}\ \bibnamefont {Bhat}},\
  and\ \bibinfo {author} {\bibfnamefont {S.~J.}\ \bibnamefont {Kapadia}},\
  }\bibfield  {title} {\bibinfo {title} {{A study of the
  inspiral-merger-ringdown consistency test with gravitational-wave signals
  from compact binaries in eccentric orbits}},\ }\href
  {https://doi.org/10.1103/PhysRevD.110.024030} {\bibfield  {journal} {\bibinfo
   {journal} {Phys. Rev. D}\ }\textbf {\bibinfo {volume} {110}},\ \bibinfo
  {pages} {024030} (\bibinfo {year} {2024})},\ \Eprint
  {https://arxiv.org/abs/2402.15110} {arXiv:2402.15110 [gr-qc]} \BibitemShut
  {NoStop}%
\bibitem [{\citenamefont {Bhat}\ \emph {et~al.}(2023)\citenamefont {Bhat},
  \citenamefont {Saini}, \citenamefont {Favata},\ and\ \citenamefont
  {Arun}}]{Bhat:2022amc}%
  \BibitemOpen
  \bibfield  {author} {\bibinfo {author} {\bibfnamefont {S.~A.}\ \bibnamefont
  {Bhat}}, \bibinfo {author} {\bibfnamefont {P.}~\bibnamefont {Saini}},
  \bibinfo {author} {\bibfnamefont {M.}~\bibnamefont {Favata}},\ and\ \bibinfo
  {author} {\bibfnamefont {K.~G.}\ \bibnamefont {Arun}},\ }\bibfield  {title}
  {\bibinfo {title} {{Systematic bias on the inspiral-merger-ringdown
  consistency test due to neglect of orbital eccentricity}},\ }\href
  {https://doi.org/10.1103/PhysRevD.107.024009} {\bibfield  {journal} {\bibinfo
   {journal} {Phys. Rev. D}\ }\textbf {\bibinfo {volume} {107}},\ \bibinfo
  {pages} {024009} (\bibinfo {year} {2023})},\ \Eprint
  {https://arxiv.org/abs/2207.13761} {arXiv:2207.13761 [gr-qc]} \BibitemShut
  {NoStop}%
\bibitem [{\citenamefont {Bhat}\ \emph {et~al.}(2024)\citenamefont {Bhat},
  \citenamefont {Saini}, \citenamefont {Favata}, \citenamefont {Gandevikar},
  \citenamefont {Mishra},\ and\ \citenamefont {Arun}}]{Bhat:2024hyb}%
  \BibitemOpen
  \bibfield  {author} {\bibinfo {author} {\bibfnamefont {S.~A.}\ \bibnamefont
  {Bhat}}, \bibinfo {author} {\bibfnamefont {P.}~\bibnamefont {Saini}},
  \bibinfo {author} {\bibfnamefont {M.}~\bibnamefont {Favata}}, \bibinfo
  {author} {\bibfnamefont {C.}~\bibnamefont {Gandevikar}}, \bibinfo {author}
  {\bibfnamefont {C.~K.}\ \bibnamefont {Mishra}},\ and\ \bibinfo {author}
  {\bibfnamefont {K.~G.}\ \bibnamefont {Arun}},\ }\bibfield  {title} {\bibinfo
  {title} {{Parametrized tests of general relativity using eccentric compact
  binaries}},\ }\href {https://doi.org/10.1103/PhysRevD.110.124062} {\bibfield
  {journal} {\bibinfo  {journal} {Phys. Rev. D}\ }\textbf {\bibinfo {volume}
  {110}},\ \bibinfo {pages} {124062} (\bibinfo {year} {2024})},\ \Eprint
  {https://arxiv.org/abs/2408.14132} {arXiv:2408.14132 [gr-qc]} \BibitemShut
  {NoStop}%
\bibitem [{\citenamefont {Blackman}\ \emph {et~al.}(2015)\citenamefont
  {Blackman}, \citenamefont {Field}, \citenamefont {Galley}, \citenamefont
  {Szil\'agyi}, \citenamefont {Scheel}, \citenamefont {Tiglio},\ and\
  \citenamefont {Hemberger}}]{Blackman:2015pia}%
  \BibitemOpen
  \bibfield  {author} {\bibinfo {author} {\bibfnamefont {J.}~\bibnamefont
  {Blackman}}, \bibinfo {author} {\bibfnamefont {S.~E.}\ \bibnamefont {Field}},
  \bibinfo {author} {\bibfnamefont {C.~R.}\ \bibnamefont {Galley}}, \bibinfo
  {author} {\bibfnamefont {B.}~\bibnamefont {Szil\'agyi}}, \bibinfo {author}
  {\bibfnamefont {M.~A.}\ \bibnamefont {Scheel}}, \bibinfo {author}
  {\bibfnamefont {M.}~\bibnamefont {Tiglio}},\ and\ \bibinfo {author}
  {\bibfnamefont {D.~A.}\ \bibnamefont {Hemberger}},\ }\bibfield  {title}
  {\bibinfo {title} {{Fast and Accurate Prediction of Numerical Relativity
  Waveforms from Binary Black Hole Coalescences Using Surrogate Models}},\
  }\href {https://doi.org/10.1103/PhysRevLett.115.121102} {\bibfield  {journal}
  {\bibinfo  {journal} {Phys. Rev. Lett.}\ }\textbf {\bibinfo {volume} {115}},\
  \bibinfo {pages} {121102} (\bibinfo {year} {2015})},\ \Eprint
  {https://arxiv.org/abs/1502.07758} {arXiv:1502.07758 [gr-qc]} \BibitemShut
  {NoStop}%
\bibitem [{\citenamefont {Varma}\ \emph
  {et~al.}(2019{\natexlab{a}})\citenamefont {Varma}, \citenamefont {Field},
  \citenamefont {Scheel}, \citenamefont {Blackman}, \citenamefont {Kidder},\
  and\ \citenamefont {Pfeiffer}}]{Varma:2018mmi}%
  \BibitemOpen
  \bibfield  {author} {\bibinfo {author} {\bibfnamefont {V.}~\bibnamefont
  {Varma}}, \bibinfo {author} {\bibfnamefont {S.~E.}\ \bibnamefont {Field}},
  \bibinfo {author} {\bibfnamefont {M.~A.}\ \bibnamefont {Scheel}}, \bibinfo
  {author} {\bibfnamefont {J.}~\bibnamefont {Blackman}}, \bibinfo {author}
  {\bibfnamefont {L.~E.}\ \bibnamefont {Kidder}},\ and\ \bibinfo {author}
  {\bibfnamefont {H.~P.}\ \bibnamefont {Pfeiffer}},\ }\bibfield  {title}
  {\bibinfo {title} {{Surrogate model of hybridized numerical relativity binary
  black hole waveforms}},\ }\href {https://doi.org/10.1103/PhysRevD.99.064045}
  {\bibfield  {journal} {\bibinfo  {journal} {Phys. Rev. D}\ }\textbf {\bibinfo
  {volume} {99}},\ \bibinfo {pages} {064045} (\bibinfo {year}
  {2019}{\natexlab{a}})},\ \Eprint {https://arxiv.org/abs/1812.07865}
  {arXiv:1812.07865 [gr-qc]} \BibitemShut {NoStop}%
\bibitem [{\citenamefont {Varma}\ \emph
  {et~al.}(2019{\natexlab{b}})\citenamefont {Varma}, \citenamefont {Field},
  \citenamefont {Scheel}, \citenamefont {Blackman}, \citenamefont {Gerosa},
  \citenamefont {Stein}, \citenamefont {Kidder},\ and\ \citenamefont
  {Pfeiffer}}]{Varma:2019csw}%
  \BibitemOpen
  \bibfield  {author} {\bibinfo {author} {\bibfnamefont {V.}~\bibnamefont
  {Varma}}, \bibinfo {author} {\bibfnamefont {S.~E.}\ \bibnamefont {Field}},
  \bibinfo {author} {\bibfnamefont {M.~A.}\ \bibnamefont {Scheel}}, \bibinfo
  {author} {\bibfnamefont {J.}~\bibnamefont {Blackman}}, \bibinfo {author}
  {\bibfnamefont {D.}~\bibnamefont {Gerosa}}, \bibinfo {author} {\bibfnamefont
  {L.~C.}\ \bibnamefont {Stein}}, \bibinfo {author} {\bibfnamefont {L.~E.}\
  \bibnamefont {Kidder}},\ and\ \bibinfo {author} {\bibfnamefont {H.~P.}\
  \bibnamefont {Pfeiffer}},\ }\bibfield  {title} {\bibinfo {title} {{Surrogate
  models for precessing binary black hole simulations with unequal masses}},\
  }\href {https://doi.org/10.1103/PhysRevResearch.1.033015} {\bibfield
  {journal} {\bibinfo  {journal} {Phys. Rev. Research.}\ }\textbf {\bibinfo
  {volume} {1}},\ \bibinfo {pages} {033015} (\bibinfo {year}
  {2019}{\natexlab{b}})},\ \Eprint {https://arxiv.org/abs/1905.09300}
  {arXiv:1905.09300 [gr-qc]} \BibitemShut {NoStop}%
\bibitem [{\citenamefont {Yoo}\ \emph {et~al.}(2023)\citenamefont {Yoo} \emph
  {et~al.}}]{Yoo:2023spi}%
  \BibitemOpen
  \bibfield  {author} {\bibinfo {author} {\bibfnamefont {J.}~\bibnamefont
  {Yoo}} \emph {et~al.},\ }\bibfield  {title} {\bibinfo {title} {{Numerical
  relativity surrogate model with memory effects and post-Newtonian
  hybridization}},\ }\href {https://doi.org/10.1103/PhysRevD.108.064027}
  {\bibfield  {journal} {\bibinfo  {journal} {Phys. Rev. D}\ }\textbf {\bibinfo
  {volume} {108}},\ \bibinfo {pages} {064027} (\bibinfo {year} {2023})},\
  \Eprint {https://arxiv.org/abs/2306.03148} {arXiv:2306.03148 [gr-qc]}
  \BibitemShut {NoStop}%
\bibitem [{\citenamefont {Rifat}\ \emph {et~al.}(2020)\citenamefont {Rifat},
  \citenamefont {Field}, \citenamefont {Khanna},\ and\ \citenamefont
  {Varma}}]{Rifat:2019ltp}%
  \BibitemOpen
  \bibfield  {author} {\bibinfo {author} {\bibfnamefont {N.~E.~M.}\
  \bibnamefont {Rifat}}, \bibinfo {author} {\bibfnamefont {S.~E.}\ \bibnamefont
  {Field}}, \bibinfo {author} {\bibfnamefont {G.}~\bibnamefont {Khanna}},\ and\
  \bibinfo {author} {\bibfnamefont {V.}~\bibnamefont {Varma}},\ }\bibfield
  {title} {\bibinfo {title} {{Surrogate model for gravitational wave signals
  from comparable and large-mass-ratio black hole binaries}},\ }\href
  {https://doi.org/10.1103/PhysRevD.101.081502} {\bibfield  {journal} {\bibinfo
   {journal} {Phys. Rev. D}\ }\textbf {\bibinfo {volume} {101}},\ \bibinfo
  {pages} {081502} (\bibinfo {year} {2020})},\ \Eprint
  {https://arxiv.org/abs/1910.10473} {arXiv:1910.10473 [gr-qc]} \BibitemShut
  {NoStop}%
\bibitem [{\citenamefont {Islam}\ \emph {et~al.}(2022)\citenamefont {Islam},
  \citenamefont {Field}, \citenamefont {Hughes}, \citenamefont {Khanna},
  \citenamefont {Varma}, \citenamefont {Giesler}, \citenamefont {Scheel},
  \citenamefont {Kidder},\ and\ \citenamefont {Pfeiffer}}]{Islam:2022laz}%
  \BibitemOpen
  \bibfield  {author} {\bibinfo {author} {\bibfnamefont {T.}~\bibnamefont
  {Islam}}, \bibinfo {author} {\bibfnamefont {S.~E.}\ \bibnamefont {Field}},
  \bibinfo {author} {\bibfnamefont {S.~A.}\ \bibnamefont {Hughes}}, \bibinfo
  {author} {\bibfnamefont {G.}~\bibnamefont {Khanna}}, \bibinfo {author}
  {\bibfnamefont {V.}~\bibnamefont {Varma}}, \bibinfo {author} {\bibfnamefont
  {M.}~\bibnamefont {Giesler}}, \bibinfo {author} {\bibfnamefont {M.~A.}\
  \bibnamefont {Scheel}}, \bibinfo {author} {\bibfnamefont {L.~E.}\
  \bibnamefont {Kidder}},\ and\ \bibinfo {author} {\bibfnamefont {H.~P.}\
  \bibnamefont {Pfeiffer}},\ }\bibfield  {title} {\bibinfo {title} {{Surrogate
  model for gravitational wave signals from nonspinning, comparable-to
  large-mass-ratio black hole binaries built on black hole perturbation theory
  waveforms calibrated to numerical relativity}},\ }\href
  {https://doi.org/10.1103/PhysRevD.106.104025} {\bibfield  {journal} {\bibinfo
   {journal} {Phys. Rev. D}\ }\textbf {\bibinfo {volume} {106}},\ \bibinfo
  {pages} {104025} (\bibinfo {year} {2022})},\ \Eprint
  {https://arxiv.org/abs/2204.01972} {arXiv:2204.01972 [gr-qc]} \BibitemShut
  {NoStop}%
\bibitem [{\citenamefont {Rink}\ \emph {et~al.}(2024)\citenamefont {Rink},
  \citenamefont {Bachhar}, \citenamefont {Islam}, \citenamefont {Rifat},
  \citenamefont {Gonzalez-Quesada}, \citenamefont {Field}, \citenamefont
  {Khanna}, \citenamefont {Hughes},\ and\ \citenamefont
  {Varma}}]{Rink:2024swg}%
  \BibitemOpen
  \bibfield  {author} {\bibinfo {author} {\bibfnamefont {K.}~\bibnamefont
  {Rink}}, \bibinfo {author} {\bibfnamefont {R.}~\bibnamefont {Bachhar}},
  \bibinfo {author} {\bibfnamefont {T.}~\bibnamefont {Islam}}, \bibinfo
  {author} {\bibfnamefont {N.~E.~M.}\ \bibnamefont {Rifat}}, \bibinfo {author}
  {\bibfnamefont {K.}~\bibnamefont {Gonzalez-Quesada}}, \bibinfo {author}
  {\bibfnamefont {S.~E.}\ \bibnamefont {Field}}, \bibinfo {author}
  {\bibfnamefont {G.}~\bibnamefont {Khanna}}, \bibinfo {author} {\bibfnamefont
  {S.~A.}\ \bibnamefont {Hughes}},\ and\ \bibinfo {author} {\bibfnamefont
  {V.}~\bibnamefont {Varma}},\ }\bibfield  {title} {\bibinfo {title}
  {{Gravitational wave surrogate model for spinning, intermediate mass ratio
  binaries based on perturbation theory and numerical relativity}},\ }\href
  {https://doi.org/10.1103/PhysRevD.110.124069} {\bibfield  {journal} {\bibinfo
   {journal} {Phys. Rev. D}\ }\textbf {\bibinfo {volume} {110}},\ \bibinfo
  {pages} {124069} (\bibinfo {year} {2024})},\ \Eprint
  {https://arxiv.org/abs/2407.18319} {arXiv:2407.18319 [gr-qc]} \BibitemShut
  {NoStop}%
\bibitem [{\citenamefont {Buonanno}\ and\ \citenamefont
  {Damour}(1999)}]{Buonanno:1998gg}%
  \BibitemOpen
  \bibfield  {author} {\bibinfo {author} {\bibfnamefont {A.}~\bibnamefont
  {Buonanno}}\ and\ \bibinfo {author} {\bibfnamefont {T.}~\bibnamefont
  {Damour}},\ }\bibfield  {title} {\bibinfo {title} {{Effective one-body
  approach to general relativistic two-body dynamics}},\ }\href
  {https://doi.org/10.1103/PhysRevD.59.084006} {\bibfield  {journal} {\bibinfo
  {journal} {Phys. Rev. D}\ }\textbf {\bibinfo {volume} {59}},\ \bibinfo
  {pages} {084006} (\bibinfo {year} {1999})},\ \Eprint
  {https://arxiv.org/abs/gr-qc/9811091} {arXiv:gr-qc/9811091} \BibitemShut
  {NoStop}%
\bibitem [{\citenamefont {Buonanno}\ and\ \citenamefont
  {Damour}(2000)}]{Buonanno:2000ef}%
  \BibitemOpen
  \bibfield  {author} {\bibinfo {author} {\bibfnamefont {A.}~\bibnamefont
  {Buonanno}}\ and\ \bibinfo {author} {\bibfnamefont {T.}~\bibnamefont
  {Damour}},\ }\bibfield  {title} {\bibinfo {title} {{Transition from inspiral
  to plunge in binary black hole coalescences}},\ }\href
  {https://doi.org/10.1103/PhysRevD.62.064015} {\bibfield  {journal} {\bibinfo
  {journal} {Phys. Rev. D}\ }\textbf {\bibinfo {volume} {62}},\ \bibinfo
  {pages} {064015} (\bibinfo {year} {2000})},\ \Eprint
  {https://arxiv.org/abs/gr-qc/0001013} {arXiv:gr-qc/0001013} \BibitemShut
  {NoStop}%
\bibitem [{\citenamefont {Damour}\ \emph {et~al.}(2009)\citenamefont {Damour},
  \citenamefont {Iyer},\ and\ \citenamefont {Nagar}}]{Damour:2008gu}%
  \BibitemOpen
  \bibfield  {author} {\bibinfo {author} {\bibfnamefont {T.}~\bibnamefont
  {Damour}}, \bibinfo {author} {\bibfnamefont {B.~R.}\ \bibnamefont {Iyer}},\
  and\ \bibinfo {author} {\bibfnamefont {A.}~\bibnamefont {Nagar}},\ }\bibfield
   {title} {\bibinfo {title} {{Improved resummation of post-Newtonian
  multipolar waveforms from circularized compact binaries}},\ }\href
  {https://doi.org/10.1103/PhysRevD.79.064004} {\bibfield  {journal} {\bibinfo
  {journal} {Phys. Rev. D}\ }\textbf {\bibinfo {volume} {79}},\ \bibinfo
  {pages} {064004} (\bibinfo {year} {2009})},\ \Eprint
  {https://arxiv.org/abs/0811.2069} {arXiv:0811.2069 [gr-qc]} \BibitemShut
  {NoStop}%
\bibitem [{\citenamefont {Pan}\ \emph {et~al.}(2011)\citenamefont {Pan},
  \citenamefont {Buonanno}, \citenamefont {Fujita}, \citenamefont {Racine},\
  and\ \citenamefont {Tagoshi}}]{Pan:2010hz}%
  \BibitemOpen
  \bibfield  {author} {\bibinfo {author} {\bibfnamefont {Y.}~\bibnamefont
  {Pan}}, \bibinfo {author} {\bibfnamefont {A.}~\bibnamefont {Buonanno}},
  \bibinfo {author} {\bibfnamefont {R.}~\bibnamefont {Fujita}}, \bibinfo
  {author} {\bibfnamefont {E.}~\bibnamefont {Racine}},\ and\ \bibinfo {author}
  {\bibfnamefont {H.}~\bibnamefont {Tagoshi}},\ }\bibfield  {title} {\bibinfo
  {title} {{Post-Newtonian factorized multipolar waveforms for spinning,
  non-precessing black-hole binaries}},\ }\href
  {https://doi.org/10.1103/PhysRevD.83.064003} {\bibfield  {journal} {\bibinfo
  {journal} {Phys. Rev. D}\ }\textbf {\bibinfo {volume} {83}},\ \bibinfo
  {pages} {064003} (\bibinfo {year} {2011})},\ \bibinfo {note} {[Erratum:
  Phys.Rev.D 87, 109901 (2013)]},\ \Eprint {https://arxiv.org/abs/1006.0431}
  {arXiv:1006.0431 [gr-qc]} \BibitemShut {NoStop}%
\bibitem [{\citenamefont {Pan}\ \emph {et~al.}(2014)\citenamefont {Pan},
  \citenamefont {Buonanno}, \citenamefont {Taracchini}, \citenamefont {Kidder},
  \citenamefont {Mrou\'e}, \citenamefont {Pfeiffer}, \citenamefont {Scheel},\
  and\ \citenamefont {Szil\'agyi}}]{Pan:2013rra}%
  \BibitemOpen
  \bibfield  {author} {\bibinfo {author} {\bibfnamefont {Y.}~\bibnamefont
  {Pan}}, \bibinfo {author} {\bibfnamefont {A.}~\bibnamefont {Buonanno}},
  \bibinfo {author} {\bibfnamefont {A.}~\bibnamefont {Taracchini}}, \bibinfo
  {author} {\bibfnamefont {L.~E.}\ \bibnamefont {Kidder}}, \bibinfo {author}
  {\bibfnamefont {A.~H.}\ \bibnamefont {Mrou\'e}}, \bibinfo {author}
  {\bibfnamefont {H.~P.}\ \bibnamefont {Pfeiffer}}, \bibinfo {author}
  {\bibfnamefont {M.~A.}\ \bibnamefont {Scheel}},\ and\ \bibinfo {author}
  {\bibfnamefont {B.}~\bibnamefont {Szil\'agyi}},\ }\bibfield  {title}
  {\bibinfo {title} {{Inspiral-merger-ringdown waveforms of spinning,
  precessing black-hole binaries in the effective-one-body formalism}},\ }\href
  {https://doi.org/10.1103/PhysRevD.89.084006} {\bibfield  {journal} {\bibinfo
  {journal} {Phys. Rev. D}\ }\textbf {\bibinfo {volume} {89}},\ \bibinfo
  {pages} {084006} (\bibinfo {year} {2014})},\ \Eprint
  {https://arxiv.org/abs/1307.6232} {arXiv:1307.6232 [gr-qc]} \BibitemShut
  {NoStop}%
\bibitem [{\citenamefont {Taracchini}\ \emph
  {et~al.}(2014{\natexlab{a}})\citenamefont {Taracchini} \emph
  {et~al.}}]{Taracchini:2013rva}%
  \BibitemOpen
  \bibfield  {author} {\bibinfo {author} {\bibfnamefont {A.}~\bibnamefont
  {Taracchini}} \emph {et~al.},\ }\bibfield  {title} {\bibinfo {title}
  {{Effective-one-body model for black-hole binaries with generic mass ratios
  and spins}},\ }\href {https://doi.org/10.1103/PhysRevD.89.061502} {\bibfield
  {journal} {\bibinfo  {journal} {Phys. Rev. D}\ }\textbf {\bibinfo {volume}
  {89}},\ \bibinfo {pages} {061502} (\bibinfo {year} {2014}{\natexlab{a}})},\
  \Eprint {https://arxiv.org/abs/1311.2544} {arXiv:1311.2544 [gr-qc]}
  \BibitemShut {NoStop}%
\bibitem [{\citenamefont {Boh\'e}\ \emph {et~al.}(2017)\citenamefont {Boh\'e}
  \emph {et~al.}}]{Bohe:2016gbl}%
  \BibitemOpen
  \bibfield  {author} {\bibinfo {author} {\bibfnamefont {A.}~\bibnamefont
  {Boh\'e}} \emph {et~al.},\ }\bibfield  {title} {\bibinfo {title} {{Improved
  effective-one-body model of spinning, nonprecessing binary black holes for
  the era of gravitational-wave astrophysics with advanced detectors}},\ }\href
  {https://doi.org/10.1103/PhysRevD.95.044028} {\bibfield  {journal} {\bibinfo
  {journal} {Phys. Rev. D}\ }\textbf {\bibinfo {volume} {95}},\ \bibinfo
  {pages} {044028} (\bibinfo {year} {2017})},\ \Eprint
  {https://arxiv.org/abs/1611.03703} {arXiv:1611.03703 [gr-qc]} \BibitemShut
  {NoStop}%
\bibitem [{\citenamefont {Nagar}\ \emph {et~al.}(2018)\citenamefont {Nagar}
  \emph {et~al.}}]{Nagar:2018zoe}%
  \BibitemOpen
  \bibfield  {author} {\bibinfo {author} {\bibfnamefont {A.}~\bibnamefont
  {Nagar}} \emph {et~al.},\ }\bibfield  {title} {\bibinfo {title} {{Time-domain
  effective-one-body gravitational waveforms for coalescing compact binaries
  with nonprecessing spins, tides and self-spin effects}},\ }\href
  {https://doi.org/10.1103/PhysRevD.98.104052} {\bibfield  {journal} {\bibinfo
  {journal} {Phys. Rev. D}\ }\textbf {\bibinfo {volume} {98}},\ \bibinfo
  {pages} {104052} (\bibinfo {year} {2018})},\ \Eprint
  {https://arxiv.org/abs/1806.01772} {arXiv:1806.01772 [gr-qc]} \BibitemShut
  {NoStop}%
\bibitem [{\citenamefont {Cotesta}\ \emph {et~al.}(2018)\citenamefont
  {Cotesta}, \citenamefont {Buonanno}, \citenamefont {Boh\'e}, \citenamefont
  {Taracchini}, \citenamefont {Hinder},\ and\ \citenamefont
  {Ossokine}}]{Cotesta:2018fcv}%
  \BibitemOpen
  \bibfield  {author} {\bibinfo {author} {\bibfnamefont {R.}~\bibnamefont
  {Cotesta}}, \bibinfo {author} {\bibfnamefont {A.}~\bibnamefont {Buonanno}},
  \bibinfo {author} {\bibfnamefont {A.}~\bibnamefont {Boh\'e}}, \bibinfo
  {author} {\bibfnamefont {A.}~\bibnamefont {Taracchini}}, \bibinfo {author}
  {\bibfnamefont {I.}~\bibnamefont {Hinder}},\ and\ \bibinfo {author}
  {\bibfnamefont {S.}~\bibnamefont {Ossokine}},\ }\bibfield  {title} {\bibinfo
  {title} {{Enriching the Symphony of Gravitational Waves from Binary Black
  Holes by Tuning Higher Harmonics}},\ }\href
  {https://doi.org/10.1103/PhysRevD.98.084028} {\bibfield  {journal} {\bibinfo
  {journal} {Phys. Rev. D}\ }\textbf {\bibinfo {volume} {98}},\ \bibinfo
  {pages} {084028} (\bibinfo {year} {2018})},\ \Eprint
  {https://arxiv.org/abs/1803.10701} {arXiv:1803.10701 [gr-qc]} \BibitemShut
  {NoStop}%
\bibitem [{\citenamefont {Babak}\ \emph {et~al.}(2017)\citenamefont {Babak},
  \citenamefont {Taracchini},\ and\ \citenamefont {Buonanno}}]{Babak:2016tgq}%
  \BibitemOpen
  \bibfield  {author} {\bibinfo {author} {\bibfnamefont {S.}~\bibnamefont
  {Babak}}, \bibinfo {author} {\bibfnamefont {A.}~\bibnamefont {Taracchini}},\
  and\ \bibinfo {author} {\bibfnamefont {A.}~\bibnamefont {Buonanno}},\
  }\bibfield  {title} {\bibinfo {title} {{Validating the effective-one-body
  model of spinning, precessing binary black holes against numerical
  relativity}},\ }\href {https://doi.org/10.1103/PhysRevD.95.024010} {\bibfield
   {journal} {\bibinfo  {journal} {Phys. Rev. D}\ }\textbf {\bibinfo {volume}
  {95}},\ \bibinfo {pages} {024010} (\bibinfo {year} {2017})},\ \Eprint
  {https://arxiv.org/abs/1607.05661} {arXiv:1607.05661 [gr-qc]} \BibitemShut
  {NoStop}%
\bibitem [{\citenamefont {Ossokine}\ \emph {et~al.}(2020)\citenamefont
  {Ossokine} \emph {et~al.}}]{Ossokine:2020kjp}%
  \BibitemOpen
  \bibfield  {author} {\bibinfo {author} {\bibfnamefont {S.}~\bibnamefont
  {Ossokine}} \emph {et~al.},\ }\bibfield  {title} {\bibinfo {title}
  {{Multipolar Effective-One-Body Waveforms for Precessing Binary Black Holes:
  Construction and Validation}},\ }\href
  {https://doi.org/10.1103/PhysRevD.102.044055} {\bibfield  {journal} {\bibinfo
   {journal} {Phys. Rev. D}\ }\textbf {\bibinfo {volume} {102}},\ \bibinfo
  {pages} {044055} (\bibinfo {year} {2020})},\ \Eprint
  {https://arxiv.org/abs/2004.09442} {arXiv:2004.09442 [gr-qc]} \BibitemShut
  {NoStop}%
\bibitem [{\citenamefont {Nagar}\ and\ \citenamefont
  {Rettegno}(2019)}]{Nagar:2018gnk}%
  \BibitemOpen
  \bibfield  {author} {\bibinfo {author} {\bibfnamefont {A.}~\bibnamefont
  {Nagar}}\ and\ \bibinfo {author} {\bibfnamefont {P.}~\bibnamefont
  {Rettegno}},\ }\bibfield  {title} {\bibinfo {title} {{Efficient effective one
  body time-domain gravitational waveforms}},\ }\href
  {https://doi.org/10.1103/PhysRevD.99.021501} {\bibfield  {journal} {\bibinfo
  {journal} {Phys. Rev. D}\ }\textbf {\bibinfo {volume} {99}},\ \bibinfo
  {pages} {021501} (\bibinfo {year} {2019})},\ \Eprint
  {https://arxiv.org/abs/1805.03891} {arXiv:1805.03891 [gr-qc]} \BibitemShut
  {NoStop}%
\bibitem [{\citenamefont {Nagar}\ \emph {et~al.}(2020)\citenamefont {Nagar},
  \citenamefont {Riemenschneider}, \citenamefont {Pratten}, \citenamefont
  {Rettegno},\ and\ \citenamefont {Messina}}]{Nagar:2020pcj}%
  \BibitemOpen
  \bibfield  {author} {\bibinfo {author} {\bibfnamefont {A.}~\bibnamefont
  {Nagar}}, \bibinfo {author} {\bibfnamefont {G.}~\bibnamefont
  {Riemenschneider}}, \bibinfo {author} {\bibfnamefont {G.}~\bibnamefont
  {Pratten}}, \bibinfo {author} {\bibfnamefont {P.}~\bibnamefont {Rettegno}},\
  and\ \bibinfo {author} {\bibfnamefont {F.}~\bibnamefont {Messina}},\
  }\bibfield  {title} {\bibinfo {title} {{Multipolar effective one body
  waveform model for spin-aligned black hole binaries}},\ }\href
  {https://doi.org/10.1103/PhysRevD.102.024077} {\bibfield  {journal} {\bibinfo
   {journal} {Phys. Rev. D}\ }\textbf {\bibinfo {volume} {102}},\ \bibinfo
  {pages} {024077} (\bibinfo {year} {2020})},\ \Eprint
  {https://arxiv.org/abs/2001.09082} {arXiv:2001.09082 [gr-qc]} \BibitemShut
  {NoStop}%
\bibitem [{\citenamefont {Riemenschneider}\ \emph {et~al.}(2021)\citenamefont
  {Riemenschneider}, \citenamefont {Rettegno}, \citenamefont {Breschi},
  \citenamefont {Albertini}, \citenamefont {Gamba}, \citenamefont {Bernuzzi},\
  and\ \citenamefont {Nagar}}]{Riemenschneider:2021ppj}%
  \BibitemOpen
  \bibfield  {author} {\bibinfo {author} {\bibfnamefont {G.}~\bibnamefont
  {Riemenschneider}}, \bibinfo {author} {\bibfnamefont {P.}~\bibnamefont
  {Rettegno}}, \bibinfo {author} {\bibfnamefont {M.}~\bibnamefont {Breschi}},
  \bibinfo {author} {\bibfnamefont {A.}~\bibnamefont {Albertini}}, \bibinfo
  {author} {\bibfnamefont {R.}~\bibnamefont {Gamba}}, \bibinfo {author}
  {\bibfnamefont {S.}~\bibnamefont {Bernuzzi}},\ and\ \bibinfo {author}
  {\bibfnamefont {A.}~\bibnamefont {Nagar}},\ }\bibfield  {title} {\bibinfo
  {title} {{Assessment of consistent next-to-quasicircular corrections and
  postadiabatic approximation in effective-one-body multipolar waveforms for
  binary black hole coalescences}},\ }\href
  {https://doi.org/10.1103/PhysRevD.104.104045} {\bibfield  {journal} {\bibinfo
   {journal} {Phys. Rev. D}\ }\textbf {\bibinfo {volume} {104}},\ \bibinfo
  {pages} {104045} (\bibinfo {year} {2021})},\ \Eprint
  {https://arxiv.org/abs/2104.07533} {arXiv:2104.07533 [gr-qc]} \BibitemShut
  {NoStop}%
\bibitem [{\citenamefont {Pompili}\ \emph {et~al.}(2023)\citenamefont {Pompili}
  \emph {et~al.}}]{Pompili:2023tna}%
  \BibitemOpen
  \bibfield  {author} {\bibinfo {author} {\bibfnamefont {L.}~\bibnamefont
  {Pompili}} \emph {et~al.},\ }\bibfield  {title} {\bibinfo {title} {{Laying
  the foundation of the effective-one-body waveform models SEOBNRv5: Improved
  accuracy and efficiency for spinning nonprecessing binary black holes}},\
  }\href {https://doi.org/10.1103/PhysRevD.108.124035} {\bibfield  {journal}
  {\bibinfo  {journal} {Phys. Rev. D}\ }\textbf {\bibinfo {volume} {108}},\
  \bibinfo {pages} {124035} (\bibinfo {year} {2023})},\ \Eprint
  {https://arxiv.org/abs/2303.18039} {arXiv:2303.18039 [gr-qc]} \BibitemShut
  {NoStop}%
\bibitem [{\citenamefont {Khalil}\ \emph {et~al.}(2021)\citenamefont {Khalil},
  \citenamefont {Buonanno}, \citenamefont {Steinhoff},\ and\ \citenamefont
  {Vines}}]{Khalil:2021txt}%
  \BibitemOpen
  \bibfield  {author} {\bibinfo {author} {\bibfnamefont {M.}~\bibnamefont
  {Khalil}}, \bibinfo {author} {\bibfnamefont {A.}~\bibnamefont {Buonanno}},
  \bibinfo {author} {\bibfnamefont {J.}~\bibnamefont {Steinhoff}},\ and\
  \bibinfo {author} {\bibfnamefont {J.}~\bibnamefont {Vines}},\ }\bibfield
  {title} {\bibinfo {title} {{Radiation-reaction force and multipolar waveforms
  for eccentric, spin-aligned binaries in the effective-one-body formalism}},\
  }\href {https://doi.org/10.1103/PhysRevD.104.024046} {\bibfield  {journal}
  {\bibinfo  {journal} {Phys. Rev. D}\ }\textbf {\bibinfo {volume} {104}},\
  \bibinfo {pages} {024046} (\bibinfo {year} {2021})},\ \Eprint
  {https://arxiv.org/abs/2104.11705} {arXiv:2104.11705 [gr-qc]} \BibitemShut
  {NoStop}%
\bibitem [{\citenamefont {Ramos-Buades}\ \emph {et~al.}(2022)\citenamefont
  {Ramos-Buades}, \citenamefont {Buonanno}, \citenamefont {Khalil},\ and\
  \citenamefont {Ossokine}}]{Ramos-Buades:2021adz}%
  \BibitemOpen
  \bibfield  {author} {\bibinfo {author} {\bibfnamefont {A.}~\bibnamefont
  {Ramos-Buades}}, \bibinfo {author} {\bibfnamefont {A.}~\bibnamefont
  {Buonanno}}, \bibinfo {author} {\bibfnamefont {M.}~\bibnamefont {Khalil}},\
  and\ \bibinfo {author} {\bibfnamefont {S.}~\bibnamefont {Ossokine}},\
  }\bibfield  {title} {\bibinfo {title} {{Effective-one-body multipolar
  waveforms for eccentric binary black holes with nonprecessing spins}},\
  }\href {https://doi.org/10.1103/PhysRevD.105.044035} {\bibfield  {journal}
  {\bibinfo  {journal} {Phys. Rev. D}\ }\textbf {\bibinfo {volume} {105}},\
  \bibinfo {pages} {044035} (\bibinfo {year} {2022})},\ \Eprint
  {https://arxiv.org/abs/2112.06952} {arXiv:2112.06952 [gr-qc]} \BibitemShut
  {NoStop}%
\bibitem [{\citenamefont {Gamboa}\ \emph
  {et~al.}(2024{\natexlab{a}})\citenamefont {Gamboa} \emph
  {et~al.}}]{Gamboa:2024hli}%
  \BibitemOpen
  \bibfield  {author} {\bibinfo {author} {\bibfnamefont {A.}~\bibnamefont
  {Gamboa}} \emph {et~al.},\ }\bibfield  {title} {\bibinfo {title} {{Accurate
  waveforms for eccentric, aligned-spin binary black holes: The multipolar
  effective-one-body model SEOBNRv5EHM}},\ }\Eprint
  {https://arxiv.org/abs/2412.12823} {arXiv:2412.12823 [gr-qc]} \BibitemShut
  {NoStop}%
\bibitem [{\citenamefont {Gamboa}\ \emph
  {et~al.}(2024{\natexlab{b}})\citenamefont {Gamboa}, \citenamefont {Khalil},\
  and\ \citenamefont {Buonanno}}]{Gamboa:2024imd}%
  \BibitemOpen
  \bibfield  {author} {\bibinfo {author} {\bibfnamefont {A.}~\bibnamefont
  {Gamboa}}, \bibinfo {author} {\bibfnamefont {M.}~\bibnamefont {Khalil}},\
  and\ \bibinfo {author} {\bibfnamefont {A.}~\bibnamefont {Buonanno}},\
  }\bibfield  {title} {\bibinfo {title} {{Third post-Newtonian dynamics for
  eccentric orbits and aligned spins in the effective-one-body waveform model
  SEOBNRv5EHM}},\ }\Eprint {https://arxiv.org/abs/2412.12831} {arXiv:2412.12831
  [gr-qc]} \BibitemShut {NoStop}%
\bibitem [{\citenamefont {Bini}\ and\ \citenamefont
  {Damour}(2012)}]{Bini:2012ji}%
  \BibitemOpen
  \bibfield  {author} {\bibinfo {author} {\bibfnamefont {D.}~\bibnamefont
  {Bini}}\ and\ \bibinfo {author} {\bibfnamefont {T.}~\bibnamefont {Damour}},\
  }\bibfield  {title} {\bibinfo {title} {{Gravitational radiation reaction
  along general orbits in the effective one-body formalism}},\ }\href
  {https://doi.org/10.1103/PhysRevD.86.124012} {\bibfield  {journal} {\bibinfo
  {journal} {Phys. Rev. D}\ }\textbf {\bibinfo {volume} {86}},\ \bibinfo
  {pages} {124012} (\bibinfo {year} {2012})},\ \Eprint
  {https://arxiv.org/abs/1210.2834} {arXiv:1210.2834 [gr-qc]} \BibitemShut
  {NoStop}%
\bibitem [{\citenamefont {Chiaramello}\ and\ \citenamefont
  {Nagar}(2020)}]{Chiaramello:2020ehz}%
  \BibitemOpen
  \bibfield  {author} {\bibinfo {author} {\bibfnamefont {D.}~\bibnamefont
  {Chiaramello}}\ and\ \bibinfo {author} {\bibfnamefont {A.}~\bibnamefont
  {Nagar}},\ }\bibfield  {title} {\bibinfo {title} {{Faithful analytical
  effective-one-body waveform model for spin-aligned, moderately eccentric,
  coalescing black hole binaries}},\ }\href
  {https://doi.org/10.1103/PhysRevD.101.101501} {\bibfield  {journal} {\bibinfo
   {journal} {Phys. Rev. D}\ }\textbf {\bibinfo {volume} {101}},\ \bibinfo
  {pages} {101501} (\bibinfo {year} {2020})},\ \Eprint
  {https://arxiv.org/abs/2001.11736} {arXiv:2001.11736 [gr-qc]} \BibitemShut
  {NoStop}%
\bibitem [{\citenamefont {Nagar}\ \emph {et~al.}(2021)\citenamefont {Nagar},
  \citenamefont {Bonino},\ and\ \citenamefont {Rettegno}}]{Nagar:2021gss}%
  \BibitemOpen
  \bibfield  {author} {\bibinfo {author} {\bibfnamefont {A.}~\bibnamefont
  {Nagar}}, \bibinfo {author} {\bibfnamefont {A.}~\bibnamefont {Bonino}},\ and\
  \bibinfo {author} {\bibfnamefont {P.}~\bibnamefont {Rettegno}},\ }\bibfield
  {title} {\bibinfo {title} {{Effective one-body multipolar waveform model for
  spin-aligned, quasicircular, eccentric, hyperbolic black hole binaries}},\
  }\href {https://doi.org/10.1103/PhysRevD.103.104021} {\bibfield  {journal}
  {\bibinfo  {journal} {Phys. Rev. D}\ }\textbf {\bibinfo {volume} {103}},\
  \bibinfo {pages} {104021} (\bibinfo {year} {2021})},\ \Eprint
  {https://arxiv.org/abs/2101.08624} {arXiv:2101.08624 [gr-qc]} \BibitemShut
  {NoStop}%
\bibitem [{\citenamefont {Albanesi}\ \emph {et~al.}(2021)\citenamefont
  {Albanesi}, \citenamefont {Nagar},\ and\ \citenamefont
  {Bernuzzi}}]{Albanesi:2021rby}%
  \BibitemOpen
  \bibfield  {author} {\bibinfo {author} {\bibfnamefont {S.}~\bibnamefont
  {Albanesi}}, \bibinfo {author} {\bibfnamefont {A.}~\bibnamefont {Nagar}},\
  and\ \bibinfo {author} {\bibfnamefont {S.}~\bibnamefont {Bernuzzi}},\
  }\bibfield  {title} {\bibinfo {title} {{Effective one-body model for
  extreme-mass-ratio spinning binaries on eccentric equatorial orbits: Testing
  radiation reaction and waveform}},\ }\href
  {https://doi.org/10.1103/PhysRevD.104.024067} {\bibfield  {journal} {\bibinfo
   {journal} {Phys. Rev. D}\ }\textbf {\bibinfo {volume} {104}},\ \bibinfo
  {pages} {024067} (\bibinfo {year} {2021})},\ \Eprint
  {https://arxiv.org/abs/2104.10559} {arXiv:2104.10559 [gr-qc]} \BibitemShut
  {NoStop}%
\bibitem [{\citenamefont {Placidi}\ \emph {et~al.}(2022)\citenamefont
  {Placidi}, \citenamefont {Albanesi}, \citenamefont {Nagar}, \citenamefont
  {Orselli}, \citenamefont {Bernuzzi},\ and\ \citenamefont
  {Grignani}}]{Placidi:2021rkh}%
  \BibitemOpen
  \bibfield  {author} {\bibinfo {author} {\bibfnamefont {A.}~\bibnamefont
  {Placidi}}, \bibinfo {author} {\bibfnamefont {S.}~\bibnamefont {Albanesi}},
  \bibinfo {author} {\bibfnamefont {A.}~\bibnamefont {Nagar}}, \bibinfo
  {author} {\bibfnamefont {M.}~\bibnamefont {Orselli}}, \bibinfo {author}
  {\bibfnamefont {S.}~\bibnamefont {Bernuzzi}},\ and\ \bibinfo {author}
  {\bibfnamefont {G.}~\bibnamefont {Grignani}},\ }\bibfield  {title} {\bibinfo
  {title} {{Exploiting Newton-factorized, 2PN-accurate waveform multipoles in
  effective-one-body models for spin-aligned noncircularized binaries}},\
  }\href {https://doi.org/10.1103/PhysRevD.105.104030} {\bibfield  {journal}
  {\bibinfo  {journal} {Phys. Rev. D}\ }\textbf {\bibinfo {volume} {105}},\
  \bibinfo {pages} {104030} (\bibinfo {year} {2022})},\ \Eprint
  {https://arxiv.org/abs/2112.05448} {arXiv:2112.05448 [gr-qc]} \BibitemShut
  {NoStop}%
\bibitem [{\citenamefont {Nagar}\ and\ \citenamefont
  {Rettegno}(2021)}]{Nagar:2021xnh}%
  \BibitemOpen
  \bibfield  {author} {\bibinfo {author} {\bibfnamefont {A.}~\bibnamefont
  {Nagar}}\ and\ \bibinfo {author} {\bibfnamefont {P.}~\bibnamefont
  {Rettegno}},\ }\bibfield  {title} {\bibinfo {title} {{Next generation: Impact
  of high-order analytical information on effective one body waveform models
  for noncircularized, spin-aligned black hole binaries}},\ }\href
  {https://doi.org/10.1103/PhysRevD.104.104004} {\bibfield  {journal} {\bibinfo
   {journal} {Phys. Rev. D}\ }\textbf {\bibinfo {volume} {104}},\ \bibinfo
  {pages} {104004} (\bibinfo {year} {2021})},\ \Eprint
  {https://arxiv.org/abs/2108.02043} {arXiv:2108.02043 [gr-qc]} \BibitemShut
  {NoStop}%
\bibitem [{\citenamefont {Albanesi}\ \emph
  {et~al.}(2022{\natexlab{a}})\citenamefont {Albanesi}, \citenamefont {Nagar},
  \citenamefont {Bernuzzi}, \citenamefont {Placidi},\ and\ \citenamefont
  {Orselli}}]{Albanesi:2022ywx}%
  \BibitemOpen
  \bibfield  {author} {\bibinfo {author} {\bibfnamefont {S.}~\bibnamefont
  {Albanesi}}, \bibinfo {author} {\bibfnamefont {A.}~\bibnamefont {Nagar}},
  \bibinfo {author} {\bibfnamefont {S.}~\bibnamefont {Bernuzzi}}, \bibinfo
  {author} {\bibfnamefont {A.}~\bibnamefont {Placidi}},\ and\ \bibinfo {author}
  {\bibfnamefont {M.}~\bibnamefont {Orselli}},\ }\bibfield  {title} {\bibinfo
  {title} {{Assessment of effective-one-body radiation reactions for generic
  planar orbits}},\ }\href {https://doi.org/10.1103/PhysRevD.105.104031}
  {\bibfield  {journal} {\bibinfo  {journal} {Phys. Rev. D}\ }\textbf {\bibinfo
  {volume} {105}},\ \bibinfo {pages} {104031} (\bibinfo {year}
  {2022}{\natexlab{a}})},\ \Eprint {https://arxiv.org/abs/2202.10063}
  {arXiv:2202.10063 [gr-qc]} \BibitemShut {NoStop}%
\bibitem [{\citenamefont {Albanesi}\ \emph
  {et~al.}(2022{\natexlab{b}})\citenamefont {Albanesi}, \citenamefont
  {Placidi}, \citenamefont {Nagar}, \citenamefont {Orselli},\ and\
  \citenamefont {Bernuzzi}}]{Albanesi:2022xge}%
  \BibitemOpen
  \bibfield  {author} {\bibinfo {author} {\bibfnamefont {S.}~\bibnamefont
  {Albanesi}}, \bibinfo {author} {\bibfnamefont {A.}~\bibnamefont {Placidi}},
  \bibinfo {author} {\bibfnamefont {A.}~\bibnamefont {Nagar}}, \bibinfo
  {author} {\bibfnamefont {M.}~\bibnamefont {Orselli}},\ and\ \bibinfo {author}
  {\bibfnamefont {S.}~\bibnamefont {Bernuzzi}},\ }\bibfield  {title} {\bibinfo
  {title} {{New avenue for accurate analytical waveforms and fluxes for
  eccentric compact binaries}},\ }\href
  {https://doi.org/10.1103/PhysRevD.105.L121503} {\bibfield  {journal}
  {\bibinfo  {journal} {Phys. Rev. D}\ }\textbf {\bibinfo {volume} {105}},\
  \bibinfo {pages} {L121503} (\bibinfo {year} {2022}{\natexlab{b}})},\ \Eprint
  {https://arxiv.org/abs/2203.16286} {arXiv:2203.16286 [gr-qc]} \BibitemShut
  {NoStop}%
\bibitem [{\citenamefont {Nagar}\ and\ \citenamefont
  {Albanesi}(2022)}]{Nagar:2022fep}%
  \BibitemOpen
  \bibfield  {author} {\bibinfo {author} {\bibfnamefont {A.}~\bibnamefont
  {Nagar}}\ and\ \bibinfo {author} {\bibfnamefont {S.}~\bibnamefont
  {Albanesi}},\ }\bibfield  {title} {\bibinfo {title} {{Toward a gravitational
  self-force-informed effective-one-body waveform model for nonprecessing,
  eccentric, large-mass-ratio inspirals}},\ }\href
  {https://doi.org/10.1103/PhysRevD.106.064049} {\bibfield  {journal} {\bibinfo
   {journal} {Phys. Rev. D}\ }\textbf {\bibinfo {volume} {106}},\ \bibinfo
  {pages} {064049} (\bibinfo {year} {2022})},\ \Eprint
  {https://arxiv.org/abs/2207.14002} {arXiv:2207.14002 [gr-qc]} \BibitemShut
  {NoStop}%
\bibitem [{\citenamefont {Albanesi}\ \emph {et~al.}(2023)\citenamefont
  {Albanesi}, \citenamefont {Bernuzzi}, \citenamefont {Damour}, \citenamefont
  {Nagar},\ and\ \citenamefont {Placidi}}]{Albanesi:2023bgi}%
  \BibitemOpen
  \bibfield  {author} {\bibinfo {author} {\bibfnamefont {S.}~\bibnamefont
  {Albanesi}}, \bibinfo {author} {\bibfnamefont {S.}~\bibnamefont {Bernuzzi}},
  \bibinfo {author} {\bibfnamefont {T.}~\bibnamefont {Damour}}, \bibinfo
  {author} {\bibfnamefont {A.}~\bibnamefont {Nagar}},\ and\ \bibinfo {author}
  {\bibfnamefont {A.}~\bibnamefont {Placidi}},\ }\bibfield  {title} {\bibinfo
  {title} {{Faithful effective-one-body waveform of small-mass-ratio coalescing
  black hole binaries: The eccentric, nonspinning case}},\ }\href
  {https://doi.org/10.1103/PhysRevD.108.084037} {\bibfield  {journal} {\bibinfo
   {journal} {Phys. Rev. D}\ }\textbf {\bibinfo {volume} {108}},\ \bibinfo
  {pages} {084037} (\bibinfo {year} {2023})},\ \Eprint
  {https://arxiv.org/abs/2305.19336} {arXiv:2305.19336 [gr-qc]} \BibitemShut
  {NoStop}%
\bibitem [{\citenamefont {Placidi}\ \emph {et~al.}(2023)\citenamefont
  {Placidi}, \citenamefont {Grignani}, \citenamefont {Harmark}, \citenamefont
  {Orselli}, \citenamefont {Gliorio},\ and\ \citenamefont
  {Nagar}}]{Placidi:2023ofj}%
  \BibitemOpen
  \bibfield  {author} {\bibinfo {author} {\bibfnamefont {A.}~\bibnamefont
  {Placidi}}, \bibinfo {author} {\bibfnamefont {G.}~\bibnamefont {Grignani}},
  \bibinfo {author} {\bibfnamefont {T.}~\bibnamefont {Harmark}}, \bibinfo
  {author} {\bibfnamefont {M.}~\bibnamefont {Orselli}}, \bibinfo {author}
  {\bibfnamefont {S.}~\bibnamefont {Gliorio}},\ and\ \bibinfo {author}
  {\bibfnamefont {A.}~\bibnamefont {Nagar}},\ }\bibfield  {title} {\bibinfo
  {title} {{2.5PN accurate waveform information for generic-planar-orbit
  binaries in effective one-body models}},\ }\href
  {https://doi.org/10.1103/PhysRevD.108.024068} {\bibfield  {journal} {\bibinfo
   {journal} {Phys. Rev. D}\ }\textbf {\bibinfo {volume} {108}},\ \bibinfo
  {pages} {024068} (\bibinfo {year} {2023})},\ \Eprint
  {https://arxiv.org/abs/2305.14440} {arXiv:2305.14440 [gr-qc]} \BibitemShut
  {NoStop}%
\bibitem [{\citenamefont {Nagar}\ \emph
  {et~al.}(2024{\natexlab{a}})\citenamefont {Nagar}, \citenamefont {Gamba},
  \citenamefont {Rettegno}, \citenamefont {Fantini},\ and\ \citenamefont
  {Bernuzzi}}]{Nagar:2024dzj}%
  \BibitemOpen
  \bibfield  {author} {\bibinfo {author} {\bibfnamefont {A.}~\bibnamefont
  {Nagar}}, \bibinfo {author} {\bibfnamefont {R.}~\bibnamefont {Gamba}},
  \bibinfo {author} {\bibfnamefont {P.}~\bibnamefont {Rettegno}}, \bibinfo
  {author} {\bibfnamefont {V.}~\bibnamefont {Fantini}},\ and\ \bibinfo {author}
  {\bibfnamefont {S.}~\bibnamefont {Bernuzzi}},\ }\bibfield  {title} {\bibinfo
  {title} {{Effective-one-body waveform model for noncircularized, planar,
  coalescing black hole binaries: The importance of radiation reaction}},\
  }\href {https://doi.org/10.1103/PhysRevD.110.084001} {\bibfield  {journal}
  {\bibinfo  {journal} {Phys. Rev. D}\ }\textbf {\bibinfo {volume} {110}},\
  \bibinfo {pages} {084001} (\bibinfo {year} {2024}{\natexlab{a}})},\ \Eprint
  {https://arxiv.org/abs/2404.05288} {arXiv:2404.05288 [gr-qc]} \BibitemShut
  {NoStop}%
\bibitem [{\citenamefont {Nagar}\ \emph
  {et~al.}(2024{\natexlab{b}})\citenamefont {Nagar}, \citenamefont {Bernuzzi},
  \citenamefont {Chiaramello}, \citenamefont {Fantini}, \citenamefont {Gamba},
  \citenamefont {Panzeri},\ and\ \citenamefont {Rettegno}}]{Nagar:2024oyk}%
  \BibitemOpen
  \bibfield  {author} {\bibinfo {author} {\bibfnamefont {A.}~\bibnamefont
  {Nagar}}, \bibinfo {author} {\bibfnamefont {S.}~\bibnamefont {Bernuzzi}},
  \bibinfo {author} {\bibfnamefont {D.}~\bibnamefont {Chiaramello}}, \bibinfo
  {author} {\bibfnamefont {V.}~\bibnamefont {Fantini}}, \bibinfo {author}
  {\bibfnamefont {R.}~\bibnamefont {Gamba}}, \bibinfo {author} {\bibfnamefont
  {M.}~\bibnamefont {Panzeri}},\ and\ \bibinfo {author} {\bibfnamefont
  {P.}~\bibnamefont {Rettegno}},\ }\bibfield  {title} {\bibinfo {title}
  {{Effective-one-body waveform model for noncircularized, planar, coalescing
  black hole binaries II: high accuracy by improving logarithmic terms in
  resummations}},\ }\href@noop {} {\  (\bibinfo {year} {2024}{\natexlab{b}})},\
  \Eprint {https://arxiv.org/abs/2407.04762} {arXiv:2407.04762 [gr-qc]}
  \BibitemShut {NoStop}%
\bibitem [{\citenamefont {Hinderer}\ and\ \citenamefont
  {Babak}(2017)}]{Hinderer:2017jcs}%
  \BibitemOpen
  \bibfield  {author} {\bibinfo {author} {\bibfnamefont {T.}~\bibnamefont
  {Hinderer}}\ and\ \bibinfo {author} {\bibfnamefont {S.}~\bibnamefont
  {Babak}},\ }\bibfield  {title} {\bibinfo {title} {{Foundations of an
  effective-one-body model for coalescing binaries on eccentric orbits}},\
  }\href {https://doi.org/10.1103/PhysRevD.96.104048} {\bibfield  {journal}
  {\bibinfo  {journal} {Phys. Rev. D}\ }\textbf {\bibinfo {volume} {96}},\
  \bibinfo {pages} {104048} (\bibinfo {year} {2017})},\ \Eprint
  {https://arxiv.org/abs/1707.08426} {arXiv:1707.08426 [gr-qc]} \BibitemShut
  {NoStop}%
\bibitem [{\citenamefont {Cao}\ and\ \citenamefont {Han}(2017)}]{Cao:2017ndf}%
  \BibitemOpen
  \bibfield  {author} {\bibinfo {author} {\bibfnamefont {Z.}~\bibnamefont
  {Cao}}\ and\ \bibinfo {author} {\bibfnamefont {W.-B.}\ \bibnamefont {Han}},\
  }\bibfield  {title} {\bibinfo {title} {{Waveform model for an eccentric
  binary black hole based on the effective-one-body-numerical-relativity
  formalism}},\ }\href {https://doi.org/10.1103/PhysRevD.96.044028} {\bibfield
  {journal} {\bibinfo  {journal} {Phys. Rev. D}\ }\textbf {\bibinfo {volume}
  {96}},\ \bibinfo {pages} {044028} (\bibinfo {year} {2017})},\ \Eprint
  {https://arxiv.org/abs/1708.00166} {arXiv:1708.00166 [gr-qc]} \BibitemShut
  {NoStop}%
\bibitem [{\citenamefont {Liu}\ \emph {et~al.}(2020)\citenamefont {Liu},
  \citenamefont {Cao},\ and\ \citenamefont {Shao}}]{Liu:2019jpg}%
  \BibitemOpen
  \bibfield  {author} {\bibinfo {author} {\bibfnamefont {X.}~\bibnamefont
  {Liu}}, \bibinfo {author} {\bibfnamefont {Z.}~\bibnamefont {Cao}},\ and\
  \bibinfo {author} {\bibfnamefont {L.}~\bibnamefont {Shao}},\ }\bibfield
  {title} {\bibinfo {title} {{Validating the Effective-One-Body
  Numerical-Relativity Waveform Models for Spin-aligned Binary Black Holes
  along Eccentric Orbits}},\ }\href
  {https://doi.org/10.1103/PhysRevD.101.044049} {\bibfield  {journal} {\bibinfo
   {journal} {Phys. Rev. D}\ }\textbf {\bibinfo {volume} {101}},\ \bibinfo
  {pages} {044049} (\bibinfo {year} {2020})},\ \Eprint
  {https://arxiv.org/abs/1910.00784} {arXiv:1910.00784 [gr-qc]} \BibitemShut
  {NoStop}%
\bibitem [{\citenamefont {Liu}\ \emph {et~al.}(2022)\citenamefont {Liu},
  \citenamefont {Cao},\ and\ \citenamefont {Zhu}}]{Liu:2021pkr}%
  \BibitemOpen
  \bibfield  {author} {\bibinfo {author} {\bibfnamefont {X.}~\bibnamefont
  {Liu}}, \bibinfo {author} {\bibfnamefont {Z.}~\bibnamefont {Cao}},\ and\
  \bibinfo {author} {\bibfnamefont {Z.-H.}\ \bibnamefont {Zhu}},\ }\bibfield
  {title} {\bibinfo {title} {{A higher-multipole gravitational waveform model
  for an eccentric binary black holes based on the
  effective-one-body-numerical-relativity formalism}},\ }\href
  {https://doi.org/10.1088/1361-6382/ac4119} {\bibfield  {journal} {\bibinfo
  {journal} {Class. Quant. Grav.}\ }\textbf {\bibinfo {volume} {39}},\ \bibinfo
  {pages} {035009} (\bibinfo {year} {2022})},\ \Eprint
  {https://arxiv.org/abs/2102.08614} {arXiv:2102.08614 [gr-qc]} \BibitemShut
  {NoStop}%
\bibitem [{\citenamefont {Liu}\ \emph {et~al.}(2023)\citenamefont {Liu},
  \citenamefont {Cao},\ and\ \citenamefont {Shao}}]{Liu:2023dgl}%
  \BibitemOpen
  \bibfield  {author} {\bibinfo {author} {\bibfnamefont {X.}~\bibnamefont
  {Liu}}, \bibinfo {author} {\bibfnamefont {Z.}~\bibnamefont {Cao}},\ and\
  \bibinfo {author} {\bibfnamefont {L.}~\bibnamefont {Shao}},\ }\bibfield
  {title} {\bibinfo {title} {{Upgraded waveform model of eccentric binary black
  hole based on effective-one-body-numerical-relativity for spin-aligned binary
  black holes}},\ }\href {https://doi.org/10.1142/S0218271823500153} {\bibfield
   {journal} {\bibinfo  {journal} {Int. J. Mod. Phys. D}\ }\textbf {\bibinfo
  {volume} {32}},\ \bibinfo {pages} {2350015} (\bibinfo {year} {2023})},\
  \Eprint {https://arxiv.org/abs/2306.15277} {arXiv:2306.15277 [gr-qc]}
  \BibitemShut {NoStop}%
\bibitem [{\citenamefont {Damour}\ and\ \citenamefont
  {Nagar}(2007)}]{Damour:2007xr}%
  \BibitemOpen
  \bibfield  {author} {\bibinfo {author} {\bibfnamefont {T.}~\bibnamefont
  {Damour}}\ and\ \bibinfo {author} {\bibfnamefont {A.}~\bibnamefont {Nagar}},\
  }\bibfield  {title} {\bibinfo {title} {{Faithful effective-one-body waveforms
  of small-mass-ratio coalescing black-hole binaries}},\ }\href
  {https://doi.org/10.1103/PhysRevD.76.064028} {\bibfield  {journal} {\bibinfo
  {journal} {Phys. Rev. D}\ }\textbf {\bibinfo {volume} {76}},\ \bibinfo
  {pages} {064028} (\bibinfo {year} {2007})},\ \Eprint
  {https://arxiv.org/abs/0705.2519} {arXiv:0705.2519 [gr-qc]} \BibitemShut
  {NoStop}%
\bibitem [{\citenamefont {Teukolsky}(1973)}]{Teukolsky:1973ha}%
  \BibitemOpen
  \bibfield  {author} {\bibinfo {author} {\bibfnamefont {S.~A.}\ \bibnamefont
  {Teukolsky}},\ }\bibfield  {title} {\bibinfo {title} {{Perturbations of a
  rotating black hole. 1. Fundamental equations for gravitational
  electromagnetic and neutrino field perturbations}},\ }\href
  {https://doi.org/10.1086/152444} {\bibfield  {journal} {\bibinfo  {journal}
  {Astrophys. J.}\ }\textbf {\bibinfo {volume} {185}},\ \bibinfo {pages} {635}
  (\bibinfo {year} {1973})}\BibitemShut {NoStop}%
\bibitem [{\citenamefont {Nagar}\ \emph {et~al.}(2007)\citenamefont {Nagar},
  \citenamefont {Damour},\ and\ \citenamefont {Tartaglia}}]{Nagar:2006xv}%
  \BibitemOpen
  \bibfield  {author} {\bibinfo {author} {\bibfnamefont {A.}~\bibnamefont
  {Nagar}}, \bibinfo {author} {\bibfnamefont {T.}~\bibnamefont {Damour}},\ and\
  \bibinfo {author} {\bibfnamefont {A.}~\bibnamefont {Tartaglia}},\ }\bibfield
  {title} {\bibinfo {title} {{Binary black hole merger in the extreme mass
  ratio limit}},\ }\href {https://doi.org/10.1088/0264-9381/24/12/S08}
  {\bibfield  {journal} {\bibinfo  {journal} {Class. Quant. Grav.}\ }\textbf
  {\bibinfo {volume} {24}},\ \bibinfo {pages} {S109} (\bibinfo {year}
  {2007})},\ \Eprint {https://arxiv.org/abs/gr-qc/0612096}
  {arXiv:gr-qc/0612096} \BibitemShut {NoStop}%
\bibitem [{\citenamefont {Barausse}\ \emph {et~al.}(2012)\citenamefont
  {Barausse}, \citenamefont {Buonanno}, \citenamefont {Hughes}, \citenamefont
  {Khanna}, \citenamefont {O'Sullivan},\ and\ \citenamefont
  {Pan}}]{Barausse:2011kb}%
  \BibitemOpen
  \bibfield  {author} {\bibinfo {author} {\bibfnamefont {E.}~\bibnamefont
  {Barausse}}, \bibinfo {author} {\bibfnamefont {A.}~\bibnamefont {Buonanno}},
  \bibinfo {author} {\bibfnamefont {S.~A.}\ \bibnamefont {Hughes}}, \bibinfo
  {author} {\bibfnamefont {G.}~\bibnamefont {Khanna}}, \bibinfo {author}
  {\bibfnamefont {S.}~\bibnamefont {O'Sullivan}},\ and\ \bibinfo {author}
  {\bibfnamefont {Y.}~\bibnamefont {Pan}},\ }\bibfield  {title} {\bibinfo
  {title} {{Modeling multipolar gravitational-wave emission from small
  mass-ratio mergers}},\ }\href {https://doi.org/10.1103/PhysRevD.85.024046}
  {\bibfield  {journal} {\bibinfo  {journal} {Phys. Rev. D}\ }\textbf {\bibinfo
  {volume} {85}},\ \bibinfo {pages} {024046} (\bibinfo {year} {2012})},\
  \Eprint {https://arxiv.org/abs/1110.3081} {arXiv:1110.3081 [gr-qc]}
  \BibitemShut {NoStop}%
\bibitem [{\citenamefont {Taracchini}\ \emph {et~al.}(2013)\citenamefont
  {Taracchini}, \citenamefont {Buonanno}, \citenamefont {Hughes},\ and\
  \citenamefont {Khanna}}]{Taracchini:2013wfa}%
  \BibitemOpen
  \bibfield  {author} {\bibinfo {author} {\bibfnamefont {A.}~\bibnamefont
  {Taracchini}}, \bibinfo {author} {\bibfnamefont {A.}~\bibnamefont
  {Buonanno}}, \bibinfo {author} {\bibfnamefont {S.~A.}\ \bibnamefont
  {Hughes}},\ and\ \bibinfo {author} {\bibfnamefont {G.}~\bibnamefont
  {Khanna}},\ }\bibfield  {title} {\bibinfo {title} {{Modeling the
  horizon-absorbed gravitational flux for equatorial-circular orbits in Kerr
  spacetime}},\ }\href {https://doi.org/10.1103/PhysRevD.88.044001} {\bibfield
  {journal} {\bibinfo  {journal} {Phys. Rev. D}\ }\textbf {\bibinfo {volume}
  {88}},\ \bibinfo {pages} {044001} (\bibinfo {year} {2013})},\ \bibinfo {note}
  {[Erratum: Phys.Rev.D 88, 109903 (2013)]},\ \Eprint
  {https://arxiv.org/abs/1305.2184} {arXiv:1305.2184 [gr-qc]} \BibitemShut
  {NoStop}%
\bibitem [{\citenamefont {Taracchini}\ \emph
  {et~al.}(2014{\natexlab{b}})\citenamefont {Taracchini}, \citenamefont
  {Buonanno}, \citenamefont {Khanna},\ and\ \citenamefont
  {Hughes}}]{Taracchini:2014zpa}%
  \BibitemOpen
  \bibfield  {author} {\bibinfo {author} {\bibfnamefont {A.}~\bibnamefont
  {Taracchini}}, \bibinfo {author} {\bibfnamefont {A.}~\bibnamefont
  {Buonanno}}, \bibinfo {author} {\bibfnamefont {G.}~\bibnamefont {Khanna}},\
  and\ \bibinfo {author} {\bibfnamefont {S.~A.}\ \bibnamefont {Hughes}},\
  }\bibfield  {title} {\bibinfo {title} {{Small mass plunging into a Kerr black
  hole: Anatomy of the inspiral-merger-ringdown waveforms}},\ }\href
  {https://doi.org/10.1103/PhysRevD.90.084025} {\bibfield  {journal} {\bibinfo
  {journal} {Phys. Rev. D}\ }\textbf {\bibinfo {volume} {90}},\ \bibinfo
  {pages} {084025} (\bibinfo {year} {2014}{\natexlab{b}})},\ \Eprint
  {https://arxiv.org/abs/1404.1819} {arXiv:1404.1819 [gr-qc]} \BibitemShut
  {NoStop}%
\bibitem [{\citenamefont {Albertini}\ \emph
  {et~al.}(2022{\natexlab{a}})\citenamefont {Albertini}, \citenamefont {Nagar},
  \citenamefont {Pound}, \citenamefont {Warburton}, \citenamefont {Wardell},
  \citenamefont {Durkan},\ and\ \citenamefont {Miller}}]{Albertini:2022rfe}%
  \BibitemOpen
  \bibfield  {author} {\bibinfo {author} {\bibfnamefont {A.}~\bibnamefont
  {Albertini}}, \bibinfo {author} {\bibfnamefont {A.}~\bibnamefont {Nagar}},
  \bibinfo {author} {\bibfnamefont {A.}~\bibnamefont {Pound}}, \bibinfo
  {author} {\bibfnamefont {N.}~\bibnamefont {Warburton}}, \bibinfo {author}
  {\bibfnamefont {B.}~\bibnamefont {Wardell}}, \bibinfo {author} {\bibfnamefont
  {L.}~\bibnamefont {Durkan}},\ and\ \bibinfo {author} {\bibfnamefont
  {J.}~\bibnamefont {Miller}},\ }\bibfield  {title} {\bibinfo {title}
  {{Comparing second-order gravitational self-force, numerical relativity, and
  effective one body waveforms from inspiralling, quasicircular, and
  nonspinning black hole binaries}},\ }\href
  {https://doi.org/10.1103/PhysRevD.106.084061} {\bibfield  {journal} {\bibinfo
   {journal} {Phys. Rev. D}\ }\textbf {\bibinfo {volume} {106}},\ \bibinfo
  {pages} {084061} (\bibinfo {year} {2022}{\natexlab{a}})},\ \Eprint
  {https://arxiv.org/abs/2208.01049} {arXiv:2208.01049 [gr-qc]} \BibitemShut
  {NoStop}%
\bibitem [{\citenamefont {Albertini}\ \emph
  {et~al.}(2022{\natexlab{b}})\citenamefont {Albertini}, \citenamefont {Nagar},
  \citenamefont {Pound}, \citenamefont {Warburton}, \citenamefont {Wardell},
  \citenamefont {Durkan},\ and\ \citenamefont {Miller}}]{Albertini:2022dmc}%
  \BibitemOpen
  \bibfield  {author} {\bibinfo {author} {\bibfnamefont {A.}~\bibnamefont
  {Albertini}}, \bibinfo {author} {\bibfnamefont {A.}~\bibnamefont {Nagar}},
  \bibinfo {author} {\bibfnamefont {A.}~\bibnamefont {Pound}}, \bibinfo
  {author} {\bibfnamefont {N.}~\bibnamefont {Warburton}}, \bibinfo {author}
  {\bibfnamefont {B.}~\bibnamefont {Wardell}}, \bibinfo {author} {\bibfnamefont
  {L.}~\bibnamefont {Durkan}},\ and\ \bibinfo {author} {\bibfnamefont
  {J.}~\bibnamefont {Miller}},\ }\bibfield  {title} {\bibinfo {title}
  {{Comparing second-order gravitational self-force and effective one body
  waveforms from inspiralling, quasicircular and nonspinning black hole
  binaries. II. The large-mass-ratio case}},\ }\href
  {https://doi.org/10.1103/PhysRevD.106.084062} {\bibfield  {journal} {\bibinfo
   {journal} {Phys. Rev. D}\ }\textbf {\bibinfo {volume} {106}},\ \bibinfo
  {pages} {084062} (\bibinfo {year} {2022}{\natexlab{b}})},\ \Eprint
  {https://arxiv.org/abs/2208.02055} {arXiv:2208.02055 [gr-qc]} \BibitemShut
  {NoStop}%
\bibitem [{\citenamefont {van~de Meent}\ \emph {et~al.}(2023)\citenamefont
  {van~de Meent}, \citenamefont {Buonanno}, \citenamefont {Mihaylov},
  \citenamefont {Ossokine}, \citenamefont {Pompili}, \citenamefont {Warburton},
  \citenamefont {Pound}, \citenamefont {Wardell}, \citenamefont {Durkan},\ and\
  \citenamefont {Miller}}]{vandeMeent:2023ols}%
  \BibitemOpen
  \bibfield  {author} {\bibinfo {author} {\bibfnamefont {M.}~\bibnamefont
  {van~de Meent}}, \bibinfo {author} {\bibfnamefont {A.}~\bibnamefont
  {Buonanno}}, \bibinfo {author} {\bibfnamefont {D.~P.}\ \bibnamefont
  {Mihaylov}}, \bibinfo {author} {\bibfnamefont {S.}~\bibnamefont {Ossokine}},
  \bibinfo {author} {\bibfnamefont {L.}~\bibnamefont {Pompili}}, \bibinfo
  {author} {\bibfnamefont {N.}~\bibnamefont {Warburton}}, \bibinfo {author}
  {\bibfnamefont {A.}~\bibnamefont {Pound}}, \bibinfo {author} {\bibfnamefont
  {B.}~\bibnamefont {Wardell}}, \bibinfo {author} {\bibfnamefont
  {L.}~\bibnamefont {Durkan}},\ and\ \bibinfo {author} {\bibfnamefont
  {J.}~\bibnamefont {Miller}},\ }\bibfield  {title} {\bibinfo {title}
  {{Enhancing the SEOBNRv5 effective-one-body waveform model with second-order
  gravitational self-force fluxes}},\ }\href
  {https://doi.org/10.1103/PhysRevD.108.124038} {\bibfield  {journal} {\bibinfo
   {journal} {Phys. Rev. D}\ }\textbf {\bibinfo {volume} {108}},\ \bibinfo
  {pages} {124038} (\bibinfo {year} {2023})},\ \Eprint
  {https://arxiv.org/abs/2303.18026} {arXiv:2303.18026 [gr-qc]} \BibitemShut
  {NoStop}%
\bibitem [{\citenamefont {Albertini}\ \emph
  {et~al.}(2024{\natexlab{a}})\citenamefont {Albertini}, \citenamefont {Gamba},
  \citenamefont {Nagar},\ and\ \citenamefont {Bernuzzi}}]{Albertini:2023aol}%
  \BibitemOpen
  \bibfield  {author} {\bibinfo {author} {\bibfnamefont {A.}~\bibnamefont
  {Albertini}}, \bibinfo {author} {\bibfnamefont {R.}~\bibnamefont {Gamba}},
  \bibinfo {author} {\bibfnamefont {A.}~\bibnamefont {Nagar}},\ and\ \bibinfo
  {author} {\bibfnamefont {S.}~\bibnamefont {Bernuzzi}},\ }\bibfield  {title}
  {\bibinfo {title} {{Effective-one-body waveforms for extreme-mass-ratio
  binaries: Consistency with second-order gravitational self-force
  quasicircular results and extension to nonprecessing spins and
  eccentricity}},\ }\href {https://doi.org/10.1103/PhysRevD.109.044022}
  {\bibfield  {journal} {\bibinfo  {journal} {Phys. Rev. D}\ }\textbf {\bibinfo
  {volume} {109}},\ \bibinfo {pages} {044022} (\bibinfo {year}
  {2024}{\natexlab{a}})},\ \Eprint {https://arxiv.org/abs/2310.13578}
  {arXiv:2310.13578 [gr-qc]} \BibitemShut {NoStop}%
\bibitem [{\citenamefont {Albertini}\ \emph
  {et~al.}(2024{\natexlab{b}})\citenamefont {Albertini}, \citenamefont {Nagar},
  \citenamefont {Mathews},\ and\ \citenamefont
  {Lukes-Gerakopoulos}}]{Albertini:2024rrs}%
  \BibitemOpen
  \bibfield  {author} {\bibinfo {author} {\bibfnamefont {A.}~\bibnamefont
  {Albertini}}, \bibinfo {author} {\bibfnamefont {A.}~\bibnamefont {Nagar}},
  \bibinfo {author} {\bibfnamefont {J.}~\bibnamefont {Mathews}},\ and\ \bibinfo
  {author} {\bibfnamefont {G.}~\bibnamefont {Lukes-Gerakopoulos}},\ }\bibfield
  {title} {\bibinfo {title} {{Comparing second-order gravitational self-force
  and effective-one-body waveforms from inspiralling, quasicircular black hole
  binaries with a nonspinning primary and a spinning secondary}},\ }\href
  {https://doi.org/10.1103/PhysRevD.110.044034} {\bibfield  {journal} {\bibinfo
   {journal} {Phys. Rev. D}\ }\textbf {\bibinfo {volume} {110}},\ \bibinfo
  {pages} {044034} (\bibinfo {year} {2024}{\natexlab{b}})},\ \Eprint
  {https://arxiv.org/abs/2406.04108} {arXiv:2406.04108 [gr-qc]} \BibitemShut
  {NoStop}%
\bibitem [{\citenamefont {Faggioli}\ \emph {et~al.}(2024)\citenamefont
  {Faggioli}, \citenamefont {van~de Meent}, \citenamefont {Buonanno},
  \citenamefont {Gamboa}, \citenamefont {Khalil},\ and\ \citenamefont
  {Khanna}}]{Faggioli:2024ugn}%
  \BibitemOpen
  \bibfield  {author} {\bibinfo {author} {\bibfnamefont {G.}~\bibnamefont
  {Faggioli}}, \bibinfo {author} {\bibfnamefont {M.}~\bibnamefont {van~de
  Meent}}, \bibinfo {author} {\bibfnamefont {A.}~\bibnamefont {Buonanno}},
  \bibinfo {author} {\bibfnamefont {A.}~\bibnamefont {Gamboa}}, \bibinfo
  {author} {\bibfnamefont {M.}~\bibnamefont {Khalil}},\ and\ \bibinfo {author}
  {\bibfnamefont {G.}~\bibnamefont {Khanna}},\ }\bibfield  {title} {\bibinfo
  {title} {{Testing eccentric corrections to the radiation-reaction force in
  the test-mass limit of effective-one-body models}},\ }\href@noop {} {\
  (\bibinfo {year} {2024})},\ \Eprint {https://arxiv.org/abs/2405.19006}
  {arXiv:2405.19006 [gr-qc]} \BibitemShut {NoStop}%
\bibitem [{\citenamefont {Albanesi}(2024)}]{Albanesi:2024fts}%
  \BibitemOpen
  \bibfield  {author} {\bibinfo {author} {\bibfnamefont {S.}~\bibnamefont
  {Albanesi}},\ }\bibfield  {title} {\bibinfo {title} {{Real modes and null
  memory contributions in effective-one-body models}},\ }\href@noop {} {\
  (\bibinfo {year} {2024})},\ \Eprint {https://arxiv.org/abs/2411.04024}
  {arXiv:2411.04024 [gr-qc]} \BibitemShut {NoStop}%
\bibitem [{\citenamefont {Leather}\ \emph {et~al.}(2025)\citenamefont
  {Leather}, \citenamefont {Buonanno},\ and\ \citenamefont {van~de
  Meent}}]{Leather:2025nhu}%
  \BibitemOpen
  \bibfield  {author} {\bibinfo {author} {\bibfnamefont {B.}~\bibnamefont
  {Leather}}, \bibinfo {author} {\bibfnamefont {A.}~\bibnamefont {Buonanno}},\
  and\ \bibinfo {author} {\bibfnamefont {M.}~\bibnamefont {van~de Meent}},\
  }\bibfield  {title} {\bibinfo {title} {{Inspiral-merger-ringdown waveforms
  with gravitational self-force results within the effective-one-body
  formalism}},\ }\Eprint {https://arxiv.org/abs/2505.11242} {arXiv:2505.11242
  [gr-qc]} \BibitemShut {NoStop}%
\bibitem [{\citenamefont {Carullo}\ \emph {et~al.}(2024)\citenamefont
  {Carullo}, \citenamefont {Albanesi}, \citenamefont {Nagar}, \citenamefont
  {Gamba}, \citenamefont {Bernuzzi}, \citenamefont {Andrade},\ and\
  \citenamefont {Trenado}}]{Carullo:2023kvj}%
  \BibitemOpen
  \bibfield  {author} {\bibinfo {author} {\bibfnamefont {G.}~\bibnamefont
  {Carullo}}, \bibinfo {author} {\bibfnamefont {S.}~\bibnamefont {Albanesi}},
  \bibinfo {author} {\bibfnamefont {A.}~\bibnamefont {Nagar}}, \bibinfo
  {author} {\bibfnamefont {R.}~\bibnamefont {Gamba}}, \bibinfo {author}
  {\bibfnamefont {S.}~\bibnamefont {Bernuzzi}}, \bibinfo {author}
  {\bibfnamefont {T.}~\bibnamefont {Andrade}},\ and\ \bibinfo {author}
  {\bibfnamefont {J.}~\bibnamefont {Trenado}},\ }\bibfield  {title} {\bibinfo
  {title} {{Unveiling the Merger Structure of Black Hole Binaries in Generic
  Planar Orbits}},\ }\href {https://doi.org/10.1103/PhysRevLett.132.101401}
  {\bibfield  {journal} {\bibinfo  {journal} {Phys. Rev. Lett.}\ }\textbf
  {\bibinfo {volume} {132}},\ \bibinfo {pages} {101401} (\bibinfo {year}
  {2024})},\ \Eprint {https://arxiv.org/abs/2309.07228} {arXiv:2309.07228
  [gr-qc]} \BibitemShut {NoStop}%
\bibitem [{\citenamefont {Carullo}(2024)}]{Carullo:2024smg}%
  \BibitemOpen
  \bibfield  {author} {\bibinfo {author} {\bibfnamefont {G.}~\bibnamefont
  {Carullo}},\ }\bibfield  {title} {\bibinfo {title} {{Ringdown amplitudes of
  nonspinning eccentric binaries}},\ }\href
  {https://doi.org/10.1088/1475-7516/2024/10/061} {\bibfield  {journal}
  {\bibinfo  {journal} {JCAP}\ }\textbf {\bibinfo {volume} {10}},\ \bibinfo
  {pages} {061}},\ \Eprint {https://arxiv.org/abs/2406.19442} {arXiv:2406.19442
  [gr-qc]} \BibitemShut {NoStop}%
\bibitem [{\citenamefont {Healy}\ and\ \citenamefont
  {Lousto}(2022)}]{Healy:2022wdn}%
  \BibitemOpen
  \bibfield  {author} {\bibinfo {author} {\bibfnamefont {J.}~\bibnamefont
  {Healy}}\ and\ \bibinfo {author} {\bibfnamefont {C.~O.}\ \bibnamefont
  {Lousto}},\ }\bibfield  {title} {\bibinfo {title} {{Fourth RIT binary black
  hole simulations catalog: Extension to eccentric orbits}},\ }\href
  {https://doi.org/10.1103/PhysRevD.105.124010} {\bibfield  {journal} {\bibinfo
   {journal} {Phys. Rev. D}\ }\textbf {\bibinfo {volume} {105}},\ \bibinfo
  {pages} {124010} (\bibinfo {year} {2022})},\ \Eprint
  {https://arxiv.org/abs/2202.00018} {arXiv:2202.00018 [gr-qc]} \BibitemShut
  {NoStop}%
\bibitem [{\citenamefont {Chu}\ \emph {et~al.}(2009)\citenamefont {Chu},
  \citenamefont {Pfeiffer},\ and\ \citenamefont {Scheel}}]{Chu:2009md}%
  \BibitemOpen
  \bibfield  {author} {\bibinfo {author} {\bibfnamefont {T.}~\bibnamefont
  {Chu}}, \bibinfo {author} {\bibfnamefont {H.~P.}\ \bibnamefont {Pfeiffer}},\
  and\ \bibinfo {author} {\bibfnamefont {M.~A.}\ \bibnamefont {Scheel}},\
  }\bibfield  {title} {\bibinfo {title} {{High accuracy simulations of black
  hole binaries: Spins anti-aligned with the orbital angular momentum}},\
  }\href {https://doi.org/10.1103/PhysRevD.80.124051} {\bibfield  {journal}
  {\bibinfo  {journal} {Phys. Rev. D}\ }\textbf {\bibinfo {volume} {80}},\
  \bibinfo {pages} {124051} (\bibinfo {year} {2009})},\ \Eprint
  {https://arxiv.org/abs/0909.1313} {arXiv:0909.1313 [gr-qc]} \BibitemShut
  {NoStop}%
\bibitem [{\citenamefont {Lovelace}\ \emph {et~al.}(2011)\citenamefont
  {Lovelace}, \citenamefont {Scheel},\ and\ \citenamefont
  {Szilagyi}}]{Lovelace:2010ne}%
  \BibitemOpen
  \bibfield  {author} {\bibinfo {author} {\bibfnamefont {G.}~\bibnamefont
  {Lovelace}}, \bibinfo {author} {\bibfnamefont {M.~A.}\ \bibnamefont
  {Scheel}},\ and\ \bibinfo {author} {\bibfnamefont {B.}~\bibnamefont
  {Szilagyi}},\ }\bibfield  {title} {\bibinfo {title} {{Simulating merging
  binary black holes with nearly extremal spins}},\ }\href
  {https://doi.org/10.1103/PhysRevD.83.024010} {\bibfield  {journal} {\bibinfo
  {journal} {Phys. Rev. D}\ }\textbf {\bibinfo {volume} {83}},\ \bibinfo
  {pages} {024010} (\bibinfo {year} {2011})},\ \Eprint
  {https://arxiv.org/abs/1010.2777} {arXiv:1010.2777 [gr-qc]} \BibitemShut
  {NoStop}%
\bibitem [{\citenamefont {Lovelace}\ \emph {et~al.}(2012)\citenamefont
  {Lovelace}, \citenamefont {Boyle}, \citenamefont {Scheel},\ and\
  \citenamefont {Szilagyi}}]{Lovelace:2011nu}%
  \BibitemOpen
  \bibfield  {author} {\bibinfo {author} {\bibfnamefont {G.}~\bibnamefont
  {Lovelace}}, \bibinfo {author} {\bibfnamefont {M.}~\bibnamefont {Boyle}},
  \bibinfo {author} {\bibfnamefont {M.~A.}\ \bibnamefont {Scheel}},\ and\
  \bibinfo {author} {\bibfnamefont {B.}~\bibnamefont {Szilagyi}},\ }\bibfield
  {title} {\bibinfo {title} {{Accurate gravitational waveforms for
  binary-black-hole mergers with nearly extremal spins}},\ }\href
  {https://doi.org/10.1088/0264-9381/29/4/045003} {\bibfield  {journal}
  {\bibinfo  {journal} {Class. Quant. Grav.}\ }\textbf {\bibinfo {volume}
  {29}},\ \bibinfo {pages} {045003} (\bibinfo {year} {2012})},\ \Eprint
  {https://arxiv.org/abs/1110.2229} {arXiv:1110.2229 [gr-qc]} \BibitemShut
  {NoStop}%
\bibitem [{\citenamefont {Buchman}\ \emph {et~al.}(2012)\citenamefont
  {Buchman}, \citenamefont {Pfeiffer}, \citenamefont {Scheel},\ and\
  \citenamefont {Szilagyi}}]{Buchman:2012dw}%
  \BibitemOpen
  \bibfield  {author} {\bibinfo {author} {\bibfnamefont {L.~T.}\ \bibnamefont
  {Buchman}}, \bibinfo {author} {\bibfnamefont {H.~P.}\ \bibnamefont
  {Pfeiffer}}, \bibinfo {author} {\bibfnamefont {M.~A.}\ \bibnamefont
  {Scheel}},\ and\ \bibinfo {author} {\bibfnamefont {B.}~\bibnamefont
  {Szilagyi}},\ }\bibfield  {title} {\bibinfo {title} {{Simulations of
  non-equal mass black hole binaries with spectral methods}},\ }\href
  {https://doi.org/10.1103/PhysRevD.86.084033} {\bibfield  {journal} {\bibinfo
  {journal} {Phys. Rev. D}\ }\textbf {\bibinfo {volume} {86}},\ \bibinfo
  {pages} {084033} (\bibinfo {year} {2012})},\ \Eprint
  {https://arxiv.org/abs/1206.3015} {arXiv:1206.3015 [gr-qc]} \BibitemShut
  {NoStop}%
\bibitem [{\citenamefont {Hemberger}\ \emph {et~al.}(2013)\citenamefont
  {Hemberger}, \citenamefont {Lovelace}, \citenamefont {Loredo}, \citenamefont
  {Kidder}, \citenamefont {Scheel}, \citenamefont {Szil\'agyi}, \citenamefont
  {Taylor},\ and\ \citenamefont {Teukolsky}}]{Hemberger:2013hsa}%
  \BibitemOpen
  \bibfield  {author} {\bibinfo {author} {\bibfnamefont {D.~A.}\ \bibnamefont
  {Hemberger}}, \bibinfo {author} {\bibfnamefont {G.}~\bibnamefont {Lovelace}},
  \bibinfo {author} {\bibfnamefont {T.~J.}\ \bibnamefont {Loredo}}, \bibinfo
  {author} {\bibfnamefont {L.~E.}\ \bibnamefont {Kidder}}, \bibinfo {author}
  {\bibfnamefont {M.~A.}\ \bibnamefont {Scheel}}, \bibinfo {author}
  {\bibfnamefont {B.}~\bibnamefont {Szil\'agyi}}, \bibinfo {author}
  {\bibfnamefont {N.~W.}\ \bibnamefont {Taylor}},\ and\ \bibinfo {author}
  {\bibfnamefont {S.~A.}\ \bibnamefont {Teukolsky}},\ }\bibfield  {title}
  {\bibinfo {title} {{Final spin and radiated energy in numerical simulations
  of binary black holes with equal masses and equal, aligned or anti-aligned
  spins}},\ }\href {https://doi.org/10.1103/PhysRevD.88.064014} {\bibfield
  {journal} {\bibinfo  {journal} {Phys. Rev. D}\ }\textbf {\bibinfo {volume}
  {88}},\ \bibinfo {pages} {064014} (\bibinfo {year} {2013})},\ \Eprint
  {https://arxiv.org/abs/1305.5991} {arXiv:1305.5991 [gr-qc]} \BibitemShut
  {NoStop}%
\bibitem [{\citenamefont {Scheel}\ \emph {et~al.}(2015)\citenamefont {Scheel},
  \citenamefont {Giesler}, \citenamefont {Hemberger}, \citenamefont {Lovelace},
  \citenamefont {Kuper}, \citenamefont {Boyle}, \citenamefont {Szil\'agyi},\
  and\ \citenamefont {Kidder}}]{Scheel:2014ina}%
  \BibitemOpen
  \bibfield  {author} {\bibinfo {author} {\bibfnamefont {M.~A.}\ \bibnamefont
  {Scheel}}, \bibinfo {author} {\bibfnamefont {M.}~\bibnamefont {Giesler}},
  \bibinfo {author} {\bibfnamefont {D.~A.}\ \bibnamefont {Hemberger}}, \bibinfo
  {author} {\bibfnamefont {G.}~\bibnamefont {Lovelace}}, \bibinfo {author}
  {\bibfnamefont {K.}~\bibnamefont {Kuper}}, \bibinfo {author} {\bibfnamefont
  {M.}~\bibnamefont {Boyle}}, \bibinfo {author} {\bibfnamefont
  {B.}~\bibnamefont {Szil\'agyi}},\ and\ \bibinfo {author} {\bibfnamefont
  {L.~E.}\ \bibnamefont {Kidder}},\ }\bibfield  {title} {\bibinfo {title}
  {{Improved methods for simulating nearly extremal binary black holes}},\
  }\href {https://doi.org/10.1088/0264-9381/32/10/105009} {\bibfield  {journal}
  {\bibinfo  {journal} {Class. Quant. Grav.}\ }\textbf {\bibinfo {volume}
  {32}},\ \bibinfo {pages} {105009} (\bibinfo {year} {2015})},\ \Eprint
  {https://arxiv.org/abs/1412.1803} {arXiv:1412.1803 [gr-qc]} \BibitemShut
  {NoStop}%
\bibitem [{\citenamefont {Lovelace}\ \emph {et~al.}(2015)\citenamefont
  {Lovelace} \emph {et~al.}}]{Lovelace:2014twa}%
  \BibitemOpen
  \bibfield  {author} {\bibinfo {author} {\bibfnamefont {G.}~\bibnamefont
  {Lovelace}} \emph {et~al.},\ }\bibfield  {title} {\bibinfo {title} {{Nearly
  extremal apparent horizons in simulations of merging black holes}},\ }\href
  {https://doi.org/10.1088/0264-9381/32/6/065007} {\bibfield  {journal}
  {\bibinfo  {journal} {Class. Quant. Grav.}\ }\textbf {\bibinfo {volume}
  {32}},\ \bibinfo {pages} {065007} (\bibinfo {year} {2015})},\ \Eprint
  {https://arxiv.org/abs/1411.7297} {arXiv:1411.7297 [gr-qc]} \BibitemShut
  {NoStop}%
\bibitem [{\citenamefont {Mroue}\ \emph {et~al.}(2013)\citenamefont {Mroue}
  \emph {et~al.}}]{Mroue:2013xna}%
  \BibitemOpen
  \bibfield  {author} {\bibinfo {author} {\bibfnamefont {A.~H.}\ \bibnamefont
  {Mroue}} \emph {et~al.},\ }\bibfield  {title} {\bibinfo {title} {{Catalog of
  174 Binary Black Hole Simulations for Gravitational Wave Astronomy}},\ }\href
  {https://doi.org/10.1103/PhysRevLett.111.241104} {\bibfield  {journal}
  {\bibinfo  {journal} {Phys. Rev. Lett.}\ }\textbf {\bibinfo {volume} {111}},\
  \bibinfo {pages} {241104} (\bibinfo {year} {2013})},\ \Eprint
  {https://arxiv.org/abs/1304.6077} {arXiv:1304.6077 [gr-qc]} \BibitemShut
  {NoStop}%
\bibitem [{\citenamefont {Kumar}\ \emph {et~al.}(2015)\citenamefont {Kumar},
  \citenamefont {Barkett}, \citenamefont {Bhagwat}, \citenamefont {Afshari},
  \citenamefont {Brown}, \citenamefont {Lovelace}, \citenamefont {Scheel},\
  and\ \citenamefont {Szil\'agyi}}]{Kumar:2015tha}%
  \BibitemOpen
  \bibfield  {author} {\bibinfo {author} {\bibfnamefont {P.}~\bibnamefont
  {Kumar}}, \bibinfo {author} {\bibfnamefont {K.}~\bibnamefont {Barkett}},
  \bibinfo {author} {\bibfnamefont {S.}~\bibnamefont {Bhagwat}}, \bibinfo
  {author} {\bibfnamefont {N.}~\bibnamefont {Afshari}}, \bibinfo {author}
  {\bibfnamefont {D.~A.}\ \bibnamefont {Brown}}, \bibinfo {author}
  {\bibfnamefont {G.}~\bibnamefont {Lovelace}}, \bibinfo {author}
  {\bibfnamefont {M.~A.}\ \bibnamefont {Scheel}},\ and\ \bibinfo {author}
  {\bibfnamefont {B.}~\bibnamefont {Szil\'agyi}},\ }\bibfield  {title}
  {\bibinfo {title} {{Accuracy and precision of gravitational-wave models of
  inspiraling neutron star-black hole binaries with spin: Comparison with
  matter-free numerical relativity in the low-frequency regime}},\ }\href
  {https://doi.org/10.1103/PhysRevD.92.102001} {\bibfield  {journal} {\bibinfo
  {journal} {Phys. Rev. D}\ }\textbf {\bibinfo {volume} {92}},\ \bibinfo
  {pages} {102001} (\bibinfo {year} {2015})},\ \Eprint
  {https://arxiv.org/abs/1507.00103} {arXiv:1507.00103 [gr-qc]} \BibitemShut
  {NoStop}%
\bibitem [{\citenamefont {Chu}\ \emph {et~al.}(2016)\citenamefont {Chu},
  \citenamefont {Fong}, \citenamefont {Kumar}, \citenamefont {Pfeiffer},
  \citenamefont {Boyle}, \citenamefont {Hemberger}, \citenamefont {Kidder},
  \citenamefont {Scheel},\ and\ \citenamefont {Szilagyi}}]{Chu:2015kft}%
  \BibitemOpen
  \bibfield  {author} {\bibinfo {author} {\bibfnamefont {T.}~\bibnamefont
  {Chu}}, \bibinfo {author} {\bibfnamefont {H.}~\bibnamefont {Fong}}, \bibinfo
  {author} {\bibfnamefont {P.}~\bibnamefont {Kumar}}, \bibinfo {author}
  {\bibfnamefont {H.~P.}\ \bibnamefont {Pfeiffer}}, \bibinfo {author}
  {\bibfnamefont {M.}~\bibnamefont {Boyle}}, \bibinfo {author} {\bibfnamefont
  {D.~A.}\ \bibnamefont {Hemberger}}, \bibinfo {author} {\bibfnamefont {L.~E.}\
  \bibnamefont {Kidder}}, \bibinfo {author} {\bibfnamefont {M.~A.}\
  \bibnamefont {Scheel}},\ and\ \bibinfo {author} {\bibfnamefont
  {B.}~\bibnamefont {Szilagyi}},\ }\bibfield  {title} {\bibinfo {title} {{On
  the accuracy and precision of numerical waveforms: Effect of waveform
  extraction methodology}},\ }\href
  {https://doi.org/10.1088/0264-9381/33/16/165001} {\bibfield  {journal}
  {\bibinfo  {journal} {Class. Quant. Grav.}\ }\textbf {\bibinfo {volume}
  {33}},\ \bibinfo {pages} {165001} (\bibinfo {year} {2016})},\ \Eprint
  {https://arxiv.org/abs/1512.06800} {arXiv:1512.06800 [gr-qc]} \BibitemShut
  {NoStop}%
\bibitem [{\citenamefont {Boyle}\ \emph {et~al.}(2019)\citenamefont {Boyle}
  \emph {et~al.}}]{Boyle:2019kee}%
  \BibitemOpen
  \bibfield  {author} {\bibinfo {author} {\bibfnamefont {M.}~\bibnamefont
  {Boyle}} \emph {et~al.},\ }\bibfield  {title} {\bibinfo {title} {{The SXS
  Collaboration catalog of binary black hole simulations}},\ }\href
  {https://doi.org/10.1088/1361-6382/ab34e2} {\bibfield  {journal} {\bibinfo
  {journal} {Class. Quant. Grav.}\ }\textbf {\bibinfo {volume} {36}},\ \bibinfo
  {pages} {195006} (\bibinfo {year} {2019})},\ \Eprint
  {https://arxiv.org/abs/1904.04831} {arXiv:1904.04831 [gr-qc]} \BibitemShut
  {NoStop}%
\bibitem [{\citenamefont {Nee}\ \emph {et~al.}(2025)\citenamefont {Nee} \emph
  {et~al.}}]{Nee:2025zdy}%
  \BibitemOpen
  \bibfield  {author} {\bibinfo {author} {\bibfnamefont {P.~J.}\ \bibnamefont
  {Nee}} \emph {et~al.},\ }\bibfield  {title} {\bibinfo {title} {{Impact of
  eccentricity and mean anomaly in numerical relativity mergers}},\ }\Eprint
  {https://arxiv.org/abs/2503.05422} {arXiv:2503.05422 [gr-qc]} \BibitemShut
  {NoStop}%
\bibitem [{\citenamefont {Becker}\ and\ \citenamefont
  {Hughes}(2024)}]{Becker:2024xdi}%
  \BibitemOpen
  \bibfield  {author} {\bibinfo {author} {\bibfnamefont {D.~R.}\ \bibnamefont
  {Becker}}\ and\ \bibinfo {author} {\bibfnamefont {S.~A.}\ \bibnamefont
  {Hughes}},\ }\bibfield  {title} {\bibinfo {title} {{Transition from adiabatic
  inspiral to plunge for eccentric binaries}},\ }\href@noop {} {\  (\bibinfo
  {year} {2024})},\ \Eprint {https://arxiv.org/abs/2410.09160}
  {arXiv:2410.09160 [gr-qc]} \BibitemShut {NoStop}%
\bibitem [{\citenamefont {Hughes}\ \emph {et~al.}(2021)\citenamefont {Hughes},
  \citenamefont {Warburton}, \citenamefont {Khanna}, \citenamefont {Chua},\
  and\ \citenamefont {Katz}}]{Hughes:2021exa}%
  \BibitemOpen
  \bibfield  {author} {\bibinfo {author} {\bibfnamefont {S.~A.}\ \bibnamefont
  {Hughes}}, \bibinfo {author} {\bibfnamefont {N.}~\bibnamefont {Warburton}},
  \bibinfo {author} {\bibfnamefont {G.}~\bibnamefont {Khanna}}, \bibinfo
  {author} {\bibfnamefont {A.~J.~K.}\ \bibnamefont {Chua}},\ and\ \bibinfo
  {author} {\bibfnamefont {M.~L.}\ \bibnamefont {Katz}},\ }\bibfield  {title}
  {\bibinfo {title} {{Adiabatic waveforms for extreme mass-ratio inspirals via
  multivoice decomposition in time and frequency}},\ }\href
  {https://doi.org/10.1103/PhysRevD.103.104014} {\bibfield  {journal} {\bibinfo
   {journal} {Phys. Rev. D}\ }\textbf {\bibinfo {volume} {103}},\ \bibinfo
  {pages} {104014} (\bibinfo {year} {2021})},\ \bibinfo {note} {[Erratum:
  Phys.Rev.D 107, 089901 (2023)]},\ \Eprint {https://arxiv.org/abs/2102.02713}
  {arXiv:2102.02713 [gr-qc]} \BibitemShut {NoStop}%
\bibitem [{\citenamefont {Baker}\ \emph {et~al.}(2008)\citenamefont {Baker},
  \citenamefont {Boggs}, \citenamefont {Centrella}, \citenamefont {Kelly},
  \citenamefont {McWilliams},\ and\ \citenamefont {van Meter}}]{Baker:2008mj}%
  \BibitemOpen
  \bibfield  {author} {\bibinfo {author} {\bibfnamefont {J.~G.}\ \bibnamefont
  {Baker}}, \bibinfo {author} {\bibfnamefont {W.~D.}\ \bibnamefont {Boggs}},
  \bibinfo {author} {\bibfnamefont {J.}~\bibnamefont {Centrella}}, \bibinfo
  {author} {\bibfnamefont {B.~J.}\ \bibnamefont {Kelly}}, \bibinfo {author}
  {\bibfnamefont {S.~T.}\ \bibnamefont {McWilliams}},\ and\ \bibinfo {author}
  {\bibfnamefont {J.~R.}\ \bibnamefont {van Meter}},\ }\bibfield  {title}
  {\bibinfo {title} {{Mergers of non-spinning black-hole binaries:
  Gravitational radiation characteristics}},\ }\href
  {https://doi.org/10.1103/PhysRevD.78.044046} {\bibfield  {journal} {\bibinfo
  {journal} {Phys. Rev. D}\ }\textbf {\bibinfo {volume} {78}},\ \bibinfo
  {pages} {044046} (\bibinfo {year} {2008})},\ \Eprint
  {https://arxiv.org/abs/0805.1428} {arXiv:0805.1428 [gr-qc]} \BibitemShut
  {NoStop}%
\bibitem [{\citenamefont {Kelly}\ \emph {et~al.}(2011)\citenamefont {Kelly},
  \citenamefont {Baker}, \citenamefont {Boggs}, \citenamefont {McWilliams},\
  and\ \citenamefont {Centrella}}]{Kelly:2011bp}%
  \BibitemOpen
  \bibfield  {author} {\bibinfo {author} {\bibfnamefont {B.~J.}\ \bibnamefont
  {Kelly}}, \bibinfo {author} {\bibfnamefont {J.~G.}\ \bibnamefont {Baker}},
  \bibinfo {author} {\bibfnamefont {W.~D.}\ \bibnamefont {Boggs}}, \bibinfo
  {author} {\bibfnamefont {S.~T.}\ \bibnamefont {McWilliams}},\ and\ \bibinfo
  {author} {\bibfnamefont {J.}~\bibnamefont {Centrella}},\ }\bibfield  {title}
  {\bibinfo {title} {{Mergers of black-hole binaries with aligned spins:
  Waveform characteristics}},\ }\href
  {https://doi.org/10.1103/PhysRevD.84.084009} {\bibfield  {journal} {\bibinfo
  {journal} {Phys. Rev. D}\ }\textbf {\bibinfo {volume} {84}},\ \bibinfo
  {pages} {084009} (\bibinfo {year} {2011})},\ \Eprint
  {https://arxiv.org/abs/1107.1181} {arXiv:1107.1181 [gr-qc]} \BibitemShut
  {NoStop}%
\bibitem [{\citenamefont {Healy}\ \emph {et~al.}(2014)\citenamefont {Healy},
  \citenamefont {Laguna},\ and\ \citenamefont {Shoemaker}}]{Healy:2014eua}%
  \BibitemOpen
  \bibfield  {author} {\bibinfo {author} {\bibfnamefont {J.}~\bibnamefont
  {Healy}}, \bibinfo {author} {\bibfnamefont {P.}~\bibnamefont {Laguna}},\ and\
  \bibinfo {author} {\bibfnamefont {D.}~\bibnamefont {Shoemaker}},\ }\bibfield
  {title} {\bibinfo {title} {{Decoding the final state in binary black hole
  mergers}},\ }\href {https://doi.org/10.1088/0264-9381/31/21/212001}
  {\bibfield  {journal} {\bibinfo  {journal} {Class. Quant. Grav.}\ }\textbf
  {\bibinfo {volume} {31}},\ \bibinfo {pages} {212001} (\bibinfo {year}
  {2014})},\ \Eprint {https://arxiv.org/abs/1407.5989} {arXiv:1407.5989
  [gr-qc]} \BibitemShut {NoStop}%
\bibitem [{\citenamefont {Healy}\ \emph {et~al.}(2017)\citenamefont {Healy},
  \citenamefont {Lousto},\ and\ \citenamefont {Zlochower}}]{Healy:2017mvh}%
  \BibitemOpen
  \bibfield  {author} {\bibinfo {author} {\bibfnamefont {J.}~\bibnamefont
  {Healy}}, \bibinfo {author} {\bibfnamefont {C.~O.}\ \bibnamefont {Lousto}},\
  and\ \bibinfo {author} {\bibfnamefont {Y.}~\bibnamefont {Zlochower}},\
  }\bibfield  {title} {\bibinfo {title} {{Nonspinning binary black hole merger
  scenario revisited}},\ }\href {https://doi.org/10.1103/PhysRevD.96.024031}
  {\bibfield  {journal} {\bibinfo  {journal} {Phys. Rev. D}\ }\textbf {\bibinfo
  {volume} {96}},\ \bibinfo {pages} {024031} (\bibinfo {year} {2017})},\
  \Eprint {https://arxiv.org/abs/1705.07034} {arXiv:1705.07034 [gr-qc]}
  \BibitemShut {NoStop}%
\bibitem [{\citenamefont {Healy}\ and\ \citenamefont
  {Lousto}(2017)}]{Healy:2016lce}%
  \BibitemOpen
  \bibfield  {author} {\bibinfo {author} {\bibfnamefont {J.}~\bibnamefont
  {Healy}}\ and\ \bibinfo {author} {\bibfnamefont {C.~O.}\ \bibnamefont
  {Lousto}},\ }\bibfield  {title} {\bibinfo {title} {{Remnant of binary
  black-hole mergers: New simulations and peak luminosity studies}},\ }\href
  {https://doi.org/10.1103/PhysRevD.95.024037} {\bibfield  {journal} {\bibinfo
  {journal} {Phys. Rev. D}\ }\textbf {\bibinfo {volume} {95}},\ \bibinfo
  {pages} {024037} (\bibinfo {year} {2017})},\ \Eprint
  {https://arxiv.org/abs/1610.09713} {arXiv:1610.09713 [gr-qc]} \BibitemShut
  {NoStop}%
\bibitem [{\citenamefont {Keitel}\ \emph {et~al.}(2017)\citenamefont {Keitel}
  \emph {et~al.}}]{Keitel:2016krm}%
  \BibitemOpen
  \bibfield  {author} {\bibinfo {author} {\bibfnamefont {D.}~\bibnamefont
  {Keitel}} \emph {et~al.},\ }\bibfield  {title} {\bibinfo {title} {{The most
  powerful astrophysical events: Gravitational-wave peak luminosity of binary
  black holes as predicted by numerical relativity}},\ }\href
  {https://doi.org/10.1103/PhysRevD.96.024006} {\bibfield  {journal} {\bibinfo
  {journal} {Phys. Rev. D}\ }\textbf {\bibinfo {volume} {96}},\ \bibinfo
  {pages} {024006} (\bibinfo {year} {2017})},\ \Eprint
  {https://arxiv.org/abs/1612.09566} {arXiv:1612.09566 [gr-qc]} \BibitemShut
  {NoStop}%
\bibitem [{\citenamefont {Gralla}\ \emph {et~al.}(2016)\citenamefont {Gralla},
  \citenamefont {Hughes},\ and\ \citenamefont {Warburton}}]{Gralla:2016qfw}%
  \BibitemOpen
  \bibfield  {author} {\bibinfo {author} {\bibfnamefont {S.~E.}\ \bibnamefont
  {Gralla}}, \bibinfo {author} {\bibfnamefont {S.~A.}\ \bibnamefont {Hughes}},\
  and\ \bibinfo {author} {\bibfnamefont {N.}~\bibnamefont {Warburton}},\
  }\bibfield  {title} {\bibinfo {title} {{Inspiral into Gargantua}},\ }\href
  {https://doi.org/10.1088/0264-9381/33/15/155002} {\bibfield  {journal}
  {\bibinfo  {journal} {Class. Quant. Grav.}\ }\textbf {\bibinfo {volume}
  {33}},\ \bibinfo {pages} {155002} (\bibinfo {year} {2016})},\ \bibinfo {note}
  {[Erratum: Class.Quant.Grav. 37, 109501 (2020)]},\ \Eprint
  {https://arxiv.org/abs/1603.01221} {arXiv:1603.01221 [gr-qc]} \BibitemShut
  {NoStop}%
\bibitem [{\citenamefont {Ori}\ and\ \citenamefont
  {Thorne}(2000)}]{Ori:2000zn}%
  \BibitemOpen
  \bibfield  {author} {\bibinfo {author} {\bibfnamefont {A.}~\bibnamefont
  {Ori}}\ and\ \bibinfo {author} {\bibfnamefont {K.~S.}\ \bibnamefont
  {Thorne}},\ }\bibfield  {title} {\bibinfo {title} {{The Transition from
  inspiral to plunge for a compact body in a circular equatorial orbit around a
  massive, spinning black hole}},\ }\href
  {https://doi.org/10.1103/PhysRevD.62.124022} {\bibfield  {journal} {\bibinfo
  {journal} {Phys. Rev. D}\ }\textbf {\bibinfo {volume} {62}},\ \bibinfo
  {pages} {124022} (\bibinfo {year} {2000})},\ \Eprint
  {https://arxiv.org/abs/gr-qc/0003032} {arXiv:gr-qc/0003032} \BibitemShut
  {NoStop}%
\bibitem [{\citenamefont {Lhost}\ and\ \citenamefont
  {Comp\`ere}(2024)}]{Lhost:2024jmw}%
  \BibitemOpen
  \bibfield  {author} {\bibinfo {author} {\bibfnamefont {G.}~\bibnamefont
  {Lhost}}\ and\ \bibinfo {author} {\bibfnamefont {G.}~\bibnamefont
  {Comp\`ere}},\ }\bibfield  {title} {\bibinfo {title} {{Approach to the
  separatrix with eccentric orbits}},\ }\Eprint
  {https://arxiv.org/abs/2412.04249} {arXiv:2412.04249 [gr-qc]} \BibitemShut
  {NoStop}%
\bibitem [{\citenamefont {Mummery}\ and\ \citenamefont
  {Balbus}(2023)}]{Mummery:2023hlo}%
  \BibitemOpen
  \bibfield  {author} {\bibinfo {author} {\bibfnamefont {A.}~\bibnamefont
  {Mummery}}\ and\ \bibinfo {author} {\bibfnamefont {S.}~\bibnamefont
  {Balbus}},\ }\bibfield  {title} {\bibinfo {title} {{Complete characterization
  of the orbital shapes of the noncircular Kerr geodesic solutions with
  circular orbit constants of motion}},\ }\href
  {https://doi.org/10.1103/PhysRevD.107.124058} {\bibfield  {journal} {\bibinfo
   {journal} {Phys. Rev. D}\ }\textbf {\bibinfo {volume} {107}},\ \bibinfo
  {pages} {124058} (\bibinfo {year} {2023})},\ \Eprint
  {https://arxiv.org/abs/2302.01159} {arXiv:2302.01159 [gr-qc]} \BibitemShut
  {NoStop}%
\bibitem [{\citenamefont {Dyson}\ and\ \citenamefont {van~de
  Meent}(2023)}]{Dyson:2023fws}%
  \BibitemOpen
  \bibfield  {author} {\bibinfo {author} {\bibfnamefont {C.}~\bibnamefont
  {Dyson}}\ and\ \bibinfo {author} {\bibfnamefont {M.}~\bibnamefont {van~de
  Meent}},\ }\bibfield  {title} {\bibinfo {title} {{Kerr-fully diving into the
  abyss: analytic solutions to plunging geodesics in Kerr}},\ }\href
  {https://doi.org/10.1088/1361-6382/acf552} {\bibfield  {journal} {\bibinfo
  {journal} {Class. Quant. Grav.}\ }\textbf {\bibinfo {volume} {40}},\ \bibinfo
  {pages} {195026} (\bibinfo {year} {2023})},\ \Eprint
  {https://arxiv.org/abs/2302.03704} {arXiv:2302.03704 [gr-qc]} \BibitemShut
  {NoStop}%
\bibitem [{\citenamefont {Sundararajan}\ \emph {et~al.}(2007)\citenamefont
  {Sundararajan}, \citenamefont {Khanna},\ and\ \citenamefont
  {Hughes}}]{Sundararajan:2007jg}%
  \BibitemOpen
  \bibfield  {author} {\bibinfo {author} {\bibfnamefont {P.~A.}\ \bibnamefont
  {Sundararajan}}, \bibinfo {author} {\bibfnamefont {G.}~\bibnamefont
  {Khanna}},\ and\ \bibinfo {author} {\bibfnamefont {S.~A.}\ \bibnamefont
  {Hughes}},\ }\bibfield  {title} {\bibinfo {title} {{Towards adiabatic
  waveforms for inspiral into Kerr black holes. I. A New model of the source
  for the time domain perturbation equation}},\ }\href
  {https://doi.org/10.1103/PhysRevD.76.104005} {\bibfield  {journal} {\bibinfo
  {journal} {Phys. Rev. D}\ }\textbf {\bibinfo {volume} {76}},\ \bibinfo
  {pages} {104005} (\bibinfo {year} {2007})},\ \Eprint
  {https://arxiv.org/abs/gr-qc/0703028} {arXiv:gr-qc/0703028} \BibitemShut
  {NoStop}%
\bibitem [{\citenamefont {Sundararajan}\ \emph {et~al.}(2008)\citenamefont
  {Sundararajan}, \citenamefont {Khanna}, \citenamefont {Hughes},\ and\
  \citenamefont {Drasco}}]{Sundararajan:2008zm}%
  \BibitemOpen
  \bibfield  {author} {\bibinfo {author} {\bibfnamefont {P.~A.}\ \bibnamefont
  {Sundararajan}}, \bibinfo {author} {\bibfnamefont {G.}~\bibnamefont
  {Khanna}}, \bibinfo {author} {\bibfnamefont {S.~A.}\ \bibnamefont {Hughes}},\
  and\ \bibinfo {author} {\bibfnamefont {S.}~\bibnamefont {Drasco}},\
  }\bibfield  {title} {\bibinfo {title} {{Towards adiabatic waveforms for
  inspiral into Kerr black holes: II. Dynamical sources and generic orbits}},\
  }\href {https://doi.org/10.1103/PhysRevD.78.024022} {\bibfield  {journal}
  {\bibinfo  {journal} {Phys. Rev. D}\ }\textbf {\bibinfo {volume} {78}},\
  \bibinfo {pages} {024022} (\bibinfo {year} {2008})},\ \Eprint
  {https://arxiv.org/abs/0803.0317} {arXiv:0803.0317 [gr-qc]} \BibitemShut
  {NoStop}%
\bibitem [{\citenamefont {Sundararajan}\ \emph {et~al.}(2010)\citenamefont
  {Sundararajan}, \citenamefont {Khanna},\ and\ \citenamefont
  {Hughes}}]{Sundararajan:2010sr}%
  \BibitemOpen
  \bibfield  {author} {\bibinfo {author} {\bibfnamefont {P.~A.}\ \bibnamefont
  {Sundararajan}}, \bibinfo {author} {\bibfnamefont {G.}~\bibnamefont
  {Khanna}},\ and\ \bibinfo {author} {\bibfnamefont {S.~A.}\ \bibnamefont
  {Hughes}},\ }\bibfield  {title} {\bibinfo {title} {{Binary black hole merger
  gravitational waves and recoil in the large mass ratio limit}},\ }\href
  {https://doi.org/10.1103/PhysRevD.81.104009} {\bibfield  {journal} {\bibinfo
  {journal} {Phys. Rev. D}\ }\textbf {\bibinfo {volume} {81}},\ \bibinfo
  {pages} {104009} (\bibinfo {year} {2010})},\ \Eprint
  {https://arxiv.org/abs/1003.0485} {arXiv:1003.0485 [gr-qc]} \BibitemShut
  {NoStop}%
\bibitem [{\citenamefont {Zenginoglu}\ and\ \citenamefont
  {Khanna}(2011)}]{Zenginoglu:2011zz}%
  \BibitemOpen
  \bibfield  {author} {\bibinfo {author} {\bibfnamefont {A.}~\bibnamefont
  {Zenginoglu}}\ and\ \bibinfo {author} {\bibfnamefont {G.}~\bibnamefont
  {Khanna}},\ }\bibfield  {title} {\bibinfo {title} {{Null infinity waveforms
  from extreme-mass-ratio inspirals in Kerr spacetime}},\ }\href
  {https://doi.org/10.1103/PhysRevX.1.021017} {\bibfield  {journal} {\bibinfo
  {journal} {Phys. Rev. X}\ }\textbf {\bibinfo {volume} {1}},\ \bibinfo {pages}
  {021017} (\bibinfo {year} {2011})},\ \Eprint
  {https://arxiv.org/abs/1108.1816} {arXiv:1108.1816 [gr-qc]} \BibitemShut
  {NoStop}%
\bibitem [{\citenamefont {Field}\ \emph {et~al.}(2023)\citenamefont {Field},
  \citenamefont {Gottlieb}, \citenamefont {Grant}, \citenamefont {Isherwood},\
  and\ \citenamefont {Khanna}}]{Field:2020rjr}%
  \BibitemOpen
  \bibfield  {author} {\bibinfo {author} {\bibfnamefont {S.~E.}\ \bibnamefont
  {Field}}, \bibinfo {author} {\bibfnamefont {S.}~\bibnamefont {Gottlieb}},
  \bibinfo {author} {\bibfnamefont {Z.~J.}\ \bibnamefont {Grant}}, \bibinfo
  {author} {\bibfnamefont {L.~F.}\ \bibnamefont {Isherwood}},\ and\ \bibinfo
  {author} {\bibfnamefont {G.}~\bibnamefont {Khanna}},\ }\bibfield  {title}
  {\bibinfo {title} {{A GPU-accelerated mixed-precision WENO method for
  extremal black hole and gravitational wave physics computations}},\ }\href
  {https://doi.org/10.1007/s42967-021-00129-2} {\bibfield  {journal} {\bibinfo
  {journal} {Appl. Math. Comput.}\ }\textbf {\bibinfo {volume} {5}},\ \bibinfo
  {pages} {97} (\bibinfo {year} {2023})},\ \Eprint
  {https://arxiv.org/abs/2010.04760} {arXiv:2010.04760 [math.NA]} \BibitemShut
  {NoStop}%
\bibitem [{\citenamefont {Kerr}(1963)}]{Kerr:1963ud}%
  \BibitemOpen
  \bibfield  {author} {\bibinfo {author} {\bibfnamefont {R.~P.}\ \bibnamefont
  {Kerr}},\ }\bibfield  {title} {\bibinfo {title} {{Gravitational field of a
  spinning mass as an example of algebraically special metrics}},\ }\href
  {https://doi.org/10.1103/PhysRevLett.11.237} {\bibfield  {journal} {\bibinfo
  {journal} {Phys. Rev. Lett.}\ }\textbf {\bibinfo {volume} {11}},\ \bibinfo
  {pages} {237} (\bibinfo {year} {1963})}\BibitemShut {NoStop}%
\bibitem [{\citenamefont {Carter}(1968)}]{Carter:1968rr}%
  \BibitemOpen
  \bibfield  {author} {\bibinfo {author} {\bibfnamefont {B.}~\bibnamefont
  {Carter}},\ }\bibfield  {title} {\bibinfo {title} {{Global structure of the
  Kerr family of gravitational fields}},\ }\href
  {https://doi.org/10.1103/PhysRev.174.1559} {\bibfield  {journal} {\bibinfo
  {journal} {Phys. Rev.}\ }\textbf {\bibinfo {volume} {174}},\ \bibinfo {pages}
  {1559} (\bibinfo {year} {1968})}\BibitemShut {NoStop}%
\bibitem [{\citenamefont {Fujita}\ and\ \citenamefont
  {Hikida}(2009)}]{Fujita:2009bp}%
  \BibitemOpen
  \bibfield  {author} {\bibinfo {author} {\bibfnamefont {R.}~\bibnamefont
  {Fujita}}\ and\ \bibinfo {author} {\bibfnamefont {W.}~\bibnamefont
  {Hikida}},\ }\bibfield  {title} {\bibinfo {title} {{Analytical solutions of
  bound timelike geodesic orbits in Kerr spacetime}},\ }\href
  {https://doi.org/10.1088/0264-9381/26/13/135002} {\bibfield  {journal}
  {\bibinfo  {journal} {Class. Quant. Grav.}\ }\textbf {\bibinfo {volume}
  {26}},\ \bibinfo {pages} {135002} (\bibinfo {year} {2009})},\ \Eprint
  {https://arxiv.org/abs/0906.1420} {arXiv:0906.1420 [gr-qc]} \BibitemShut
  {NoStop}%
\bibitem [{Dar(1959)}]{Darwin:1959}%
  \BibitemOpen
  \bibfield  {title} {\bibinfo {title} {The gravity field of a particle},\
  }\href {https://doi.org/10.1098/rspa.1959.0015} {\bibfield  {journal}
  {\bibinfo  {journal} {Proceedings of the Royal Society of London A:
  Mathematical, Physical and Engineering Sciences}\ }\textbf {\bibinfo {volume}
  {249}},\ \bibinfo {pages} {180} (\bibinfo {year} {1959})}\BibitemShut
  {NoStop}%
\bibitem [{Dar(1961)}]{Darwin:1961}%
  \BibitemOpen
  \bibfield  {title} {\bibinfo {title} {The gravity field of a particle. ii},\
  }\href {https://doi.org/10.1098/rspa.1961.0142} {\bibfield  {journal}
  {\bibinfo  {journal} {Proceedings of the Royal Society of London A:
  Mathematical, Physical and Engineering Sciences}\ }\textbf {\bibinfo {volume}
  {263}},\ \bibinfo {pages} {39} (\bibinfo {year} {1961})}\BibitemShut
  {NoStop}%
\bibitem [{\citenamefont {Levin}\ and\ \citenamefont
  {Perez-Giz}(2009)}]{Levin:2008yp}%
  \BibitemOpen
  \bibfield  {author} {\bibinfo {author} {\bibfnamefont {J.}~\bibnamefont
  {Levin}}\ and\ \bibinfo {author} {\bibfnamefont {G.}~\bibnamefont
  {Perez-Giz}},\ }\bibfield  {title} {\bibinfo {title} {{Homoclinic Orbits
  around Spinning Black Holes. I. Exact Solution for the Kerr Separatrix}},\
  }\href {https://doi.org/10.1103/PhysRevD.79.124013} {\bibfield  {journal}
  {\bibinfo  {journal} {Phys. Rev. D}\ }\textbf {\bibinfo {volume} {79}},\
  \bibinfo {pages} {124013} (\bibinfo {year} {2009})},\ \Eprint
  {https://arxiv.org/abs/0811.3814} {arXiv:0811.3814 [gr-qc]} \BibitemShut
  {NoStop}%
\bibitem [{\citenamefont {Perez-Giz}\ and\ \citenamefont
  {Levin}(2009)}]{Perez-Giz:2008ajn}%
  \BibitemOpen
  \bibfield  {author} {\bibinfo {author} {\bibfnamefont {G.}~\bibnamefont
  {Perez-Giz}}\ and\ \bibinfo {author} {\bibfnamefont {J.}~\bibnamefont
  {Levin}},\ }\bibfield  {title} {\bibinfo {title} {{Homoclinic Orbits around
  Spinning Black Holes II: The Phase Space Portrait}},\ }\href
  {https://doi.org/10.1103/PhysRevD.79.124014} {\bibfield  {journal} {\bibinfo
  {journal} {Phys. Rev. D}\ }\textbf {\bibinfo {volume} {79}},\ \bibinfo
  {pages} {124014} (\bibinfo {year} {2009})},\ \Eprint
  {https://arxiv.org/abs/0811.3815} {arXiv:0811.3815 [gr-qc]} \BibitemShut
  {NoStop}%
\bibitem [{\citenamefont {Bardeen}\ \emph {et~al.}(1972)\citenamefont
  {Bardeen}, \citenamefont {Press},\ and\ \citenamefont
  {Teukolsky}}]{Bardeen:1972fi}%
  \BibitemOpen
  \bibfield  {author} {\bibinfo {author} {\bibfnamefont {J.~M.}\ \bibnamefont
  {Bardeen}}, \bibinfo {author} {\bibfnamefont {W.~H.}\ \bibnamefont {Press}},\
  and\ \bibinfo {author} {\bibfnamefont {S.~A.}\ \bibnamefont {Teukolsky}},\
  }\bibfield  {title} {\bibinfo {title} {{Rotating black holes: Locally
  nonrotating frames, energy extraction, and scalar synchrotron radiation}},\
  }\href {https://doi.org/10.1086/151796} {\bibfield  {journal} {\bibinfo
  {journal} {Astrophys. J.}\ }\textbf {\bibinfo {volume} {178}},\ \bibinfo
  {pages} {347} (\bibinfo {year} {1972})}\BibitemShut {NoStop}%
\bibitem [{\citenamefont {Thorne}(1980)}]{Thorne:1980ru}%
  \BibitemOpen
  \bibfield  {author} {\bibinfo {author} {\bibfnamefont {K.~S.}\ \bibnamefont
  {Thorne}},\ }\bibfield  {title} {\bibinfo {title} {{Multipole Expansions of
  Gravitational Radiation}},\ }\href
  {https://doi.org/10.1103/RevModPhys.52.299} {\bibfield  {journal} {\bibinfo
  {journal} {Rev. Mod. Phys.}\ }\textbf {\bibinfo {volume} {52}},\ \bibinfo
  {pages} {299} (\bibinfo {year} {1980})}\BibitemShut {NoStop}%
\bibitem [{\citenamefont {Blanchet}(1998)}]{Blanchet:1998in}%
  \BibitemOpen
  \bibfield  {author} {\bibinfo {author} {\bibfnamefont {L.}~\bibnamefont
  {Blanchet}},\ }\bibfield  {title} {\bibinfo {title} {{On the multipole
  expansion of the gravitational field}},\ }\href
  {https://doi.org/10.1088/0264-9381/15/7/013} {\bibfield  {journal} {\bibinfo
  {journal} {Class. Quant. Grav.}\ }\textbf {\bibinfo {volume} {15}},\ \bibinfo
  {pages} {1971} (\bibinfo {year} {1998})},\ \Eprint
  {https://arxiv.org/abs/gr-qc/9801101} {arXiv:gr-qc/9801101} \BibitemShut
  {NoStop}%
\bibitem [{\citenamefont {Krivan}\ \emph {et~al.}(1997)\citenamefont {Krivan},
  \citenamefont {Laguna}, \citenamefont {Papadopoulos},\ and\ \citenamefont
  {Andersson}}]{Krivan:1997hc}%
  \BibitemOpen
  \bibfield  {author} {\bibinfo {author} {\bibfnamefont {W.}~\bibnamefont
  {Krivan}}, \bibinfo {author} {\bibfnamefont {P.}~\bibnamefont {Laguna}},
  \bibinfo {author} {\bibfnamefont {P.}~\bibnamefont {Papadopoulos}},\ and\
  \bibinfo {author} {\bibfnamefont {N.}~\bibnamefont {Andersson}},\ }\bibfield
  {title} {\bibinfo {title} {{Dynamics of perturbations of rotating black
  holes}},\ }\href {https://doi.org/10.1103/PhysRevD.56.3395} {\bibfield
  {journal} {\bibinfo  {journal} {Phys. Rev. D}\ }\textbf {\bibinfo {volume}
  {56}},\ \bibinfo {pages} {3395} (\bibinfo {year} {1997})},\ \Eprint
  {https://arxiv.org/abs/gr-qc/9702048} {arXiv:gr-qc/9702048} \BibitemShut
  {NoStop}%
\bibitem [{BHP()}]{BHPToolkit}%
  \BibitemOpen
  \href@noop {} {\bibinfo {title} {{Black Hole Perturbation Toolkit}}},\
  \bibinfo {howpublished}
  {(\href{http://bhptoolkit.org/}{bhptoolkit.org})}\BibitemShut {NoStop}%
\bibitem [{\citenamefont {Gundlach}\ \emph {et~al.}(2012)\citenamefont
  {Gundlach}, \citenamefont {Akcay}, \citenamefont {Barack},\ and\
  \citenamefont {Nagar}}]{Gundlach:2012aj}%
  \BibitemOpen
  \bibfield  {author} {\bibinfo {author} {\bibfnamefont {C.}~\bibnamefont
  {Gundlach}}, \bibinfo {author} {\bibfnamefont {S.}~\bibnamefont {Akcay}},
  \bibinfo {author} {\bibfnamefont {L.}~\bibnamefont {Barack}},\ and\ \bibinfo
  {author} {\bibfnamefont {A.}~\bibnamefont {Nagar}},\ }\bibfield  {title}
  {\bibinfo {title} {{Critical phenomena at the threshold of immediate merger
  in binary black hole systems: the extreme mass ratio case}},\ }\href
  {https://doi.org/10.1103/PhysRevD.86.084022} {\bibfield  {journal} {\bibinfo
  {journal} {Phys. Rev. D}\ }\textbf {\bibinfo {volume} {86}},\ \bibinfo
  {pages} {084022} (\bibinfo {year} {2012})},\ \Eprint
  {https://arxiv.org/abs/1207.5167} {arXiv:1207.5167 [gr-qc]} \BibitemShut
  {NoStop}%
\end{thebibliography}%
